\documentclass[%
 preprint, 
showpacs,preprintnumbers,
 amsmath,amssymb,
 aps,color,superscriptaddress
]{revtex4-1}
\usepackage[pdftex]{graphicx,color}
\graphicspath{{./Diagrams/}}
\usepackage{subfigure,
mathrsfs,braket,amsmath,amssymb,dcolumn,bm,comment}
\newlength{\subfigwidth}
\newlength{\subfigcolsep}
\setkeys{Gin}{width=\subfigwidth}
\makeatletter 
 
\@addtoreset{figure}{section} 
\makeatother 
\begin{document}
\preprint{HUPD1608}
\def\b{\begin{eqnarray}}
\def\e{\end{eqnarray}}
\def\aTr{{\rm Tr}}
\def\bpi{\bar{\pi}}
\def\tbr{\textcolor{red}}
\def\tcr{\textcolor{red}}
\def\ov{\overline}
\def\dprime{{\prime \prime}}
\def\nn{\nonumber}
\def\f{\frac}
\def\p{\partial}
\def\H{\mathcal{H}}
\def\beq{\begin{equation}}
\def\eeq{\end{equation}}
\def\bea{\begin{eqnarray}}
\def\eea{\end{eqnarray}}
\def\bsub{\begin{subequations}}
\def\esub{\end{subequations}}
\def\dc{\stackrel{\leftrightarrow}{\partial}}
\def\d{\partial}
\def\sla#1{\rlap/#1}
\def\mH{\mathscr{H}}
\def\tD{\tilde{D}}
\def\Q{{\cal Q}}
\def\mpim{m_{\pi^-}}
\def\mpi0{m_{\pi^0}}
\def\meta{m_\eta}
\def\tauepp{\tau^- \to \nu_\tau \eta \pi^- \pi^0}
\def\Vcurrent{\bar{d}\gamma_\mu u}
\def\Acurrent{\bar{d} \gamma_\mu \gamma_5 u}
\def\VmA{\bar{d}\gamma_\mu(1-\gamma_5)u}
\def\epp{\eta \pi^- \pi^0}
\def\hg{{G}_{2p}}
\def\IP{\mathrm{IP}}
\def\hc{\mathrm{h}.\mathrm{c}.}
\def\cM{\mathcal{M}}
%%%%%%    TEXT START    %%%%%%
\def\nn{\nonumber}
\def\beq{\begin{equation}}
\def\eeq{\end{equation}}
\def\bei{\begin{itemize}}
\def\eei{\end{itemize}}
\def\bea{\begin{eqnarray}}
\def\eea{\end{eqnarray}}
\def\s{\partial \hspace{-.47em}/}
\def\ad{\overleftrightarrow{\partial}}
\def\para{%
\setlength{\unitlength}{1pt}%
\thinlines %
\begin{picture}(12, 12)%
\put(0,0){/}
\put(2,0){/}
\end{picture}%
}%
%%%%%%%%%%%%%%%%%%
% inserted
%%%%%%%%%%%%%%%%%%
%\newlength{\subfigwidth}
%\newlength{\subfigcolsep}
%\setlength{\subfigcolsep}{2\tabcolsep}
%\setkeys{Gin}{width=\subfigwidth}
%\makeatletter 
%\renewcommand{\thefigure}{% 
%\thesection.\arabic{figure}} 
%\@addtoreset{figure}{section} 
%\makeatother
%%%%%%%%%%%%%%%%%
%%%%%%    TEXT START    %%%%%%
\title{
Analysis of Dalitz decays
with intrinsic parity violating\vspace{-2.0mm}
interactions in resonance chiral
perturbation theory
%%%%%%%%%%%%%%%
%%% version  %%%%%
%%%%%%%%%%%%%%%
}
\author{Daiji Kimura}
\email[E-mail: ]{kimurad@ube-k.ac.jp}
\affiliation{National Institute of Technology, Ube College, Ube Yamaguchi 755-8555, Japan}
\author{Takuya Morozumi}
\email[E-mail: ]{morozumi@hiroshima-u.ac.jp}
\affiliation{
Graduate  School of  Science, Hiroshima University, Higashi-Hiroshima 739-8526, Japan}
\affiliation{
Core  of  Research  for  the  Energetic Universe, Hiroshima University,
Higashi-Hiroshima 739-8526, Japan}
\author{Hiroyuki Umeeda}
\email[E-mail: ]{umeeda@gate.sinica.edu.tw}
\affiliation{
Graduate  School of  Science, Hiroshima University, Higashi-Hiroshima 739-8526, Japan}
\affiliation{
Institute of Science and Engineering, Shimane University, Matsue 690-8504, Japan}
\affiliation{
Institute of Physics, Academia Sinica, Nangang, Taipei 11529, Taiwan}
\date{\today}
%%%%%%%%%%%%%%%
%   Abst %%%%%%%%%
%%%%%%%%%%%%%%%
\begin{abstract}
\vspace{-0.5mm}
Observables of light hadron decays
are analyzed in a model of chiral Lagrangian
which includes resonance fields of vector mesons.
In particular, transition form factors
are investigated for Dalitz decays of  $V\to Pl^+l^-$
and $P\to \gamma l^+l^-$
$(V=1^-, P=0^-)$.
Moreover,
the differential decay width of $P\to\pi^+\pi^-\gamma$
and the partial widths of $P\to2\gamma, V\to P\gamma,
\eta^\prime\to V\gamma, \phi(1020)\to\omega(782)\pi^0$ and
$V\to 3P$ are also calculated.
In this study, we consider a model which
contains octet and singlet fields
as representation of SU(3).
As an extension of chiral perturbation theory,
we include 1-loop ordered interaction terms.
For both pseudoscalar and vector meson,
we evaluate mixing matrices in which
isospin/SU(3) breaking is taken into account.
Furthermore, intrinsic parity violating interactions
are considered with singlet fields.
For parameter estimation,
we carry out $\chi^2$ fittings in which
a spectral function of $\tau$ decays,
vector meson masses,
decay widths of $V\to P\gamma$ and
transition form factor of $V\to Pl^+l^-$ are
utilized as input data.
Using the estimated parameter region in the model,
we give predictions for decay widths
and transition form factors of intrinsic parity
violating decays.
As further model predictions,
we calculate the transition form factors of
$\phi(1020)\to\pi^0 l^+l^-$ and
$\eta^\prime(958)\to\gamma l^+l^-$
in the vicinity of resonance regions,
taking account of the contribution
for intermediate $\rho(770)$ and $\omega(782)$.
\end{abstract}
%%%%%%%%%%%%%%%
%%  Abst. end  %%%%
%%%%%%%%%%%%%%%
\maketitle
\tableofcontents
\clearpage
%%%%%%%%%%%%%%%%
%%%  Intro %%%%%%%%
%%%%%%%%%%%%%%%%
\section{Introduction}
Decays of light hadrons play a crucial role
to investigate low-energy behavior of
quantum chromodynamics (QCD),
and are measured extensively in experiments.
In particular, Dalitz decays such as $P\to\gamma l^+l^-$
and $V\to P l^+l^-$ provide rich resources as hadronic
observables.
Using these experimental data,
we can test the validity of QCD effective theories
which include resonances of vector meson.
As a recent result, high-precision data of transition form
factors (TFFs) for $\omega\to\pi^0 \mu^+\mu^-$ and
$\eta\to\gamma\mu^+\mu^-$
are measured by
the NA60 collaboration\ \cite{Arnaldi:2016pzu}
in proton-nucleus (p-A) collisions.
Moreover, the measurement of the branching ratio
and the TFF of $\eta^\prime\to\gamma e^+e^-$
has been carried out by
the BES III collaboration\ \cite{Ablikim:2015wnx}.
\par
In order to describe dynamics of light hadrons,
we adopt a model of chiral Lagrangian which
includes vector mesons.
In this model,
chiral octets and singlets are introduced
as representation of SU(3).
There are some models
\cite{Bando:1984ej, Ecker:1988te, Ecker:1989yg},
which incorporate vector mesons
and the other resonances.
In this study, we develop
the framework so that one can include
chiral correction to processes in which
vector mesons and/or pseudoscalars
are involved.
On the basis of
power counting of superficial degree of divergence,
1-loop order counter terms,
which correspond to $O(p^4)$,
are introduced \cite{Kimura:2014wsa}.
Finite parts of the coefficients of the counter terms
are estimated in the fitting procedure
with the experimental observables
in the same way as the chiral perturbation
theory (ChPT).
Once those parameters are determined,
one can predict other observables
such as TFFs and decay widths.
\par
This effective dynamics of hadrons
is applicable to a variety of phenomena,
{\it e.g.}, hadronic $\tau$ decays.
As experimental results,
a spectral function of $\tau^-\to \pi^0\pi^-\nu$
decay is measured in the experiments
\cite{Barate:1997hv, Anderson:1999ui, Fujikawa:2008ma}.
As for decays including kaons,
a mass spectrum of $\tau^-\to K^-\pi^0\nu$
is observed in the BaBar experiment\ \cite{Aubert:2007jh}
while one of $\tau^-\to K_S\pi^-\nu$ 
is measured in the Belle
experiment\ \cite{Epifanov:2007rf}.
In Ref.\ \cite{Inami:2008ar, delAmoSanchez:2010pc},
the branching ratios of $\tau$ decays including $\eta$
are reported. As theoretical study,
the spectral function of $\tau^-\to K_S\pi^-\nu$ decay
is fitted with a resonance field of $K^*(892)^{-}$
\cite{Kimura:2014wsa}.
The review for $\tau$ decays
is given in Ref.\ \cite{Pich:2013lsa}.
\par
For vector mesons,
we calculate quantum correction to self-energies
to obtain a 1-loop corrected
mass matrix.
The mixing matrix, which is an orthogonal matrix
to diagonalize the mass matrix,
is determined in the procedure of diagonalization.
After diagonalizing the mass matrix,
the relevant mass eigenstates play the role as resonance
fields of $\rho(770), \omega(782)$ and $\phi(1020)$.
In our formulation,
SU(3)/isospin breaking contribution in the self-energies
is taken into account
in the mixing matrix for vector mesons.
We also consider kinetic mixing of
neutral vector mesons, which arises from
1-loop correction to the self-energies.
Including such mixing contribution,
we obtain analytic formulae of the widths for
$V\to PP$ decays.
Furthermore, we consider mixing between
vector meson and photon, which
also comes from 1-loop correction to self-energies.
As analyzed explicitly in this paper,
we find that the $V-\gamma$ mixing
plays a crucial role in processes
such as radiative decays.
For pseudoscalars, we also take account of
quantum correction to self-energies.
We use parametrization in which
the 1-loop correction to mass matrix elements
is accounted.
The mixing matrix for pseudoscalars
is determined so as to diagonalize
the 1-loop corrected mass matrix.
Using this formulation, we consider
SU(3)/isospin breaking in the
$3\times3$ mixing matrix
for $\pi^0, \eta$ and $\eta^\prime$.
\par
In processes such as $P\to 2\gamma$
and radiative decays of $V\to P\gamma$,
intrinsic parity (IP)\ \cite{Wick:1952nb} is violated.
It is well-known that intrinsic parity violation (IPV) in
models with vector mesons
is categorized as two types: The first one is
the Wess-Zumino-Witten (WZW) term,
which results from quantum anomaly of SU(3)
symmetry\ \cite{Wess:1971yu, Witten:1983tw}.
The second one comes from the presence of resonance fields for vector mesons, as originally suggested in the framework of hidden local symmetry (HLS)\ \cite{Fujiwara:1984mp, Bando:1987br}.
\par
For IP violating interactions in the model,
we introduce operators including
SU(3) singlet fields,
in addition to ones suggested in
Refs.\ \cite{Fujiwara:1984mp, Bando:1987br, Hashimoto:1996ny}.
As shown in our numerical result,
inclusion of the singlet-induced IP violating operators
plays an important role in the framework of
the octet$+$singlet scheme,
typically for $\eta^\prime\to2\gamma$.
\par
Using the introduced operators, we write formulae
of the IP violating decays of hadrons.
In particular, the expressions of (differential) decay widths
and electromagnetic TFFs are shown.
These formulae are useful for thorough analysis
to test the validity of the model.
\par
%%%%%%%%%%%%%%%%%%%%%%%%%%%
% and it  corresponds to two loop ordered effect. Since we treat the IP conserving part and IP violating part in the same power counting, we do not introduce the chiral breaking effect in IP violating interactions and will study how the present scheme
%with less fitting parameters can reproduce the experimental results.
%%%%%%%%%%%%%%%%%%%%%%%%%%%%%%%%%
Since IPV interactions include an anti-symmetric tensor, one needs the fourth derivatives
on the chiral fields so that the Lagrangian is Lorentz invariant. It is  O($p^4$) contribution.
In contrast to IP conserving part, SU(3) breaking effect for IPV  interactions 
is $O(p^4 m_\pi^2)$ and it is one loop effect.
To this accuracy, we need to include  SU(3) breaking both in IP conserving part and in IPV part.
In the first part of our paper, we include  SU(3) breaking effect for IP conserving part up to one loop
order without introducing SU(3) breaking for IPV  interaction. 
In the last part of the
paper, we incorporate  SU(3) breaking interactions for IPV part in the tree level.
\par
In this paper, the observables of IP violating 
decays are analyzed in our model.
For the HLS model, a numerical result for
IP violating decay widths is given in
Refs.\ \cite{Bramon:1994pq, Hashimoto:1996ny}, 
with SU(3) breaking effect in IP violating interactions.
Radiative decays
are analyzed with $VP\gamma$ vertices in
Ref.\ \cite{Bramon:2000fr}.
Moreover, numerical analyses of the TFFs are given in
Refs.\ \cite{Terschluesen:2010ik, Terschlusen:2012xw, Schneider:2012ez, Chen:2012vw, Roig:2014uja, Escribano:2015vjz}.
\par
In our analysis,
$\chi^2$ fittings are carried out to
estimate parameters in the model.
As input data in the fittings,
the spectral function in
$\tau^-\to K_S\pi^-\nu$ decay
measured by the Belle
collaboration \cite{Epifanov:2007rf} are used.
Furthermore, we also utilize the data of
masses for the vector mesons,
which are precisely determined in experiments.
For parameter estimation of coefficients of
IP violating operators, the data of partial widths for
radiative decays and TFFs of $V\to Pl^+l^-$ decays
are used.
As shown in the numerical result,
one can find a parameter region which is consistent
with experimental data for the TFFs.
\par
Using the estimated parameter region,
model prediction for hadron decays is presented
in this work.
Specifically, we give predictions for
(1) the electromagnetic TFFs of $P\to\gamma l^+l^-$,
(2) the partial decay widths of $P\to\gamma l^+l^-,
P\to\pi^+\pi^-\gamma, \phi\to\omega\pi^0$
and $V\to Pl^+l^-$,
(3) the differential decay widths of
$P\to\pi^+\pi^-\gamma$,
(4) the TFFs of $\rho^0\to\pi^0 l^+l^-,
\rho^0\to\eta l^+l^-, \omega\to\eta l^+l^-$
and $\phi\to\eta^\prime l^+l^-$
and (5) the branching ratio and the partial widths of
$V\to \pi^0\pi^+\pi^-$.
As discussed in the latter part
of this paper, the TFFs for
$\phi\to\pi^0l^+l^-$ and $\eta^\prime\to\gamma l^+l^-$ have a peak region around
which di-lepton invariant mass is close to
the pole of $\omega$.
\par
Remaining part of this paper is organized as follows:
In Sec.\ \ref{secII}, the model is introduced
and 1-loop ordered interactions are given
with SU(3) octets and singlets.
The quantum correction to self-energies of vector mesons
are also shown.
Using the 1-loop corrected propagators,
we write the width of $V\to PP$ decay,
including the contribution of kinetic mixing.
The $V-\gamma$ vertex, which arises from
1-loop order interactions, is also shown.
The mixing matrix for $\pi^0, \eta$
and $\eta^\prime$, in which
1-loop correction is accounted,
is introduced.
In Sec.\ \ref{secIII}, IP violating interaction terms are given.
The formulae of decay widths for IP violating
modes are explicitly shown.
In Sec.\ \ref{sec4},
the results of numerical analysis are presented.
We show the fitting result of the invariant mass
distribution of $\tau$ decay.
Physical masses of vector mesons are also fitted
in this section.
Moreover, we estimate coefficients of the
IP violating operators,
via experimental data of hadron decays.
We give the model prediction for decay widths,
TFFs and differential decay widths
for IP violating decays.
In Sec.\ \ref{secV}, SU(3) breaking effect of IPV interaction
is studied.
Finally, Sec.\ \ref{sec6} is devoted to summary
and discussion.
%%%%%%%%%%%%%%%%%%%%%%
%%%%%%%%%%%%%%%%%%%%%%
%%%%%%%%%%%%%%%%%%%%%%
\section{The model with SU(3) octets and singlets}
\label{secII}
%%%%%%%%%%%%%%%%%%%%%%
In this section, we introduce a model of chiral Lagrangian with vector mesons \cite{Kimura:2014wsa}.
In this paper, we extend the previous one so that it includes
$\phi$ meson and electromagnetic mass of pseudoscalar mesons as follows,
\bea
\mathcal{L}_\chi
&=&\mathcal{L}_P+\mathcal{L}_V+\mathcal{L}_c,
\label{chiLag}\\
\mathcal{L}_P&=&
\frac{f^2}{4}\mathrm{Tr}(D_{\mu}U D^\mu
U^\dagger)+B\mathrm{Tr}[M(U+U^\dagger)]
+C\mathrm{Tr}QUQU^\dagger\nn\\
&&+\frac{1}{2}\partial_\mu\eta_0\partial^\mu\eta_0
-\frac{1}{2}M_{00}^2\eta_0^2
-ig_{2p}\eta_0\mathrm{Tr}[M(U-U^\dagger)],
\label{Ppart}\\
\mathcal{L}_V&=&
-\frac{1}{2}\mathrm{Tr}F_{V}^{\mu\nu}F_{V\mu\nu}
+M_V^2\mathrm{Tr}\left(V_\mu-\frac{\alpha_\mu}{g}\right)^2
+g_{1V}\phi^0_\mu\mathrm{Tr}\left\{
\left(V^\mu-\frac{\alpha^\mu}{g}\right)
\left(\frac{\xi M\xi+\xi^\dagger M\xi^\dagger}{2}
\right)\right\}\nn\\
&&-\frac{1}{4}F_{V\mu\nu}^0F^{0\mu\nu}_V
+\frac{1}{2}M_{0V}^2\phi^0_\mu \phi^{0\mu},
\label{eq:V}
\eea
where
\bea
\alpha_{\mu} &=& \frac{1}{2i}(\xi^\dagger D_{L\mu}\xi+\xi D_{R\mu }\xi^\dagger)\label{def3} , \\
D_{L(R)\mu}&=&\p_\mu+iA_{L(R)\mu},\\
U &=& \xi^2 = \exp\left(\frac{2i\pi}{f}\right)\label{def5} , \\
D_{\mu}U &=&
\p_\mu U+iA_{L\mu}U-iUA_{R\mu},\\
M&=&\mathrm{diag}(m_u, m_d, m_s),\\
F_{V\mu\nu}&=&\partial_\mu V_\nu-\p_\nu V_\mu+ig[V_\mu, V_\nu],\label{def6}\\
F_{V\mu\nu}^0&=&\p_\mu \phi_\nu-\p_\nu \phi_\mu,\\
Q&=&\mathrm{diag}\left(\frac{2}{3}, -\frac{1}{3}, -\frac{1}{3}\right).
\label{def77}
\eea
The Lagrangian is divided into three parts
in Eq.\ (\ref{chiLag}), which consist of the parts of 
pseudoscalars, vector mesons, and 1-loop order 
counter terms.
As fields of pseudoscalar, the octet matrix and the singlet
field are contained in Eq.\ (\ref{Ppart}).
$\eta_{0}$ is U(1)$_{A}$ pseudoscalar and its
mass is given as $M_{00}$.
The term denoted as $C\mathrm{Tr}QUQU^\dagger$
in Eq.\ (\ref{Ppart}) is the electromagnetic correction
to ChPT.
This term describes the effect of virtual photon \cite{Urech:1994hd}, and affects the mass of the charged
pseudoscalar. 
Vector mesons are introduced
as SU(3) octet and singlet
in Eq.\ (\ref{eq:V}).
Vector meson matrix for octet
is denoted by $V_{\mu}$,
and its mass is given as $M_{V}$, while
the field $\phi^0_\mu$ denotes SU(3) singlet vector meson.
\par
In the following, we present how 1-loop counter terms given as ${\cal L}_c$ are
introduced for chiral Lagrangian with vector mesons and pseudoscalar singlet. 
The form of 1-loop counter terms depends on the tree-level Lagrangian and 
is obtained with power counting of the
superficial degree of divergence in the loop calculation.
The tree-level Lagrangian is constructed based on
the expansion with respect to derivatives and
chiral SU(3) breaking.
The Lagrangian
includes either the second derivatives or an insertion of chiral SU(3) 
breaking.
The interaction Lagrangian which satisfies such criteria is extracted from
Eqs.\ (\ref{Ppart}, \ref{eq:V}),
\bea
{\cal L}_0&=&
\frac{f^2}{4} {\rm Tr} D_\mu U D^\mu U^\dagger+M_V^2{\rm Tr}\left(V_\mu-\frac{\alpha_\mu}{g}\right)^2\nn \\
&+& B {\rm Tr}(M(U+U^\dagger))
- i g_{2p} \eta_0 {\rm Tr}(M(U-U^\dagger)). 
\label{eq:L0}
\eea
Note that ${\cal L}_0$ does not include the 
parts which are written only with vector mesons and
singlet pseudoscalars. With ${\cal L}_0$, the divergent parts 
of the 1-loop correction is extracted and the counter terms 
are given in Eq.\ (\ref{Count1}). 
As proven in App.\ \ref{Power},
the counter terms satisfy the power counting rule,
which enables us to specify
the structure of them.
Based on the discussion, in Eq.\ (\ref{chiLag}),
we have included
a singlet-octets vector mesons mixing term as a finite counter term. 
\par
%%%%%%%%%%%%%%%%%%%%%Effective Counter terms%%%%%%%%%%%%%%%%%%%%%%%%
The counter terms for the self-energy for vector mesons and $V-\gamma$
mixing can be summarized as the effective counter terms
\cite{Kimura:2014wsa},
\bea
{\mathcal L}^{eff}_c&=&-\frac{1}{2}
Z_V^{(1)} {\rm Tr} 
(\mathcal{F}_{V\mu \nu} \mathcal{F}_{V}^{ \mu \nu})
\nn \\
&+& C_1 {\rm Tr}\left[\frac{\xi \chi \xi + \xi^{\dagger} \chi^{\dagger}
 \xi^{\dagger}}{2}
\left(V_\mu-\frac{\alpha_{\mu}}{g}\right)^2 \right] 
+ C_2 {\rm Tr}\left(\frac{\xi \chi \xi + \xi^{\dagger} \chi^{\dagger}
 \xi^{\dagger}}{2}\right)
 {\rm Tr}\left[\left(V_{\mu}-\frac{\alpha_{\mu}}{g}\right)^2 \right] \nn \\
&+& C_4 {\rm Tr}\mathcal{F}_V^{\mu \nu}( F^0_{L \mu \nu}+F^0_{R \mu \nu}),\label{eq:counter}\\
\chi&=&\frac{4BM}{f^2}.
\eea
where all the field strength are Abelian part defined by,
$\mathcal{F}_{V\mu\nu} = \partial_\mu V_\nu - \partial_\nu V_\mu$ 
and $F^0_{L(R)\mu\nu} = \partial_\mu A_{L(R) \nu} 
- \partial_\nu A_{L(R)\mu}$. 
$Z^{(1)}_V$ and $C_i$ ($i=1,2,4$) are 
renormalization constants and they are  written in terms of the coefficients
in Eq.\ (\ref{Count1}),
\bea
&Z^{(r)(1)}_V= K^{(r)}_3 (g_{\rho\pi\pi})_{\mathrm{tr}}^2, \quad
C^{(r)}_1=2 K^{(r)}_4 (g_{\rho\pi\pi})_{\mathrm{tr}}^2, &\nn \\
&C^{(r)}_2=2 K^{(r)}_5 (g_{\rho\pi\pi})_{\mathrm{tr}}^2, \quad
C^{(r)}_4=-\displaystyle\frac{(g_{\rho\pi\pi})_{\mathrm{tr}}}{2}\left(K^{(r)}_2-K^{(r)}_3 \frac{M_V^2}{2 g^2 f^2}
\right),&
\label{eq:rel}\\
&(g_{\rho\pi\pi})_{\mathrm{tree}}=
\displaystyle\frac{M_V^2}{2gf^2}.&
\label{eq:gpptree}\eea
The coefficients of the effective counter terms
in Eq.\ (\ref{eq:rel}) include 
the divergent part and the finite part. 
The finite parts are denoted with suffix $(r)$.
Both divergent and finite parts of $K_i$ are recorded in
Eq.\ (\ref{Count4}).
\par
In Eq.\ (\ref{eq:gpptree}), $(g_{\rho\pi\pi})_{\mathrm{tree}}$
denotes a tree-level vertex for $\rho \pi \pi$
coupling.
We also define the 1-loop ordered $\rho\pi\pi$ coupling,
\bea
g_{\rho\pi\pi}=\frac{M_V^2}{2gf_\pi^2}.
\label{Defgpp}
\eea
%%%%%%%%%%%%%%%%%%%
%%%%%%%%%%%%%%%%%%%
%%%%%%%%%%%%%%%%%%%%%%%%%%%%%%%%%%%%%%%%%%%%%%%%%%%%%%%%%
\subsection{Neutral vector meson}
%%%%%%%%%%%%%%%%%%%%%%%%%%%%%%%%%%%%%%%%%%%%%%%%%%%%%%%%%
In this subsection, we diagonalize the mass matrix
for neutral vector mesons and obtain
the mass eigenstates which correspond to
($\rho$, $\omega$, $\phi$).
The mixing matrix between
$(\rho_0, \omega_8, \phi_0)$
and mass eigenstates
determines the interaction
among the physical states.
The inverse propagator for
the vector mesons is,
\bea
\frac{1}{2}V^{\mu I} D^{-1}_{\mu \nu IJ} V^{\nu J }, 
\label{neutralVM-1}
\eea
where $V^{I}$ denotes the eigenstate for the mass matrix, 
$V_\mu^T =(V^1_\mu, V^2_\mu, V^3_\mu)
=(\rho_\mu, \omega_\mu, \phi_\mu)$.
The mixing matrix $O_V$ relates the
mass eigenstates to SU(3) basis
in the following,
\bea
V_\mu^{0} = \begin{pmatrix} \rho_\mu^0 \\ \omega_\mu^8 \\ \phi_\mu^0 
\end{pmatrix}
=O_V V_\mu .
\eea
In Eq.\ (\ref{neutralVM-1}), $D_{\mu\nu}^{-1}=O_V^T D^{0-1}_{\mu\nu} O_V$
contains the self-energy correction, 
\bea
D^{0-1}_{\mu \nu}&=&g_{\mu \nu} \begin{pmatrix} M^2_{\rho} & M^2_{V \rho 8} & M^2_{V0\rho} \\
M^2_{V \rho 8}& M^2_{V88} & M^2_{V08}\\
M^2_{V0\rho} & M^2_{V08} & M^2_{0V}\\
\end{pmatrix} +Q_{\mu \nu} \begin{pmatrix} \delta B_\rho(Q^2) & 
\delta B_{\rho 8}(Q^2) & 0\\
                                             \delta B_{\rho 8}(Q^2) & \delta B_{88}(Q^2) & 0 \\
                                             0 & 0 & 1 \end{pmatrix},
\label{massIsobase}\\
Q_{\mu\nu} &=&Q_\mu Q_\nu - g_{\mu\nu}Q^2, \label{Qmunudef}\\
\delta B_{\rho}&=&Z^r_V(\mu)+g_{\rho\pi\pi}^2
\left(4M^r_\pi+M^r_{K^+}+M^r_{K^0}\right) , \label{Zvr}\\
\delta B_{\rho 8}&=&\sqrt{3}g_{\rho\pi\pi}^2
\Delta M^r_{K^+ K^0} ,  \\
\delta B_{88}&=&Z^r_V(\mu)+3g_{\rho\pi\pi}^2
(M^r_{K^+}+M^r_{K^0})\label{Zvr2},
\eea
where $\Delta M^r_{K^+ K^0}=M^r_{K^+}-M^r_{K^0}$.
In Eqs.\ (\ref{Zvr}, \ref{Zvr2}), $Z_V^r$ denotes
the coefficient of kinetic term of octet vector meson
defined as $1+Z_V^{r(1)}$.
$M_P^r$ are the loop functions of vector mesons,
\bea
M^r_{P}&=&\frac{1}{12}
\left[\left( 1-\frac{4 M_P^2}{Q^2} \right) \bar{J}_P-\frac{1}{16 \pi^2}
\ln \frac{M_P^2}{\mu^2}-\frac{1}{4 8 \pi^2} \right], \\
\bar{J}_P&=& \left\{ \begin{array}{c}
-\displaystyle\frac{1}{16 \pi^2} \sqrt{1-\frac{4 M_P^2}{Q^2}}
\ln \frac{1+\sqrt{1-\frac{4 M_P^2}{Q^2}}}{1-\sqrt{1-\frac{4 M_P^2}{Q^2}}}
+\frac{1}{8 \pi^2}+ i \frac{1}{16 \pi} \sqrt{1-\frac{4 M_P^2}{Q^2}},
 \quad (Q^2 \geq 4M_P^2), \nn \\
\displaystyle\frac{1}{8 \pi^2} \left(1-\sqrt{\frac{4 M_P^2}{Q^2}-1} \arctan
\frac{1}{\sqrt{\frac{4 M_P^2}{Q^2}-1}}\right),
\quad (Q^2 \leq 4 M_P^2) ,
\end{array} \right.   
\eea
where $\mu$ is a renormalization scale.
In the numerical analysis, we fix it as
$\mu=m_{K^{*+}}$.
The elements in the mass matrix (\ref{massIsobase}) are given by,
\def\barmk{\overline{m}_K^2}
\bea
M_\rho^2
&=& M_V^2+C^r_1M_\pi^2+C^r_2  (2\bar{M}_K^2+M_\pi^2)
-4g_{\rho\pi\pi}^2 \left(\mu_\pi+\frac{\bar{\mu}_K}{2} \right) f^2 ,\label{mrhosq}\\
M_{V88}^2
&=& M_V^2+C^r_1\frac{4 \bar{M}_K^2-M_\pi^2}{3}+C_2^r(2\bar{M}_K^2+M_\pi^2)-
6g_{\rho\pi\pi}^2
\bar{\mu}_Kf^2, \label{mV88sq}\\
M^2_{V \rho 8}&=&\frac{1}{\sqrt{3}} \left\{ C^r_1 \Delta_{K^+ K^0}-3
g_{\rho\pi\pi}^2 \Delta \mu_K f^2 \right\},\label{mvrho8sq}\\
M^2_{V0\rho}&=&\frac{\hat{g}_{1V}}{4} \Delta_{K^+ K^0} , \label{mv0rhosq}
\\
M^2_{V08}&=&-\frac{\hat{g}_{1V}}{2 \sqrt{3}} \Delta_{K \pi}, \label{mv08sq} 
\eea
with
\bea
&\mu_P=\displaystyle\frac{M_P^2}{32\pi^2f^2}\ln\left(\displaystyle\frac{M_P^2}{\mu^2}\right),\quad
\bar{\mu}_K=\displaystyle\frac{\mu_{K^+}+ \mu_{K^0}}{2}&, \label{Eq:Defmu}\\
&\bar{M}_K^2 = \displaystyle\frac{M^2_{K^+}+M^2_{K^0}}{2}, \quad
M^2_\pi = M^2_{\pi^+}=M^2_{\pi^0},&\label{MKbardef}\\
&\hat{g}_{1V}=\displaystyle\frac{f^2 g_{1V}}{B}, \quad
\Delta\mu_K = \mu_{K^+}-\mu_{K^0},&\\
&\Delta_{PQ}=M_P^2-M_Q^2,\quad
\Delta_{K\pi}=\bar{M}_K^2-M_\pi^2.& 
\eea
We calculate the mass of the neutral vector mesons, $\rho, \omega$ and $\phi$.
The first term in 
Eq.\ (\ref{massIsobase}) is diagonalized as,
\bea
D^{-1}_{\mu \nu}&=&g_{\mu \nu}\cM^2
+ \delta B_V(Q^2) Q_{\mu \nu},\label{BVterm}\\
\cM^2&=&\mathrm{diag}(\cM_1^2, \cM_2^2, \cM_3^2),
\eea
where $\delta B_V(Q^2)$ is a $3 \times 3$ matrix,
\bea
\delta B_V(Q^2)=O_V^T \delta B(Q^2) O_V . 
\label{RotatedB}\eea
The propagator for the neutral vector mesons is denoted as,
\bea
D^{\mu \nu}&=&g^{\mu \nu} D_0 + Q^{\mu} Q^{\nu} D_L,
\label{proDmunu}\\
D_0 &=& (\cM^2-Q^2 \delta B_V)^{-1} , \label{proD} \\ 
D_L &=& \frac{1}{Q^2}\left(\frac{1}{\cM^2}-\frac{1}{\cM^2-Q^2 \delta B_V}\right).
\eea
In the following, we expand the propagator in
Eq.\ (\ref{proDmunu}) with respect to the off-diagonal parts
of $\delta B_V$,
\bea
(D_{\mu\nu })_{IJ} &=&
\begin{cases}
\displaystyle\frac{g_{\mu\nu}-\displaystyle\frac{Q_\mu Q_\nu}{\cM_I^2}\delta B_{VI}}{\cM_I^2 - Q^2 \delta B_{VI}},\quad (I=J)\\
-Q_{\mu\nu}\displaystyle\frac1{\cM_I^2 - Q^2 \delta B_{VI}} \delta 
B_{VIJ}\displaystyle\frac1{\cM_J^2 - Q^2 \delta B_{VJ}},\quad (I\neq J)
\end{cases}
\label{propVdiag}
\eea
where $I, J=1,2,3$ and $\delta B_{VI}$ denotes the diagonal part in 
the matrix in Eq.\ (\ref{RotatedB}).
If one neglects the off-diagonal parts of $\delta B_V$,
the above propagator becomes diagonal matrix
given in the first line in Eq.\ (\ref{propVdiag}).
The pole mass squared is defined as the momentum squared where the real part of the denominator of 
the propagator vanishes
in the following as,
\bea
\cM_I^2-m_I^2 {\rm Re}\delta B_{VI}(m_I^2)=0. 
\eea
The denominator of the propagator in Eq.\ (\ref{propVdiag})
is expanded in the vicinity of the pole mass,
\bea
\cM_I^2-Q^2 {\rm Re}\delta B_{VI}(Q^2)&=&m_I^2 {\rm Re}\delta B_{VI}(m_I^2) -
Q^2 {\rm Re}\delta B_{VI}(Q^2) \nn \\
&\simeq& (m_I^2-Q^2)\frac{d Q^2 {\rm Re}\delta B_{VI}(Q^2)}{d Q^2}
\Bigr{|}_{Q^2=m_I^2}.
\eea 
We define the wave function renormalization of neutral vector meson,
\bea
Z_I^{-1}=\frac{d Q^2 {\rm Re}\delta B_{VI}(Q^2)}{d Q^2}\Bigr{|}_{Q^2=m_I^2}={\rm Re}\delta B_{VI}(m_I^2)+
m_I^2 \frac{d {\rm Re}\delta B_{VI}(Q^2)}{d Q^2}\Bigr{|}_{Q^2=m_I^2}.
\label{WaveFun}
\eea
Thus, in the vicinity of the pole mass, the propagator takes the following form,
\bea
(D_{\mu\nu})_{II}
&\simeq&Z_I \frac{g_{\mu\nu}-\displaystyle\frac{Q_\mu Q_\nu}{m_I^2}\left(1+i
\frac{{\rm Im} \delta B_{VI}(m_I^2)}{{\rm Re}\delta B_{VI}(m_I^2)}\right)}
{m_I^2-Q^2-i m_I \Gamma_I},
\eea
where the definitions of the pole mass and width are given as,
\bea
m_I^2&=&\frac{\cM_I^2}{\mathrm{Re}\delta B_{VI}(m_I^2)},\label{poleMv}\\
\Gamma_I&=&m_I Z_I {\rm Im} \delta B_{VI}(m_I^2).
\eea
%%%%%%%%%%%%%
%%%%%%%%%%%%%
\subsection{
1-loop correction to decay width of
$V\to PP$}
\begin{figure}[!h]
\begin{center}
\includegraphics[width=12cm]{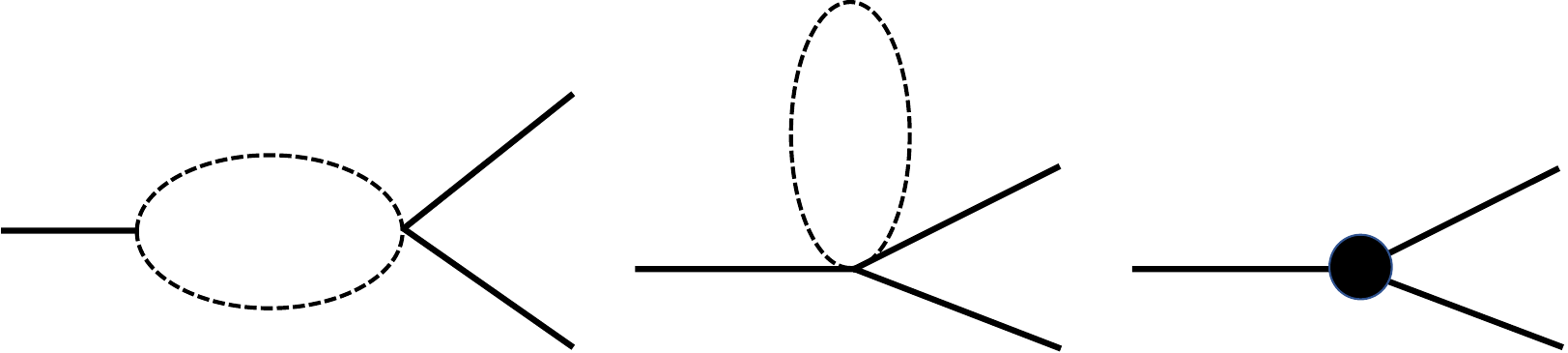}
\caption{1-loop ordered Feynman diagrams
for$V \to PP$ decays.
A black circle indicates a vertex of
the 1-loop ordered counter term.}
\label{figvpp}
\end{center}
\end{figure}

%%%%%%%%%%%%%
In this subsection,
we derive the formulae for width of 
vector meson decay into two pseudoscalars.
We compute
the 1-loop diagrams which are shown in Fig.\ \ref{figvpp}.
For neutral vector mesons, $\rho, \omega$ and $\phi$,
the contribution of the kinetic mixing is also taken into
account, in addition to the diagrams in Fig.\ \ref{figvpp}.
%%%%%%%%%%%%%
\subsubsection{
$K^*\to K \pi$ and $\rho^\pm \to \pi^\pm \pi^0$}
The amplitude for $V \to P \pi$ is written as the sum of the tree-level amplitude and 1-loop correction,
\bea
\mathcal{M}(V \to P \pi)=\epsilon^{\mu \ast} q_\mu (\hat{g}_{V P \pi}+ \Delta g_{V P \pi}), 
\eea
where $\hat{g}_{V P \pi}$ denotes the tree-level coupling and $\Delta g_{V P \pi}$ is 1-loop correction.
We denote $q=p_P-p_\pi$.
Firstly, we consider the case that $V$ and $P$ consist of the same quark flavor contents. 
For  $V=K^{\ast +(0)}$ and $P=K^{+ (0)}$, they are given as,
\bea
\hat{g}_{K^{\ast+} K^+ \pi^0}&=&\frac{(g_{\rho \pi \pi})_{\rm tree}}{2}\sqrt{Z_{K^+}Z_{\pi^+}}=\frac{M_V^2}{4 g f_K f_\pi}\nn \\
\Delta g_{K^{\ast+} K^+ \pi^0}&=&\frac{{C_2^r}(2 {M_K}^2+{M_\pi}^2)+{C_1^r} {M_K}^2}{4 g f^2} \nn \\
&-&\frac{3 (g_{\rho \pi \pi})_{\rm tree}}{4 f^2}
\left(1-\frac{M_V^2}{2 g^2 f^2}\right)\{-(M_{K \pi}^r+M_{K \eta_8}^r) {m^2_{K^\ast}}+(L_{K \pi}^r+L_{K \eta_8}^r)\}  \nn \\
&-& \frac{3 (g_{\rho \pi \pi})_{\rm tree}^2}{8 g} 
(2\mu_K+\mu_\pi+\mu_\eta)+ \frac{C_3^r}{8 f^2}{m^2_{K^\ast}}.
\label{Eq:DeltagKKP}
\eea
In the isospin limit, one can find the relations,
$\hat{g}_{K^{\ast+} K^0 \pi^+}=\sqrt{2}
\hat{g}_{K^{\ast+} K^+ \pi^0}$ and
$\Delta g_{K^{\ast+} K^0 \pi^+}=\sqrt{2}
\Delta g_{K^{\ast+} K^+ \pi^0}$,
are satisfied.
Therefore,
the decay width of $K^{*+}\to K^0\pi^+$ 
is two times larger than that of $K^{*+}\to K^+\pi^0$.
For $V=\rho^+$ and $P=\pi^+$, the couplings
are given as, 
\bea
\hat{g}_{\rho \pi \pi}&=&(g_{\rho \pi \pi})_{\rm tree} Z_{\pi^+} \sqrt{Z_{\rho^+}}=\frac{M_V^2}{2 g f_\pi^2}  \sqrt{Z_{\rho^+}},\nn \\
\Delta g_{\rho \pi \pi}&=& \frac{{C_2^r}(2 {M_K}^2+{M_\pi}^2)+{C_1^r} {M_\pi}^2}{2 g f^2}  \nn \\
&+& \frac{(g_{\rho \pi \pi})_{\rm tree}}{f^2}
\left(1-\frac{M_V^2}{2 g^2 f^2}\right)
(2 M_{\pi \pi}^r+M_{K K}^r) {m^2_{\rho}} \nn \\
&-&\frac{(g_{\rho \pi \pi})_{\rm tree}}{g}(\mu_K+2\mu_\pi)+\frac{C_3^r}{4 f^2} m^2_\rho. \label{Eq:DeltaRPP}
\eea
In the above calculation, the isospin breaking effect is not taken into account. Using the 
1-loop corrected couplings,
we obtain the partial decay width for $V \to P \pi$,
\bea
\Gamma[V \to P \pi]&=&\frac{\nu_{P \pi}(m_V^2)^3}{48 \pi }\frac{\hat{g}^2_{V P \pi }}{m_V^5}
\left(1+2 \mathrm{Re}
\left(\frac{\Delta g_{V P \pi}}{\hat{g}_{V P \pi}}\right)\right),
\label{Eq:rhopipi1loop}\\
\nu_{P \pi}(Q^2) &=& \sqrt{Q^4-Q^2(M_P^2+M_\pi^2)+(M_P^2-M_\pi^2)^2}
\label{Eq:nuDef}.
\eea
Using the isospin relation of $K^*$ decays,
one can find,
\bea
\Gamma[K^{\ast +} \to K^+ \pi^0]+\Gamma[K^{\ast +} \to K^0 \pi^+]=\frac{\nu_{K \pi}(m_{K^\ast}^2)^3}{16 \pi }\frac{\hat{g}^2_{K^{\ast+} K^+ \pi^0}}{m_{K^\ast}^5}
\left(1+2 \mathrm{Re}\left(\frac{\Delta g_{K^{\ast+} K^+ \pi^0}}{\hat{g}_{K^{\ast+} K^+ \pi^0}}\right)\right).\quad\qquad\label{Eq:KstKpi1loop}
\eea
\subsubsection{$V \to \pi^+ \pi^-$
$(V= \omega, \phi)$ and $\phi \to K^+ K^- (K^0 \bar{K}^0)$ }
%%%%%%%%%%%%%%%%%%%%%%%%%%%%%%%%%%%%%%%%%%%%%%%%%%%%%%%%%  
In this subsection, we study the decay width of
$V \to PP$, including the effect of kinetic mixing.
First, $V \to \pi^+ \pi^-$ ($V=\omega, \phi$)
is investigated.
Since the two pions in the final state are
p wave and form an isotriplet,
the decays of $\omega$ and $\phi$ occur due to isospin breaking.
There are two major contributions to the isospin breaking amplitude. The first one is due to a partial component of  isotriplet state ($\rho^0$) in the mass eigenstate of $\omega (\phi)$. 
This effect is incorporated as the mixing matrix of the neutral vector mesons.
Another contribution
comes from the non-vanishing decay amplitude for isosinglet due to isospin breaking.
In our model, incomplete cancellation between 1-loop diagram of charged kaon  and one of neutral kaon
leads to such contribution.
The decay amplitudes for  octet  states $\rho^0$,  $\omega^8$ and a singlet state $\phi^0$ are given as,  
\bea
T(\rho^0 \to \pi^+ \pi^-)&=&-({g_{\rho \pi \pi})_{\mathrm{tree}}} q_\mu \epsilon^{\mu \ast}, \nn \\
T(\omega^8 \to \pi^+ \pi^-)&=&\frac{\sqrt{3}}{2} 
(g_{\rho \pi \pi})_{\mathrm{tree}} \epsilon^{\mu \ast} q_\mu
\Bigl(-(H_{K^+}-H_{K^0})
\nn \\
&+&\left. \frac{M_V^2}{2 g^2 f^2}
(H_{K^+}-H_{K^0}+\mu_{K^+}-\mu_{K^0}) 
-\frac{2 C_1^r}{3}\frac{M_{K^+}^2-M_{K^0}^2}{M_V^2} \right),  \nn \\
T(\phi^0 \to \pi^+ \pi^-)&=& -\frac{\hat{g}_{1V}}{8g f^2} (M_{K^+}^2-M_{K^0}^2)=-(g_{\rho \pi \pi})_{\mathrm{tree}}\ \hat{g}_{1V}\frac{M_{K^+}^2-M_{K^0}^2}{4 M_V^2}
q_\mu \epsilon^{\mu \ast}.\label{Eq:PhiPP}
\eea
The effective Lagrangian for the singlet and  octet states is given as,   
\bea
\mathcal{L}&=&i ((g_{\rho \pi \pi})_{\mathrm{tree}}\rho^{0 \mu}+g_{\omega \pi \pi} \omega^{8 \mu}+g_{\phi \pi \pi} \phi^{0 \mu}) 
\left(\pi^+\overleftrightarrow{\partial}_\mu\pi^-\right),
\eea
where coupling constants are defined as,
\bea
g_{\omega \pi \pi}&=&  \frac{\sqrt{3}}{2} 
(g_{\rho \pi \pi})_{\mathrm{tree}} 
\Bigl(H_{K^+}-H_{K^0}
\nn \\
&&\left. \qquad 
-\frac{M_V^2}{2 g^2 f^2}
(H_{K^+}-H_{K^0}+\mu_{K^+}-\mu_{K^0})
+ \frac{2 C_1^r}{3}\frac{M_{K^+}^2-M_{K^0}^2}{M_V^2} \right),
\label{Eq:GOPP}  \\
g_{\phi \pi \pi}&=&  (g_{\rho \pi \pi})_{\mathrm{tree}}\ \hat{g}_{1V}\frac{M_{K^+}^2-M_{K^0}^2}{4 M_V^2}.\nn
\eea
Next one can rewrite the Lagrangian in terms of the mass eigenstates using their relations with the octet and singlet states,
\bea
\begin{pmatrix} \rho_\mu^0 \\ \omega_\mu^8 \\ \phi_\mu^0 \end{pmatrix}=O_V \begin{pmatrix} \sqrt{Z_{1}} &0 & 0 \\
0 & \sqrt{Z_{2}} &0  \\
0 & 0 &  \sqrt{Z_{3}}
\end{pmatrix}
\begin{pmatrix} \rho_{\mu R} \\ \omega_{\mu R} \\ \phi_{\mu R} \end{pmatrix}.
\eea
Substituting the above equation, one obtains the effective Lagrangian for renormalized mass eigenstates 
$\begin{pmatrix} V_\mu^1 & V_\mu^2 &V_\mu^3 \end{pmatrix}=
\begin{pmatrix} \rho_{R \mu} & \omega_{R \mu} &\phi_{R \mu} \end{pmatrix}$,
\bea
\mathcal{L}|_{V\pi^+\pi^-}&=& i ((g_{\rho \pi \pi})_{\mathrm{tree}} O_{V1I} + g_{\omega \pi \pi}O_{V2I} + g_{\phi \pi \pi} O_{V 3I}) \sqrt{Z_I} V_\mu^I
\left(\pi^+\overleftrightarrow{\partial}^\mu\pi^-\right)
\nn \\
&=&i(g_{\rho\pi\pi})_{\mathrm{tree}} \Pi_I \sqrt{Z_I} V_\mu^I\left(\pi^+\overleftrightarrow{\partial}^\mu\pi^-\right) 
\label{V0pi+pi-},\label{Eq:LVPP}\\
\Pi_I&=& O_{V1I}+ \frac{g_{\omega \pi \pi}}{(g_{\rho \pi \pi})_{\mathrm{tree}} }O_{V2I}+ \frac{g_{\phi \pi \pi}}{(g_{\rho \pi \pi})_{\mathrm{tree}}}O_{V3I}.\label{Eq:PII}
\eea
To evaluate the partial decay width for
$V^I \to PP$, kinetic
mixing in the decay process, i.e., $V^I \to V^J \to PP$,
should be taken into account.
Using the renormalized fields, we can express the kinetic mixing terms,
\bea
\mathcal{L}_{\mathrm{KM}} = \frac12 \begin{pmatrix} \rho_\mu^R, & \omega_\mu^R, & \phi_\mu^R 
 \end{pmatrix}
 Q^{\mu\nu} 
\begin{pmatrix} 0 & \delta B_{V12} & \delta B_{V13} \\
                     \delta B_{V12} & 0 & \delta B_{V23} \\
                     \delta B_{V13} & \delta B_{V23} & 0 \\  
\end{pmatrix}
\begin{pmatrix} \rho_\nu^R \\ \omega_\nu^R \\ \phi_\nu^R 
\end{pmatrix}.
\eea
In the above Lagrangian, we set the wave function
renormalization $Z_I = 1$, since $\delta B_{VIJ}\ (I\not=J)$ is 1-loop order contribution.\par
The $T$-matrix elements for
$V\to PP$ decays are,
\bea
T[V^I\to\pi^+\pi^-]&=&-g_{V^I\pi^+\pi^-}^{\mathrm{eff}}\ (q\cdot \epsilon^*),\quad (I=2, 3)\nn\\
T[\phi\to K^+K^-]&=&-g_{\phi K^+K^-}^\mathrm{eff}\ (q\cdot \epsilon^*),\label{T-VPP}\\
T[\phi\to K^0\bar{K^0}]&=&-g_{\phi K^0\bar{K^0}}^\mathrm{eff}\ (q\cdot \epsilon^*)\nn,
\eea
where $q=p_{+(0)}-p_{-(\bar{0})}$.
Including the contribution of kinetic mixing,
the effective couplings in Eq.\ (\ref{T-VPP})
are given as,
\bea
g_{V^I\pi^{+}\pi^-}^{\mathrm{eff}}
&=& g_{\rho\pi\pi}
\left[\Pi_I \sqrt{Z_I}+m_{I}^2\displaystyle\sum_{J\neq I}\Pi_J
\frac{\delta B_{VJI}}{\mathcal{M}_J^2-m_I^2\delta B_{VJ}(m_I^2)}\right],\label{eq:gvpp1}\\
g_{\phi K^+K^-}^{\mathrm{eff}}
&=&g_{\rho\pi\pi} \left(\frac{f_\pi}{f_K}\right)^2\left[\sqrt{Z_3}\Pi_3^{K^+}+m_{\phi}^2\displaystyle\sum_{J\neq 3}\Pi_J^{K^+}
\frac{\delta B_{VJ3}}{\mathcal{M}_J^2-m_\phi^2\delta B_{VJ}(m_3^2)}
\right],\label{eq:gvpp2}\\
g_{\phi K^0\bar{K^0}}^{\mathrm{eff}}
&=&g_{\rho\pi\pi} \left(\frac{f_\pi}{f_K}\right)^2\left[\sqrt{Z_3}\Pi_3^{K^0}+m_{\phi}^2\displaystyle\sum_{J\neq 3}\Pi_J^{K^0}
\frac{\delta B_{VJ3}}{\mathcal{M}_J^2-m_\phi^2\delta B_{VJ}(m_3^2)}
\right],\label{eq:gvpp3}\\
\Pi_I^{K^+}&=&\frac{O_{V1I}}{2}+\frac{\sqrt{3}}{2}O_{V2I}+\frac{\hat{g}_{1V}(M_{\pi^+}^2-M_{K^0}^2)}{4M_V^2}O_{V3I},\\
\Pi_I^{K^0}&=&-\frac{O_{V1I}}{2}+\frac{\sqrt{3}}{2}O_{V2I}+\frac{\hat{g}_{1V}(M_{\pi^+}^2-M_{K^+}^2)}{4M_V^2}O_{V3I}.
\eea
Ignoring the isospin breaking effect,
we note that 
$T[\rho^0 \to\pi^+\pi^-]$ is the same as 
the amplitude of $\rho^+\to \pi^+\pi^0$
which was studied in the previous subsection.
In Eqs.\ (\ref{eq:gvpp1}-\ref{eq:gvpp3}),
the second terms denote the kinetic mixing effects 
for $V^I \to V^J \to PP$ decay process
and
$\cM_J\ (J=1,2,3)$ is the eigenvalue for the vector meson mass matrix,
which differs from the physical masses, $m_\rho, m_\omega$ or $m_\phi$.
However, within the accuracy, one can set  $\cM_J=m_J$
since their difference arises from only the wavefunction
renormalization.
One can obtain the partial widths for $V\to PP$ decay,
\b
\Gamma[V^I\to PP]=\frac{m_I|g_{VPP}^{\mathrm{eff}}|^2}{48 \pi}\left(1-\frac{4M_P^2}{m_I^2}\right)^{\frac{3}{2}},
\label{vpp}
\e
where $g_{VPP}^{\mathrm{eff}}$
is the coupling associated with
Eqs.\ (\ref{eq:gvpp1}-\ref{eq:gvpp3}).
%%%%%%%%%%%%%%%%%%%%%%
\subsection{Mixing between photon and vector meson}
\begin{figure}[hbtp]
\includegraphics[clip,width=13cm]{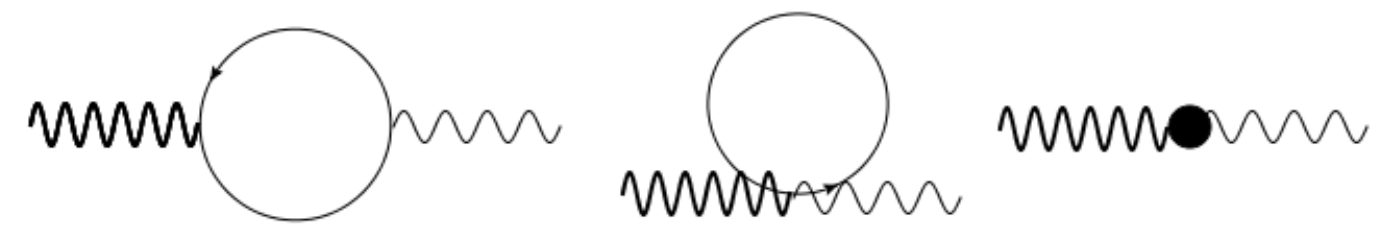}
  \caption{Feynman diagrams for the two-point
function of the mixing of photon and vector meson.
Wavy lines imply vector meson while bold wavy lines indicate the vector mesons.
}
\label{fig:VAMIX}
\end{figure}

%%%%%%%%%%%%%%%%%%%%%%
\label{CVA}
In this subsection, the mixing between photon and vector mesons is analyzed.
The contributing diagrams for $V-A$ mixing
in 1-loop order are
exhibited in Fig.\ \ref{fig:VAMIX}.
%%%%%%%%%%%%%%
%%  Figures   %%%%
%%%%%%%%%%%%%%
%%%%%%%%%%%%%%%%
%%% End of Figure %%
%%%%%%%%%%%%%%%%
The $V-\gamma$ conversion vertex
is denoted as,
\bea
\mathcal{L}_\chi|_{V\gamma}&=& V^{0I}_\mu
\Pi^{\mu \nu  V^{0I} A} A_\nu=
V^I_\mu \Pi^{\mu \nu  V^I A} A_\nu,\label{MC1}\\
\Pi_{\mu \nu}^{V^I A}&=&O_V^T\Pi_{\mu\nu}^{V^{0I} A}.
\label{MC2}
\eea
In the basis of $SU(3)$, the two-point functions
in the l.h.s. of Eq.\ (\ref{MC1}) are given as,
\bea
\Pi_{\mu \nu}^{V^{0I} A}&=&e g_{\mu \nu} \Pi^{V^{0I} A}
+eQ_{\mu \nu} \Pi_T^{V^{0I}A},\label{MVAm}\\
\Pi^{\rho_0 A}&=&\frac{1}{g}\left\{-M_V^2+4g_{\rho \pi \pi}^2
\left(\mu_\pi+\frac{\mu_{K^+}}{2}\right)f^2\right.\nn\\
&&\left.-C_1^r \left(M^2_\pi+\frac{\Delta_{K^+ K^0}}{3}\right)
-C_2^r
(2 \bar{M}_K^2+M_\pi^2)\right\},\\
\Pi_{T}^{\rho_0 A}&=&
g_{\rho \pi \pi}\left(1-\frac{M_V^2}{2g^2 f^2}\right)(4 M_\pi^r+2M_{K^+}^r)
-2C_4^r,\\
\Pi^{\omega_8 A}&=&\frac{1}{\sqrt{3}g}
\left\{-M_V^2+6
g_{\rho \pi \pi}^2\mu_{K^+} f^2\right.\nn\\
&&\left.-C_1^r
\left(\frac{4 \bar{M}_K^2-M_\pi^2}{3}+\Delta_{K^+ K^0}\right)-C_2^r
(2\bar{M}_K^2+M_\pi^2)\right\},\\
\Pi_T^{\omega_8 A}&=&2 \sqrt{3} g_{\rho \pi \pi}\left(1-\frac{M_V^2}{2 g^2 f^2}\right) M_{K^+}^r
-\frac{2}{\sqrt{3}} C_4^r,\\
\Pi^{\phi_0 A}&=&\frac{\hat{g}_{1V}}{g} 
\left(\frac{\Delta_{K\pi}}{6}-\frac{\Delta_{K^+ K^0}}{4}\right),\\
\Pi_T^{\phi_0 A}&=&0.
\eea
One can find that $g^{\mu \nu}$ part
in Eq.\ (\ref{MVAm}) is related to the matrix elements
of the 1-loop corrected neutral vector meson masses
in Eqs.\ (\ref{mrhosq}, \ref{mV88sq}, \ref{mv08sq}),
\bea
\Pi^{V^0 A}=-\frac{1}{g} \begin{pmatrix} M_\rho^2+
\displaystyle\frac{1}{\sqrt{3}}M^2_{V \rho 8} \\
M_{V \rho 8}^2+\displaystyle\frac{1}{\sqrt{3}}M_{V88}^2 \\
M_{V0 \rho}^2+\displaystyle\frac{1}{\sqrt{3}}M_{V08}^2
\end{pmatrix}.\label{tobeproven}
\eea
%%%%%%%%%%%%%
%%%%%%%%%%%%%
One can write the two-point functions in Eq.\ (\ref{MC2}),
\bea
\Pi^{VA}=O_V^T \Pi^{V^0 A}
=-\frac{1}{g}
\begin{pmatrix}\cM_1^2 & 0 & 0 \\
0 & \cM_2^2 & 0 \\
0 & 0 & \cM_3^2 
  \end{pmatrix}\begin{pmatrix} 
O_{V11} +\displaystyle\frac{1}{\sqrt{3}}O_{V21} \\
O_{V12}+\displaystyle\frac{1}{\sqrt{3}}O_{V22} \\
O_{V13}+\displaystyle\frac{1}{\sqrt{3}}O_{V23} \end{pmatrix}
.\label{MassTwoP}
\eea
The derivation of Eq.\ (\ref{MassTwoP}) is shown in
App.\ \ref{Proof}.
Thus, the mixing vertices for $V-\gamma$ in
Eq.\ (\ref{MC1})
are expressed as,
\bea
\mathcal{L}_\chi|_{V\gamma}&=&-\frac{e\cM_I^2}{g}\eta_I
V^I_\mu A^\mu, \quad\label{Vgamma}\\
\eta_I&=&O_{V1I}+\frac{1}{\sqrt{3}}O_{V2I}
.\eea
%%%%%%%%%%%%%%%%%%%%%%%
\subsection{Pseudoscalar}
%%%%%%%%%%%%%%%%%%%%%%%
In this subsection, the structures of
a mixing matrix and decay constants
for pseudoscalars are given.
We take account of 1-loop correction to
both mixing and the decay constants.
\par
The basis for an SU(3) eigenstate is
written in terms of mass eigenstates as,
\b
\begin{pmatrix}
\pi_3 \\
\eta_8\\
\eta_0
\end{pmatrix}
=\sqrt{Z} O
\begin{pmatrix}
\pi^0 \\
\eta\\
\eta^\prime
\end{pmatrix},\label{pseudomixM}
\e
where $O$ denotes an orthogonal matrix which
diagonalizes a mass matrix of pseudoscalars.
$\sqrt{Z}$ in Eq.\ (\ref{pseudomixM}) is a matrix which canonically rescales
1-loop corrected kinetic terms for pseudoscalars.
The result of 1-loop correction to the mass terms
for charged particles is summarized in App.\ \ref{APPcharged},
while one to the mass matrix for neutral particles
is shown in App.\ \ref{APPneutral}.
The 1-loop expression of $\sqrt{Z}$ is recorded
in Eq.\ (\ref{Dp}).
We denote the mixing matrix as,
\b
O&=&
\begin{pmatrix}
\cos\theta_2& \sin\theta_2& 0\\
-\sin\theta_2& \cos\theta_2& 0\\
0& 0& 1
\end{pmatrix}
\begin{pmatrix}
1& 0& 0\\
0& \cos\theta_1& \sin\theta_1\\
0& -\sin\theta_1& \cos\theta_1
\end{pmatrix}
\begin{pmatrix}
\cos\theta_3& \sin\theta_3& 0\\
-\sin\theta_3& \cos\theta_3& 0\\
0& 0& 1
\end{pmatrix}
\nn\\
&=&
\begin{pmatrix}
\cos\theta_2\cos\theta_3-\cos\theta_1\sin\theta_2\sin\theta_3 & \cos\theta_2\sin\theta_3+\cos\theta_1\sin\theta_2\cos\theta_3 &\sin\theta_1\sin\theta_2\\
-\sin\theta_2\cos\theta_3-\cos\theta_1\cos\theta_2\sin\theta_3 & -\sin\theta_2\sin\theta_3+\cos\theta_1\cos\theta_2\cos\theta_3 & \sin\theta_1\cos\theta_2\\
\sin\theta_1\sin\theta_3 &-\sin\theta_1\cos\theta_3 & \cos\theta_1
\end{pmatrix}, \quad\quad\quad
\label{mixO}
\e
where ranges of the mixing angles in Eq.\ (\ref{mixO})
are defined as,
\bea
-\pi\leq\theta_1\leq0,
\quad
-\pi\leq\theta_2\leq\pi,
\quad
-\pi\leq\theta_3\leq\pi.
\eea
The mixing angles denoted as $\theta_2$ and $\theta_3$
are almost $0$ or $\pi$ due to isospin breaking.
In Eq.\ (\ref{mixO}), if we take the limit where
$\theta_{2, 3}\to 0$ or $\pi$,
one can find that $\theta_1$ corresponds to a $2\times 2$ mixing angle for $\eta_8-\eta_0$.
Hence, in order to calculate a mixing angle
for $\eta-\eta^\prime$ in the $3\times3$ mixing
matrix, we use the value of $\theta_1$ in Eq.\ (\ref{mixO}).
\par
For decay constants of $\pi^+$ and $K^+$,
we also consider the 1-loop quantum correction.
As stated in App.\ \ref{decayC},
the ratio of a pion decay constant to one for kaon
is determined with wave function renormalization
of pseudoscalars \cite{Kimura:2014wsa},
\b
\frac{f_{K^+}}{f_{\pi^+}}&=&
\sqrt{\frac{Z_{\pi^+}}{Z_{K^+}}}
\sim1+4\frac{M_{K^+}^2-M_{\pi^+}^2}{f^2}L_5^r
+\frac{c}{4}\left
(5\mu_{\pi^+}-3\mu_{88}-2\mu_{K^+}\right)
\label{rafKfpi},
\e
where $c$ is defined as,
\bea
c=1-\frac{M_V^2}{g^2f^2}.\label{Eq:cexp}
\eea
%%%%%%%%%%%%%%%%%%%%%%%
%%  Intrinsic Parity Violation  %%
%%%%%%%%%%%%%%%%%%%%%%%
\section{Intrinsic parity violation}
\label{secIII}
In this section, 
we discuss IPV in the model.
As well as ChPT, quantum anomaly of chiral symmetry
causes an IP violating interaction.
The expression of the WZW term is given in
Eq.\ (\ref{eq:A-1}).
In addition to this operator,
IP violating interaction terms,
which come from the resonance field of
vector mesons, are introduced.
Subsequently, we write the formula of
widths, TFFs, differential widths for
IP violating decays.
\subsection{Intrinsic parity violating operators
with vector mesons}
Since SU(3) singlet fields are contained in the model,
IP violating operators with singlets should be taken into account.
We consider such singlet-induced operators
within invariance of SU(3) symmetry.
Imposing the charge conjugation (C) symmetry,
one can obtain the operators
in the model,
\bea
\mathcal{L}_1&=&i\epsilon^{\mu\nu\rho\sigma}\mathrm{Tr}[\alpha_{L\mu}
\alpha_{L\nu}\alpha_{L\rho}\alpha_{R\sigma}-(R\leftrightarrow L)],
\label{eq2-2}\\
\mathcal{L}_2&=&i\epsilon^{\mu\nu\rho\sigma}\mathrm{Tr}[\alpha_{L\mu}
\alpha_{R\nu}\alpha_{L\rho}\alpha_{R\sigma}],\label{eq2-3}\\
\mathcal{L}_3&=&\epsilon^{\mu\nu\rho\sigma}\mathrm{Tr}[gF_{V\mu\nu}
\{ \alpha_{L\rho}\alpha_{R\sigma}-(R\leftrightarrow L)\}],
\label{eq2-4}\\
\mathcal{L}_4
&=&\frac{1}{2}\epsilon^{\mu\nu\rho\sigma}\mathrm{Tr}[
(\hat{F}_{L\mu\nu}+\hat{F}_{R\mu\nu})\{ \alpha_{L\rho}, \alpha_{R\sigma}\}] ,
\label{eq2-4pp}\\
\mathcal{L}_5&=&
\epsilon^{\mu\nu\rho\sigma}F^0_{V\mu\nu}
\mathrm{Tr}[\alpha_{L\rho}\alpha_{R\sigma}-(R\leftrightarrow L)] ,
\label{eq2-4p}\\
\mathcal{L}_6&=&\frac{\eta_0}{f}
\epsilon^{\mu\nu\rho\sigma}
\mathrm{Tr}F_{V\mu\nu}F_{V\rho\sigma} ,
\label{eq2-4p2}\\
\mathcal{L}_7&=&\frac{\eta_0}{f}
\epsilon^{\mu\nu\rho\sigma}
F^0_{V\mu\nu}F^0_{V\rho\sigma} , \label{eq2-4p3}\\
\mathcal{L}_8&=&\epsilon^{\mu\nu\rho\sigma}
\mathrm{Tr}(\hat{F}_{L\mu\nu}+\hat{F}_{R\mu\nu})\phi^0_\rho
\frac{\alpha_{L\sigma}-\alpha_{R\sigma}}{2} , \\
\mathcal{L}_9&=&\frac{\eta_0}{f}
\epsilon^{\mu\nu\rho\sigma}
\mathrm{Tr}(\hat{F}_{L\mu\nu}+\hat{F}_{R\mu\nu})F_{V\rho\sigma} ,
\label{eq2-4p4}\\
\mathcal{L}_{10}&=&\frac{\eta_0}{f}
\epsilon^{\mu\nu\rho\sigma}
\mathrm{Tr}(\hat{F}_{L\mu\nu}+\hat{F}_{R\mu\nu})
(\hat{F}_{L\rho\sigma}+\hat{F}_{R\rho\sigma}) , \label{eq2-4p5}
\eea
where $\epsilon^{0123}=-\epsilon_{0123}=+1$ and,
\bea
\hat{F}_{L\mu\nu}&=&\xi^\dagger F_{L\mu\nu}\xi,\\
\hat{F}_{R\mu\nu}&=&\xi F_{R\mu\nu}\xi^\dagger,\\
F_{L(R)\mu\nu}&=&\partial_\mu A_{L(R)\nu}-\partial_\nu
A_{L(R)\mu}+i[A_{L(R)\mu}, A_{L(R)\nu}],\\
\alpha_{L\mu}&=&\alpha_\mu+\alpha_{\perp\mu}-gV_\mu,\label{def1}\\
\alpha_{R\mu}&=&\alpha_\mu-\alpha_{\perp\mu}-gV_\mu,\label{def2}\\
\alpha_{\perp\mu}&=&\frac{1}{2i}(\xi^\dagger D_{L\mu}\xi
-\xi D_{R\mu}\xi^\dagger).
\eea
In Eqs.\ (\ref{eq2-2}-\ref{eq2-4}),
$\mathcal{L}_{1-3}$
are introduced in Refs. \cite{Fujiwara:1984mp, Bando:1987br} while
$\mathcal{L}_{4}$ is considered in
Ref. \cite{Hashimoto:1996ny}.
We introduced $\mathcal{L}_{5-10}$,
which are written with singlets of $\eta_0$ or $\phi_0$.
In Eqs.\ (\ref{eq2-2}-\ref{eq2-4p5}), we required that
the operators should be Hermite.
\par
In contrast to our work, the singlet fields are
contained as a component of chiral nonet matrix in
Ref. \cite{Hashimoto:1996ny} and
$\mathcal{L}_i (i=5-10)$ is not
included in that work.
Chiral SU(3) breaking effect in IP violating interactions
is introduced with spurion field method in
Ref. \cite{Hashimoto:1996ny},
while the operators in Eqs.\ (\ref{eq2-2}-\ref{eq2-4p5}) are invariant under SU(3) transformation.
%%%%%%%%%%%%%%%%%%%%%%%%%%%%%%%%%%%%%%%%%%%%%%%%%%%%%%%%%%%%%%%%%%%%%%%%
The number of the derivatives and vector fields included in the IP violating terms
$\mathcal{L}_i (i=1-10)$ are four by applying the same  power counting rule
as that of IP conserving part.  The interaction Lagrangian  $\mathcal{L}_{6-7}$ and $\mathcal{L}_{9-10}$
include a  SU(3) singlet pseudo-scalar meson $\eta_0$.
%%%%%%%%%%%%%%%%%%%%%%%%%%%%%%%%%%%%%%%%%%%%%%%%%%%%%%%%%%%%%%%%%%%%%%%%%
\par
The IP violating interactions in our model are denoted as,
\bea
\mathcal{L}_{\mathrm{IPV}}=\mathcal{L}_{\mathrm{WZ}}
+\displaystyle\sum_{i=1}^{10}c_{i}^{\mathrm{IP}}\mathcal{L}_i
.\label{ipvio}
\eea
In Eq.\ (\ref{ipvio}), the coefficients of
the operators, $c_{i}^{\mathrm{IP}}\ (i=1-10)$, are free parameters.
As carried out in Sec.\ \ref{sec4},
these parameters are estimated via experimental data
which are sensitive to IPV.
In the following subsections, with the interaction in Eq.\ (\ref{ipvio}),
the formulae of IP violating decay modes are explicitly written.
\subsection{Intrinsic parity violating decays}
%%%%%%%%%%%%%%%%
%%%  V  -> P  gamma %%%
%%%%%%%%%%%%%%%%
\subsubsection{$V\to P\gamma$ and $P\to V\gamma$}
\begin{figure}[h]
  \setlength{\subfigwidth}{.246\linewidth}
  \addtolength{\subfigwidth}{-.246\subfigcolsep}
  \begin{minipage}[b]{\subfigwidth}
    \subfigure[]{\includegraphics[width=4.4cm]{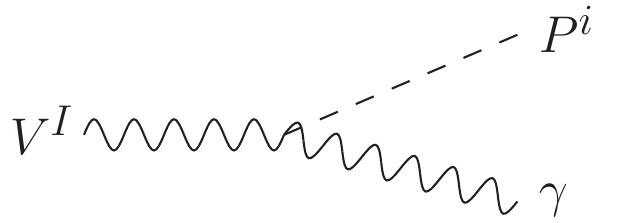}
   \label{VPg1}}
  \end{minipage}
  \begin{minipage}[b]{\subfigwidth}
    \subfigure[]{\includegraphics[width=4.4cm]{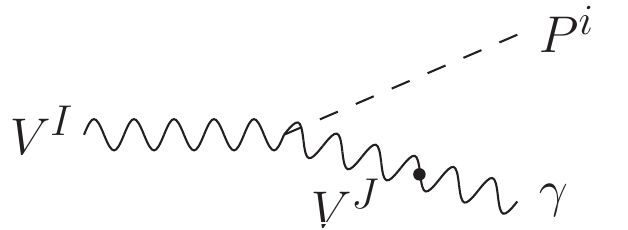}
   \label{VPg2}}
  \end{minipage}
  \begin{minipage}[b]{\subfigwidth}
    \subfigure[]{\includegraphics[width=4.35cm]{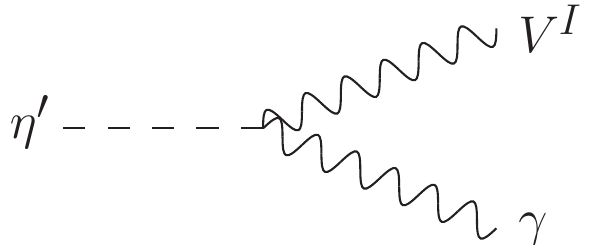}
   \label{eta1}}
  \end{minipage}
  \begin{minipage}[b]{\subfigwidth}
    \subfigure[]{\includegraphics[width=4.35cm]{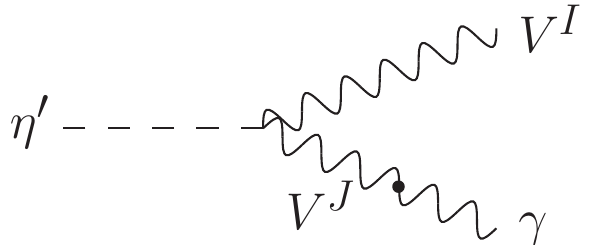}
   \label{eta2}}
  \end{minipage}
  \caption{Diagrams contributing to the decay width for:
  (a)-(b) $V^I\to P^i\gamma$ and (c)-(d) 
  $\eta^\prime\rightarrow V\gamma$.}
\label{fig:VPgamma}
\end{figure}

In this subsection,
IP violating decays of $V\rightarrow P \gamma$
and $P\to V\gamma$ are investigated.
Diagrams contributing to $V\to P\gamma$ and $P\to V\gamma$
are listed in Fig.\ \ref{fig:VPgamma}.
%%%%%%%%%%%%%%%%%%%%%%
%%% Figures V-> P \gamma %
%%%%%%%%%%%%%%%%%%%%%%
%%%%%%%%%%%%%
%% End of Figures  %
%%%%%%%%%%%%%
Interaction vertices of vector meson are shown as,
\bea
\mathcal{L}_{\mathrm{IPV}}
|_{VP\gamma}
&=&-\displaystyle\frac{e}{f_{\pi}}\chi_{iI}\epsilon^{\mu\nu\rho\sigma}
\partial_\mu
V^I_\nu\partial_\rho P^iA_\sigma,
\label{VPgamma}\\
\mathcal{L}_{\mathrm{IPV}}|_{VVP}
&=&\displaystyle\frac{g}{f_{\pi}}
\theta_{iIJ}\epsilon^{\mu\nu\rho\sigma}
\partial_\mu V^I_\nu\partial_\rho P^i
V^J_\sigma,
\label{VVP}\\
\mathcal{L}_{\mathrm{IPV}}|_{V^+P^-\gamma\:+\;\mathrm{h}.
\mathrm{c}.}&=&
\frac{2eg}{3f_{\pi}}c_{34}^-\epsilon^{\mu\nu\rho\sigma}
(\partial_\mu\rho^+_\nu\partial_\rho\pi^-
+\partial_\mu\rho^-_\nu\partial_\rho\pi^+)A_\sigma\nn\\
&&
+\frac{2eg}{3f_{K}}c_{34}^-\epsilon^{\mu\nu\rho\sigma}(\partial_\mu K^{*+}_\nu\partial_\rho K^-
+\partial_\mu K^{*-}_\nu\partial_\rho K^+)A_\sigma\nn\\
&&-\frac{4eg}{3f_{K}}c_{34}^-\epsilon^{\mu\nu\rho\sigma}
(\partial_\mu K^{*}_\nu \partial_\rho\bar{K^0}
+\partial_\mu \bar{K^{*}}_\nu \partial_\rho K^0)A_\sigma,
\label{rho+pi-gamma+hc}\\
\mathcal{L}_{\mathrm{IPV}}|_{V^-P^+V^0\:+\;\mathrm{h}.
\mathrm{c}.}&=&
\frac{g}{f_{\pi}}\gamma_I\epsilon^{\mu\nu\rho\sigma}(
\partial_\mu\rho^+_\nu\partial_\rho\pi^-
+\partial_\mu\rho^-_\nu\partial_\rho\pi^+ 
)V_{\sigma}^I\nn\\
&&+\frac{g}{f_{K}}L_I\epsilon^{\mu\nu\rho\sigma}(
\partial_\mu K^{*+}_\nu\partial_\rho K^-
+\partial_\mu K^{*-}_\nu\partial_\rho K^+
)V_{\sigma}^I\nn\\
&&-\frac{g}{f_{K}}\varphi_I
\epsilon^{\mu\nu\rho\sigma}
(\partial_\mu K^{*}_\nu \partial_\rho\bar{K^0}
+\partial_\mu \bar{K^{*}}_\nu \partial_\rho K^0)V^I_\sigma,\label{K*K0V0} 
\eea
where $i, I$ and $J$ run from $1$ to $3$
and $c_{34}^- = c_3^{\mathrm{IP}}-c_4^{\mathrm{IP}}$.
In Eqs.\ (\ref{VPgamma}-\ref{VVP}, \ref{K*K0V0}),
fields of mass eigenstate are denoted for vector mesons
as $(V^1, V^2, V^3)=(\rho, \omega, \phi)$ and
for pseudoscalars as
$(P^1, P^2, P^3)=(\pi^0, \eta, \eta^\prime)$, respectively.
The coefficients, $\chi_{iI}$ in Eq.\ (\ref{VPgamma}), describes the vertex of each component,
{\it e.g.}, $\chi_{11}$ for $\rho \pi^0\gamma$ vertex and
$\chi_{12}$ for $\omega \pi^0\gamma$ vertex.
Note that the vertex coefficient of $\rho \omega \pi^0$
is proportional to $\theta_{121}+\theta_{112}$ in
Eq.\ (\ref{VVP}) since an operator with $I=1,\:J=2$
and another operator with $I=2,\: J=1$
give the same amplitude.
The coefficients of the vertices in
Eqs.\ (\ref{VPgamma}-\ref{K*K0V0}) are given as,
\bea
V^IP^i\gamma : \ \chi_{iI} &=& -\frac{2g}{3}c_{34}^-\left[\left(O_{1i}+
\sqrt{3}\sqrt{\frac{Z_{2}^\pi}{Z_{1}^\pi}}O_{2i}\right)O_{V1I}+\left(\sqrt{3}O_{1i}-\sqrt{\frac{Z_{2}^\pi}{Z_{1}^\pi}}O_{2i}\right)O_{V2I}\right] \nn\\
&& -4c_5^{\mathrm{IP}}\left(O_{1i}+\frac{1}
{\sqrt{3}}\sqrt{\frac{Z_{2}^\pi}{Z_{1}^\pi}}O_{2i}\right)O_{V3I}
+2c_8^{\IP}\left(O_{1i}+\frac{1}{\sqrt{3}}\sqrt{\frac{Z_{2}^\pi}{Z_{1}^\pi}}O_{2i}\right)O_{V3I}\nn\\
&&\ +2c_9^{\IP}\sqrt{\frac{1}{Z_{1}^\pi}}O_{3i}\left(O_{V1I}+\frac{1}{\sqrt{3}}O_{V2I}\right)
,\label{VPgver} \\
V^IV^JP^i : \ \theta_{iIJ} &=&
-\frac{2gc_3^{\mathrm{IP}}}{\sqrt{3}}\left[\left(2O_{1i}O_{V1I}-
\sqrt{\frac{Z_{2}^\pi}{Z_{1}^\pi}}O_{2i}O_{V2I}\right)O_{V2J}
+\sqrt{\frac{Z_{2}^\pi}{Z_{1}^\pi}}O_{2i}O_{V1I}O_{V1J}\right]
\nn\\
&&
-4c_5^{\mathrm{IP}} \left(O_{1i}O_{V3I}O_{V1J}+\sqrt{\frac{Z_{2}^\pi}{Z_{1}^\pi}}O_{2i}O_{V3I}O_{V2J}\right)\nn\\
&&
-\frac{2c_6^{\mathrm{IP}}}{g}\sqrt{\frac{1}{Z_{1}^\pi}}O_{3i}(O_{V1I}O_{V1J}+O_{V2I}O_{V2J})-\frac{4c_7^{\mathrm{IP}}}{g}\sqrt{\frac{1}{Z_{1}^\pi}}O_{3i}O_{V3I}O_{V3J},
\ \label{VVPvert}
\eea
\vspace{-12mm}
\bea
\rho^+\pi^-V^I+\mathrm{h}.\mathrm{c}. : \ \gamma_I &=&
-\frac{4gc_3^{\mathrm{IP}}}{\sqrt{3}}O_{V2I}
-4c_5^{\mathrm{IP}}O_{V3I},\\
K^{*+}K^-V^I+\mathrm{h}.\mathrm{c}. : \ L_I &=&
-2gc_3^\IP\left(O_{V1I}-\frac{1}{\sqrt{3}}O_{V2I}\right)
-4c_5^{\IP}O_{V3I},\\
K^{*0}\bar{K^0}V^I+\mathrm{h}.\mathrm{c}. : \ 
\varphi_I&=&-2gc_3^\IP\left(O_{V1I}+\frac{1}{\sqrt{3}}O_{V2I}\right)
+4c_5^{\IP}O_{V3I}.
\eea\par
%%%%%%%%%%
Vector mesons can decay into $P\gamma$ directly with
the operator in Eq.\ (\ref{VPgamma}).
$VP\gamma$ vertex is absent in Ref. \cite{Hashimoto:1996ny}
since the relation $c_3^\IP=c_4^\IP$ is adopted.
Meanwhile, IP violating $VVP$ operator in Eq.\ (\ref{VVP})
also causes $V\rightarrow P\gamma$ with the $V-\gamma$ conversion
vertex in Eq.\ (\ref{Vgamma}).
The notation of propagators for neutral vector meson is given as,
\bea
iD^J_{\mu\nu}(Q)=ig_{\mu\nu}D^J(Q^2)+iQ_\mu Q_\nu D_L^J(Q^2), 
\quad (J=1, 2, 3),
\eea
where $J=1,2,3$ is assigned with the propagator
of $\rho, \omega$ and $\phi$, respectively.
In the calculation of the $V-\gamma$ conversion
decay $V\rightarrow PV^*\to P\gamma$,
the term proportional to $D_L^J(0)$ vanishes since
the momentum product $Q_\mu Q_\nu$
is eliminated when multiplied with antisymmetric tensor.
Consequently,
the conversion process $V\rightarrow PV^*\to P\gamma$
is proportional to the contribution from
the metric tensor part of
intermediate vector mesons.
With Eq.\ (\ref{MassTwoP}), one can find that the following
relation is satisfied,
\b
D^J(0)\cdot \left(-\frac{e\cM_{J}^2}{g}\eta_J\right)=-\frac{e}{g}\eta_J.\label{proV}
\e
Although vector meson propagator is shown
apparently in Fig.\ \ref{fig:VPgamma}(b),
the dependence on the mass cancels out
in Eq.\ (\ref{proV}).
\par
Decay amplitudes are obtained from the operators in
Eqs.\ (\ref{VPgamma}-\ref{K*K0V0}) as,
\bea
\mathcal{M}_{V^I\rightarrow P^i\gamma}&=&X_{iI}\epsilon^{\mu\nu\rho\sigma}
p_\mu^\gamma p_\nu^P\epsilon^V_\rho \epsilon_\sigma^{\gamma *}\label{MiI}\\
\mathcal{M}_{\rho^+\rightarrow \pi^+\gamma}&=&
X_{\rho^+}\epsilon^{\mu\nu\rho\sigma}
p^{\gamma}_\mu p^{\pi^+}_\nu \epsilon^{\rho^+}_\rho\epsilon^{\gamma*}_\sigma,\quad\\
\mathcal{M}_{K^{*+}\rightarrow K^+\gamma}&=&
X_{K^{*+}}
\epsilon^{\mu\nu\rho\sigma}p^{\gamma}_\mu p^{K^+}_\nu \epsilon^{K^{*+}}_\rho\epsilon^{\gamma*}_\sigma\\
\mathcal{M}_{K^{*0}\rightarrow K^0\gamma}
&=&X_{K^{*0}}\epsilon^{\mu\nu\rho\sigma}
p_\mu^\gamma p_\nu^{K^0}\epsilon_\rho^{K^*}\epsilon_\sigma^{\gamma*},\\
\mathcal{M}_{\eta^\prime\rightarrow V^I\gamma}&=&X_{3I}\epsilon^{\mu\nu\rho\sigma}p^V_\mu
p^\gamma_\nu
\epsilon^{V*}_\rho\epsilon^{\gamma*}_\sigma,\quad (I=1,2)\\
X_{iI}&=&\displaystyle\frac{e\sqrt{Z_I}}{f_{\pi}}\bar{\chi}_{iI},\quad\label{XiIa}\\
\bar{\chi}_{iI}&=&\chi_{iI}-\displaystyle\sum_{J=1}^3
\bar{\theta}_{iIJ}\eta_J, \quad (\bar{\theta}_{iIJ}=\theta_{iIJ}+\theta_{iJI})\\
&=&\frac{2gc_{34}^+}{\sqrt{3}}
\left[\left(\sqrt{\frac{Z^{\pi}_2}{Z^{\pi}_1}}O_{2i}+\frac{1}{\sqrt{3}}O_{1i}\right) O_{V1I}+\left(O_{1i}-\frac{1}{\sqrt{3}}\sqrt{\frac{Z^{\pi}_2}{Z^{\pi}_1}}O_{2i}\right)O_{V 2I}\right]\nn \\
&+&\frac{2 c_{69}}{g}\sqrt{\frac{1}{Z^{\pi}_1}}O_{3i}(O_{V1I}+\frac{1}{\sqrt{3}}O_{V2I})
+2c_8^{\IP}\left(O_{1i}+\frac{1}{\sqrt{3}}\sqrt{\frac{Z^{\pi}_2}{Z^{\pi}_1}}O_{2i}\right)O_{V3I},\;\\
X_{\rho^+}&=&\displaystyle\frac{e\sqrt{Z_{\rho}}}{f_{\pi}}\left(-\frac{2g}{3}c_{34}^--\displaystyle\sum_{J=1}^3\gamma_J \eta_J\right)
=\displaystyle\frac{2eg\sqrt{Z_{\rho}}}{3f_{\pi}}
c_{34}^+,\label{rhopra}\\
X_{K^{*+}}&=&\displaystyle\frac{e\sqrt{Z_{K^*}}}{f_{K}}\left(-\frac{2g}{3}c_{34}^-
-\displaystyle\sum_{J=1}^3L_J \eta_J\right)=\displaystyle\frac{2eg\sqrt{Z_{K^\ast}}}{3f_{K}}c_{34}^+,\label{Xrho+}\quad\\
X_{K^{*0}}&=&\displaystyle\frac{e\sqrt{Z_{K^*}}}{f_{K}}\left(\frac{4g}{3}c_{34}^-
+\displaystyle\sum_{J=1}^3 \varphi_J\eta_J\right)
=-\frac{4eg\sqrt{Z_{K^\ast}}}{3f_{K}}c_{34}^+,\label{XKstver}
\eea
where $c_{34}^+ = c_3^{\IP}+c_4^\IP$ and $c_{69}=2c_{6}^\IP+gc_{9}^\IP$.
The coefficient of neutral meson decay amplitude denoted as $X_{iI}$
in Eq.\ (\ref{XiIa})
includes the factor $(\sqrt{Z_1}, \sqrt{Z_2}, \sqrt{Z_3})=
(\sqrt{Z_\rho}, \sqrt{Z_\omega}, \sqrt{Z_\phi})$,
which comes from the wave function renormalization of an external vector line
in Fig.\ \ref{fig:VPgamma}.
In Eqs.\ (\ref{rhopra} ,\ref{Xrho+}), we assume that the wave function renormalization of charged vector meson is 
equal to one for neutral vector meson,
{\it i.e.}, $\sqrt{Z_{\rho^+}}=\sqrt{Z_{\rho}}$ and $\sqrt{Z_{K^*}}=\sqrt{Z_{K^{*+}}}$,
which is valid in the isospin limit.
One can write the partial decay width of
$V\rightarrow P\gamma$ and $P\to V\gamma$
with $X$'s in Eqs.\ (\ref{XiIa}, 
\ref{rhopra}-\ref{XKstver}) as,
\bea
\Gamma[V^I\rightarrow P^i\gamma]
&=&\frac{1}{96\pi}X_{iI}^2m_{I}^3\left(1-\frac{M_{P^i}^2}{m_{I}^2}\right)^3,\label{partialdecay}\\
\Gamma[\rho^+\rightarrow \pi^+\gamma]
&=&\frac{1}{96\pi}X_{\rho^+}^2m_{\rho^+}^3\left(1-
\frac{M_{\pi^+}^2}{m_{\rho^+}^2}\right)^3,\label{RhopWid}\\
\Gamma[K^{*+}\rightarrow K^+\gamma]
&=&\frac{1}{96\pi}X_{K^{*+}}^2m_{K^{*+}}^3\left(1-
\frac{M_{K^+}^2}{m_{K^{*+}}^2}\right)^3,\label{KstpWid}\\
\Gamma[K^{*0} \rightarrow K^0 \gamma]&=&\frac{1}{96\pi}X_{K^{*0}}^2m_{K^{*0}}^3\left(1-\frac{M_{K^0}^2}{m_{K^{*0}}^2}\right)^3,\label{Kstraddecay}\\
\Gamma[\eta^\prime\rightarrow V^I\gamma]
&=&\frac{1}{32\pi}X_{3I}^2M_{\eta^\prime}^3\left(1-\frac{m_{I}^2}{M_{\eta^\prime}^2}\right)^3\label{eta'partialdecay}.
\quad (I=1, 2)
\eea
The pseudoscalar decay width
in Eq.\ (\ref{eta'partialdecay}) is analogous to
one for $V\rightarrow P\gamma$ given in
Eqs.\ (\ref{partialdecay}-\ref{Kstraddecay}) and its
coefficient is different by a factor $1/3$ which comes from
spin average of vector meson.
One can find that 
$\Gamma[V^I\to P^i\gamma]$ and $\Gamma[\rho^+\to \pi^+\gamma]$
in Eqs.\ (\ref{partialdecay}-\ref{RhopWid})
provide the relation,
\b
\left|\frac{X_{iI}}{X_{\rho^+}}\right|=\sqrt{\frac
{\Gamma_{V^I\rightarrow P^i\gamma}}{\Gamma_{\rho^+\rightarrow \pi^+\gamma}}
\frac{m_{I}^3}{m_{\rho^+}^3}
\left(\displaystyle\frac{m_{\rho^+}^2-M_{\pi^+}^2}
{m_{I}^2-M_{P^i}^2}\right)^3}.\label{7-1}
\e
In the above relation, the ratio of the effective coupling for
$V^I\to P^i\gamma$
to one for $\rho^+\to\pi^+\gamma$ is
written in terms experimental data on r.h.s.
Using Eqs.\ (\ref{partialdecay}-\ref{RhopWid}), we can rewrite l.h.s. in
Eq.\ (\ref{7-1}),
\b
\left|\frac{X_{iI}}{X_{\rho^+}}\right|&=&\sqrt{\frac{Z_{I}}{Z_{\rho^+}}}
\left|O_{1i}(O_{V1I}+\sqrt{3}O_{V2I})+
\sqrt{\frac{Z^\pi_2}{Z^\pi_1}}O_{2i}
(\sqrt{3}O_{V1I}-O_{V2I}) \right.\nn\\
&&+\frac{\sqrt{3}c_{69}}{g^2c_{34}^+}\sqrt{\frac{1}{Z^\pi_1}}O_{3i}
(\sqrt{3}O_{V1I}+O_{V2I})
 \left.+\frac{\sqrt{3}c_8^\IP}{gc_{34}^+}
\left(\sqrt{3}O_{1i}+\sqrt{\frac{Z^\pi_2}{Z^\pi_1}}O_{2i}\right)O_{V3I}
\right|.\qquad\qquad\label{7-2}
\e
In the above relation, the effective coupling is
written in terms of model parameters.
We use the relations in
Eqs.\ (\ref{7-1}, \ref{7-2}) for the numerical
analysis of $\chi^2$ fitting.
%%%%%%%%%%%%%%%%%%
%%  phi -> omega pi  %%
%%%%%%%%%%%%%%%%%%
\subsubsection{$\phi\rightarrow \omega \pi^0$}
\begin{figure*}[h]
%\captionsetup{labelformat=empty,labelsep=none}
  \setlength{\subfigwidth}{.246\linewidth}
  \addtolength{\subfigwidth}{-.246\subfigcolsep}
  \begin{minipage}[b]{\subfigwidth}
    \subfigure{\includegraphics[width=4.35cm]{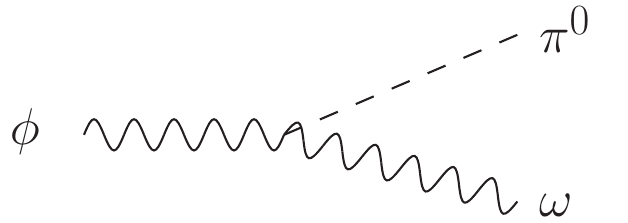}
   \label{phi1}}
  \end{minipage}
  \caption{Diagram contributing to the IP violating decay 
of $\phi\rightarrow \omega\pi^0$.}
\label{fig:phiomegapi}
\end{figure*}

In this subsection, an IP violating process of $\phi\rightarrow \omega \pi^0$ is analyzed.
The contributing operator to $\phi\rightarrow \omega \pi^0$
is given in Eq.\ (\ref{VVP}), and
the diagram is shown in Fig.\ \ref{fig:phiomegapi}.
%%%%%%%%%%%%%%
%%  Figures   %%%%
%%%%%%%%%%%%%%
%%%%%%%%%%%%%%%%
%%% End of Figure %%
%%%%%%%%%%%%%%%%
The transition amplitude of $\phi\rightarrow \omega \pi^0$ is written as,
\bea
\mathcal{M}_{\phi \rightarrow \omega \pi^0}
&=&X_{\phi\rightarrow \omega\pi^0}
\epsilon^{\mu\nu\rho\sigma}p^\omega_\mu
p^{\pi^0}_\nu \epsilon^\phi_\rho \epsilon^{\omega*}_\sigma
,\label{Mphiomegapi}\\
X_{\phi\rightarrow \omega\pi^0}
&=&-\frac{g\sqrt{Z_\phi Z_\omega}}{f_{\pi}}\bar{\theta}_{123}
\label{phiomegapi}.\eea
The contribution coming from $V-\gamma$ conversion
is negligible since it gives rise to $\mathcal{O}(\alpha)$
correction.
In Eq.\ (\ref{phiomegapi}), the factor of wave function renormalization of external vectors is included.
Thus, the partial decay width
of $\phi\rightarrow \omega\pi^0$ is,
\bea
\Gamma[\phi\rightarrow \omega\pi^0]&=&\frac{1}{96\pi}
X_{\phi\rightarrow \omega\pi^0}^2\nn\\
&&\times
\left(\frac{\sqrt{m_\phi^4+m_\omega^4+M_{\pi^0}^4
-2(m_{\phi}^2m_\omega^2+m_\omega^2M_{\pi^0}^2
+M_{\pi^0}^2m_{\phi}^2)}
}{m_\phi}
\right)^3.\label{Eq:phiomagapi0}
\eea
%%%%%%%%%%%%%
%%  P -> 2 gamma %
%%%%%%%%%%%%%
\subsubsection{$P\rightarrow 2\gamma$}
\begin{figure}[ht]
  \setlength{\subfigwidth}{.246\linewidth}
  \addtolength{\subfigwidth}{-.246\subfigcolsep}
  \begin{minipage}[b]{\subfigwidth}
    \subfigure{\includegraphics[width=4.1cm]{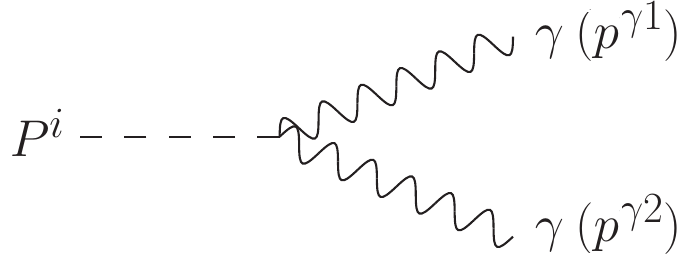}
   \label{pi2gamma1}}
  \end{minipage}
    \begin{minipage}[b]{\subfigwidth}
    \subfigure{\includegraphics[width=4.1cm]{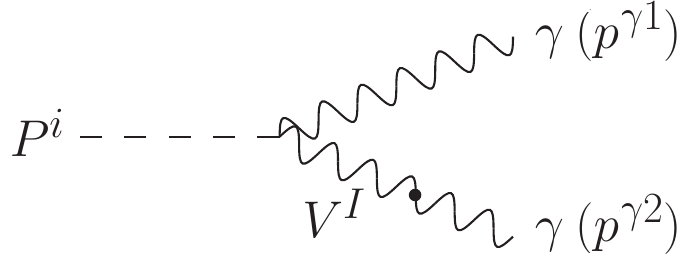}
   \label{pi2gamma2}}
  \end{minipage}
    \begin{minipage}[b]{\subfigwidth}
    \subfigure{\includegraphics[width=4.1cm]{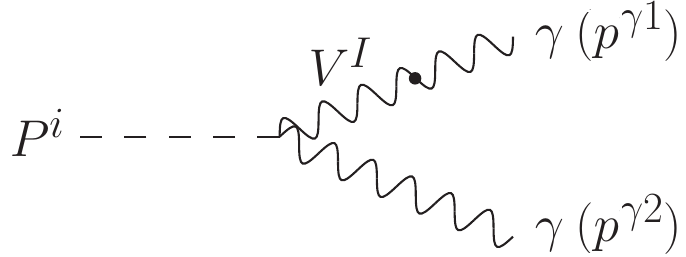}
   \label{pi2gamma3}}
  \end{minipage}
    \begin{minipage}[b]{\subfigwidth}
    \subfigure{\includegraphics[width=4.1cm]{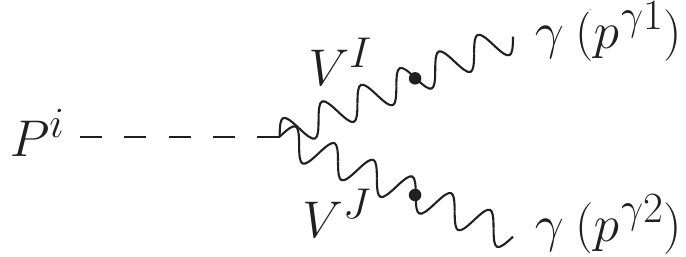}
   \label{pi2gamma4}}
  \end{minipage}
  \caption{Diagrams contributing to the
  decay width of $P^i\to 2\gamma$.}
\label{fig:pi2gamma}
\end{figure}

In this subsection, we evaluate partial decay widths
of the IP violating process given as $P^i\rightarrow 2\gamma$.
The IP violating interaction terms yield contribution to $P\gamma\gamma$
vertex as,
\bea
\mathcal{L}_{\mathrm{IPV}}
|_{P^i\gamma\gamma}
&=&-\displaystyle\frac{e^2}{f_{\pi}}h_i\epsilon^{\mu\nu\rho\sigma}P^i\partial_\mu A_\nu
\partial_\rho A_\sigma,\label{WZp02gamma} \\
h_i &=&
\left(\displaystyle\frac{1}{8\pi^2}+\displaystyle\frac{4c_4^\IP}{3}
\right)\left(O_{1i}+\displaystyle\frac{1}{\sqrt{3}}
\sqrt{\frac{Z_{2}^{\pi}}{Z_{1}^{\pi}}}O_{2i}
\right)-\displaystyle\frac{32}{3}c_{10}^\IP \sqrt{\frac{1}{Z_{1}^{\pi}}}O_{3i},\label{hiver}
\eea
where the first term proportional to $1/8\pi^2$
in Eq.\ (\ref{hiver}) implies the contribution from
the WZW term. 
Diagrams of the decay of $P^i\rightarrow 2\gamma$
are given in Fig.\ \ref{fig:pi2gamma}.
%%%%%%%%%%%%%%
%%  Figures   %%%%
%%%%%%%%%%%%%%
%%%%%%%%%%%%%%%%
%%% End of Figure %%
%%%%%%%%%%%%%%%%
With the operators in Eqs.\ (\ref{Vgamma}, \ref{VPgamma},
\ref{WZp02gamma}), the transition amplitude of
$P^i\rightarrow 2\gamma$ is written as,
\bea
\mathcal{M}_{P^i\rightarrow 2\gamma}
&=&R_i\epsilon^{\mu\nu\rho\sigma}p^{\gamma1}_\mu
p^{\gamma2}_\nu \epsilon^{\gamma1*}_\rho
\epsilon^{\gamma2*}_\sigma,\\
R_i&=&-\frac{e^2}{f_{\pi}}\left[2h_i-\frac{1}{g}\left(2\displaystyle\sum_{I=1}^3 \chi_{iI}\eta_I
-\displaystyle\sum_{I, J=1}^3 
\bar{\theta}_{iIJ}\eta_I\eta_J\right)
\right]\nn\\
&=&-\frac{e^2}{f_{\pi}}\left[\frac{1}{4\pi^2}\left(O_{1i}+\frac{1}{\sqrt{3}}\sqrt{\frac{Z_{2}^{\pi}}{Z_{1}^{\pi}}}O_{2i}\right)-\frac{16}{3}c_{6-9-10}\sqrt{\frac{1}{Z_{1}^{\pi}}}O_{3i}
\right] . \label{Ri} 
\eea
where $c_{6-9-10}=c_6^\IP/g^2+c_9^\IP+4c_{10}^\IP$.
%%%%%%%%%%%%%%%%%
%% 
%%%%%%%%%%%%%%%%%
The partial decay
width of $P^i\rightarrow 2\gamma$ is given as,
\bea
\Gamma[P^i\rightarrow 2\gamma]=
\frac{1}{64\pi}R_i^2M_{P^i}^3\label{pi2gamma}.
\eea
%%%%%%%%%%%%%%%%%%%
%%  P -> gamma l+ l- %%%
%%%%%%%%%%%%%%%%%%%
\subsubsection{$P\rightarrow \gamma l^+l^-$}
\begin{figure}[ht]
  \setlength{\subfigwidth}{.24\linewidth}
  \addtolength{\subfigwidth}{-.24\subfigcolsep}
  \begin{minipage}[b]{\subfigwidth}
    \subfigure{\includegraphics[width=4cm]{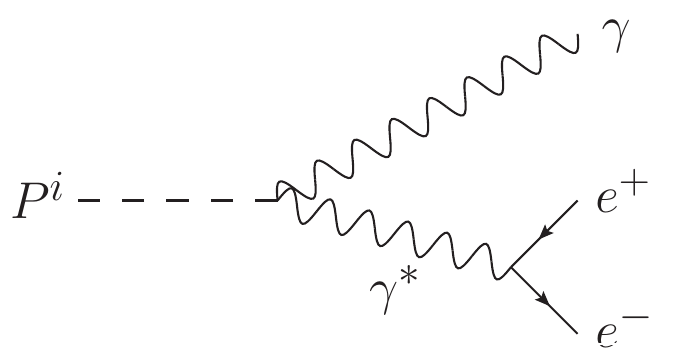}
   \label{Ptogll1}}
    \end{minipage}
    \begin{minipage}[b]{\subfigwidth}
    \subfigure{\includegraphics[width=4cm]{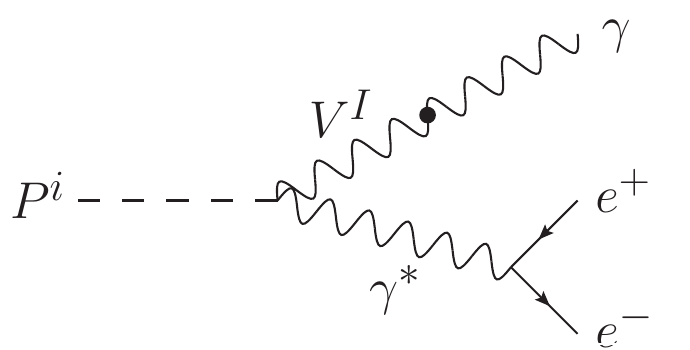}
   \label{Ptogll2}}
  \end{minipage}
    \begin{minipage}[b]{\subfigwidth}
    \subfigure{\includegraphics[width=4cm]{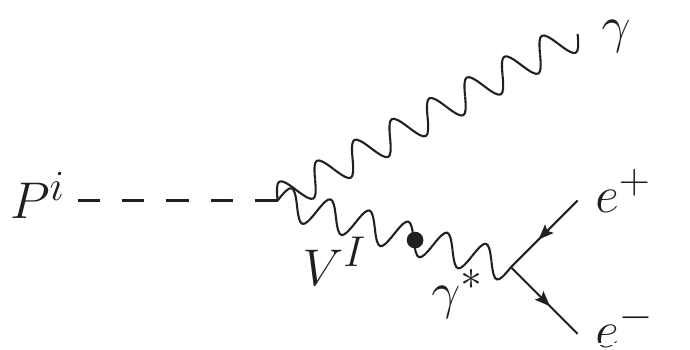}
   \label{Ptogll3}}
  \end{minipage}
    \begin{minipage}[b]{\subfigwidth}
    \subfigure{\includegraphics[width=4cm]{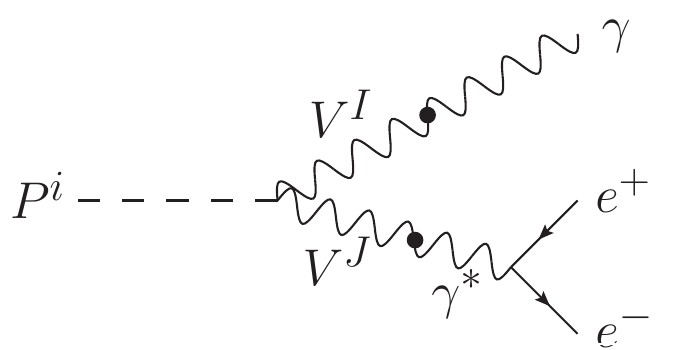}
   \label{Ptogll4}}
  \end{minipage}
        \caption{Diagrams contributing to
        the decay width of $P^i\to\gamma l^+l^-$.}
\label{Ptogll}
\end{figure}

In this subsection, a form factor for
IP violating modes $P^i\rightarrow \gamma l^+l^-$ is
obtained.
The contributing diagrams are displayed in Fig.\ \ref{Ptogll}.
%%%%%%%%%%%%
%%   Figures   %%
%%%%%%%%%%%%
%%%%%%%%%%%%%%%%
%%  Figures end   %%%
%%%%%%%%%%%%%%%%
Following the notations used in experiments,
the differential decay width is written in terms of
the TFF as,
\bea
\frac{\mathrm{d}\Gamma(P^i\to\gamma l^+ l^-)}{\mathrm{d}s\mathrm{d}\cos\theta}&=&
\frac{\alpha}{4\pi}
\Gamma(P^i\to2\gamma)\frac{\beta_l}{s}(2-\beta^2_l\sin^2\theta)
\left(1-\frac{s}{M_{P^i}^2}\right)^3
|F_{P^i}(s)|^2,\quad\quad\label{Not0}\\
\frac{\mathrm{d}\Gamma(P^i\to\gamma l^+ l^-)}{\mathrm{d}s}&=&\frac{2\alpha}{3\pi}
\Gamma(P^i\to2\gamma)
\frac{\beta_l}{s}
\left(1+\frac{2m_{l}^2}{s}\right)
\left(1-\frac{s}{M_{P^i}^2}\right)^3
|F_{P^i}(s)|^2, \label{Not1}\\
\beta_l&=&\sqrt{1-4m_l^2 /s},
\eea
where $s$ denotes the squared invariant mass
in di-lepton system while $\theta$ indicates
an angle between $P^i$ and $l^+$ in the di-lepton
rest frame.
The model prediction for the TFF is,
\b
|F_{P^i}(s)|^2=\left|1+\frac{e^2 s}{gf_{\pi}R_i}\displaystyle\sum_{I=1}^3\bar{\chi}_{iI}\eta_I\delta B_{VII}D_I(s)\right|^2
.\e
%%%%%%%%%%%%%
%%  V -> P l+ l- %
%%%%%%%%%%%%%
\subsubsection{$V\rightarrow P l^+l^-$}
\begin{figure}[ht]
  \setlength{\subfigwidth}{.5\linewidth}
  \addtolength{\subfigwidth}{-.5\subfigcolsep}
  \begin{minipage}[b]{\subfigwidth}
    \subfigure{\includegraphics[width=6cm]{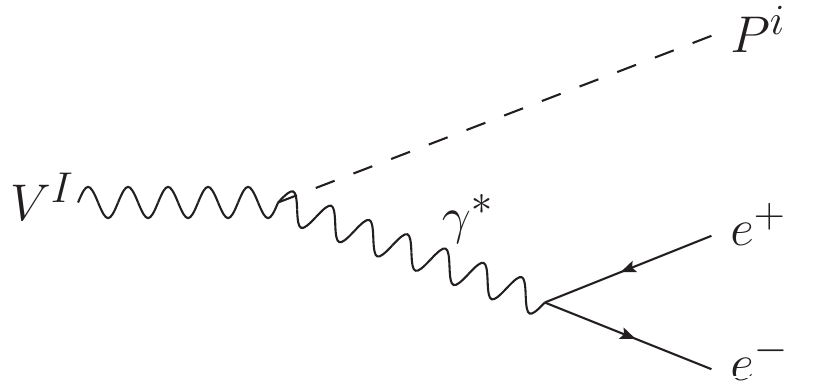}
   \label{VtoPll1}}
  \end{minipage}
    \begin{minipage}[b]{\subfigwidth}
    \subfigure{\includegraphics[width=6cm]{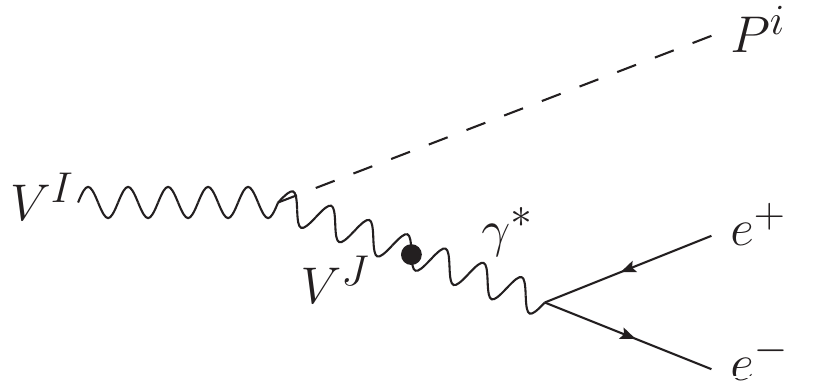}
   \label{VtoPll2}}
  \end{minipage}
        \caption{Diagrams contributing to
        the decay width of $V^I\to P^i\l^+l^-$.}
\label{VtoPll}
\end{figure}

In this subsection, a form factor for IP violating electromagnetic decays
for neutral vector mesons is analyzed.
Contributing diagrams are exhibited in Fig.\ \ref{VtoPll}.
%%%%%%%%%%%%
%%   Figures   %%
%%%%%%%%%%%%
%%%%%%%%%%%%%%%%
%%  Figures end   %%%
%%%%%%%%%%%%%%%%
The differential decay width for $V\to Pl^+l^-$ is
written in terms of the TFF in the following form,
\bea
\frac{\mathrm{d}^2\Gamma(V^I\to P^i l^+ l^-)}{\mathrm{d}s\mathrm{d}\cos\theta}&=&
\frac{\alpha}{8\pi}
\Gamma(V^I\to P^i\gamma)
\frac{\beta_l}{s}
(2-\beta^2_l\sin^2\theta)\nn\\
&&\times
\left[\left(1+\frac{s}{m_{I}^2-M_{P^i}^2}\right)^2-
\frac{4m_{I}^2 s}{(m_{I}^2-M_{P^i}^2)^2}
\right]^{\frac{3}{2}}
|F_{V^IP^i}(s)|^2,\label{Nt1}\\
\frac{\mathrm{d}\Gamma(V^I\to P^i l^+ l^-)}{\mathrm{d}s}&=&\frac{\alpha}{3\pi}
\Gamma(V^I\to P^i\gamma)
\frac{\beta_l}{s}\left(1+\frac{2m_{l}^2}{s}\right)\nn\\
&&\times
\left[\left(1+\frac{s}{m_{I}^2-M_{P^i}^2}\right)^2-
\frac{4m_{I}^2 s}{(m_{I}^2-M_{P^i}^2)^2}
\right]^{\frac{3}{2}}
|F_{V^IP^i}(s)|^2,\label{Not2}
\eea
where
$\theta$ is an angle between
$V^I$ and $l^+$ in the di-lepton rest frame.
As the model prediction, the TFF is obtained as,
\bea
|F_{V^{I}P^i}(s)|^2=\left|1+\frac{s}{\bar{\chi}_{iI}}\displaystyle\sum_{J=1}^3
\bar{\theta}_{iIJ}\eta_J\delta B_{VJJ}D_J(s)\right|^2.
\label{TFFVECTORM}
\eea
The TFF in the above equation
are normalized as unity in the limit where 
virtual photon goes on-shell.
%%%%%%%%%%%%%%
%%   V -> 3 pi   %%
%%%%%%%%%%%%%%
\subsubsection{$V\rightarrow P\pi^+\pi^-$}
\begin{figure}[ht]
  \setlength{\subfigwidth}{.246\linewidth}
  \addtolength{\subfigwidth}{-.246\subfigcolsep}
  \begin{minipage}[b]{\subfigwidth}
    \subfigure{\includegraphics[width=4.1cm]{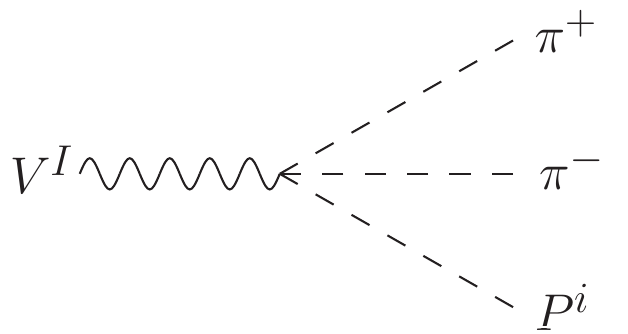}
   \label{Vto3pi1}}
  \end{minipage}
    \begin{minipage}[b]{\subfigwidth}
    \subfigure{\includegraphics[width=4.1cm]{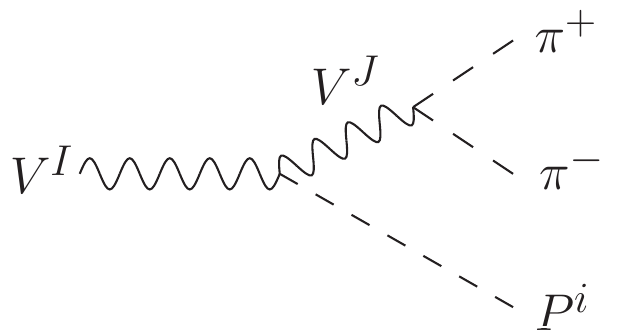}
   \label{Vto3pi2}}
  \end{minipage}
    \begin{minipage}[b]{\subfigwidth}
    \subfigure{\includegraphics[width=4.1cm]{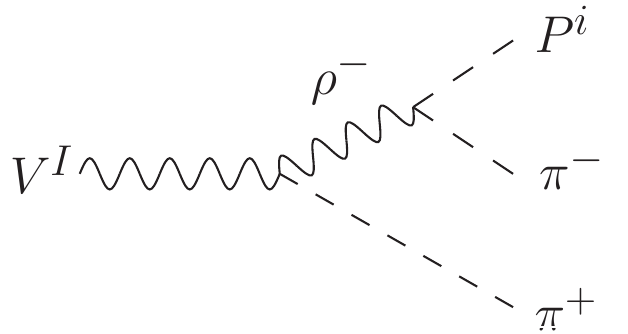}
   \label{Vto3pi3}}
  \end{minipage}
    \begin{minipage}[b]{\subfigwidth}
    \subfigure{\includegraphics[width=4.1cm]{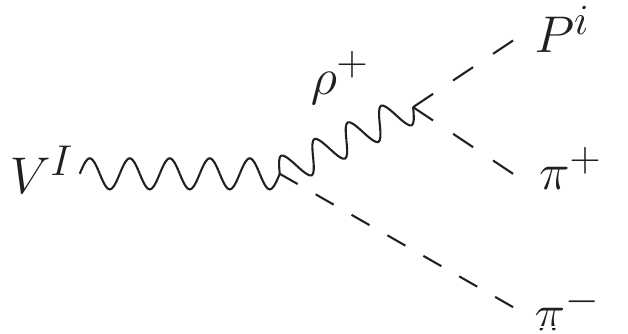}
   \label{Vto3pi4}}
  \end{minipage}
  \caption{Diagrams contributing to the decay width of $V^I\to P^i\pi^+\pi^-$.}
\label{fig:Vto3pi}
\end{figure}

In this subsection, partial decay widths
for $V\rightarrow P\pi^+\pi^-$
are analyzed.
Interaction terms for the process are,
\bea
\mathcal{L}_{\mathrm{IPV}}
|_{V^IP^i\pi^+\pi^-}
&=&
i\frac{J_{iI}}{f^3_{\pi}}\epsilon^{\mu\nu\rho\sigma}V_\mu^I\partial_\nu P^i
\partial_\rho\pi^+\partial_\sigma\pi^-,\label{IPVoP0pi+pi-} \\
\mathcal{L}_{\chi}|_{\rho^-P^i\pi^++\hc}&=&ig_{\rho\pi\pi}O_{1i}
\left[\rho^-_\mu \left(P^i\overleftrightarrow{\partial}^\mu
\pi^+\right)-\rho^+_\mu \left(P^i\overleftrightarrow{\partial}^\mu
\pi^-\right)\right]
\label{rho-rho+P0} , \\
J_{iI}&=&\frac{gc_{123}}{\sqrt{3}}
\left(3O_{1i}O_{V2I}+
\sqrt{\frac{Z^\pi_2}{Z^\pi_1}}O_{2i}O_{V1I}\right)
+12c_5^{\IP}O_{1i}O_{V3I},
\eea
where
$c_{123} = c_{12}^-+2c_3^{\mathrm{IP}}$.
The diagrams for the decay of $V^I\rightarrow P^i\pi^+\pi^-$ are given in Fig.\ \ref{fig:Vto3pi}.
%%%%%%%%%%%%%%
%%  Figures   %%%%
%%%%%%%%%%%%%%
Propagators for $\rho^\pm$ are formulated in the following form as,
\bea
iD^{\mu\nu}_{\pm}(Q)=ig^{\mu\nu}D_{\pm}(Q^2)+
iQ^\mu Q^\nu D_{L\pm}(Q^2).\label{propVc}
\eea
%%%%%%%%%%%%%%%%
%%% End of Figure %%%%%
%%%%%%%%%%%%%%%%
The transition amplitude is given as,
\bea
\mathcal{M}&=&Y_{iI}\epsilon^{\mu\nu\rho\sigma}
\epsilon^V_\mu p^-_\nu p^+_\rho p^0_\sigma,\\
Y_{iI}&=&\frac{\sqrt{Z_I}}{f^3_{\pi}}\left[J_{iI}+\displaystyle\sum_{J=1}^{3}\zeta_{iIJ}D^J(s_{+-})
+\kappa_{iI}(D_+(s_{+0})+D_-(s_{-0}))\right],\;\;\\
\zeta_{iIJ}&=&M_V^2\bar{\theta}_{iIJ}\Pi_J,
\quad\kappa_{iI}=M_V^2O_{1i}\gamma_I,\label{kappa}
\eea
where $s_{+0}, s_{-0}$ and $s_{+-}$ are squared invariant masses for the $\pi^+ P^i, \pi^- P^i$ and $\pi^+\pi^-$
system, respectively.
$s_{+-}$ is kinematically related with the other variables
as $s_{+-}=m_{I}^2+2M_{\pi^+}^2+M_{P^i}^2-s_{+0}-s_{-0}$.
The formula of the partial decay width is
obtained as,
\bea
\Gamma[V^I&\rightarrow& P^i \pi^+\pi^-]=
\displaystyle\frac{1}{3072\pi^3 m_{I}^3}\displaystyle\iint^{(m_{I}-M_{\pi^+})^2}_{(M_{\pi^+}+M_{P^i})^2}
|Y_{iI}|^2H\theta(H)ds_{+0}ds_{-0},\label{omega3pi}\\
H&=&s_{+-}[s_{+0}s_{-0}+(m_{I}^2-M_{\pi^+}^2)(M_{\pi^+}^2-M_{P^i}^2)]-M_{\pi^+}^2(m_{I}^2-M_{P^i}^2)^2 ,
\eea
where
$\theta(H)$ denotes step function,
and the integral regions are common
for $s_{+0}$ and $s_{-0}$.
%%%%%%%%%%%%%%%%%%
%%  Pi -> pipi gamma %%
%%%%%%%%%%%%%%%%%%
\subsubsection{$P\rightarrow \pi^+\pi^-\gamma$}
\begin{figure}[ht]
  \setlength{\subfigwidth}{.246\linewidth}
  \addtolength{\subfigwidth}{-.246\subfigcolsep}
  \begin{minipage}[b]{\subfigwidth}
    \subfigure{\includegraphics[width=4.1cm]{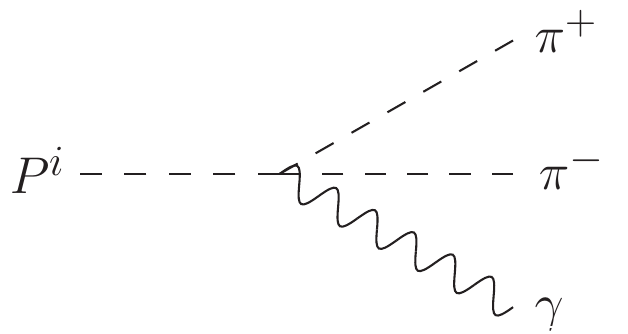}
   \label{Pto2pigamma1}}
  \end{minipage}
    \begin{minipage}[b]{\subfigwidth}
    \subfigure{\includegraphics[width=4.1cm]{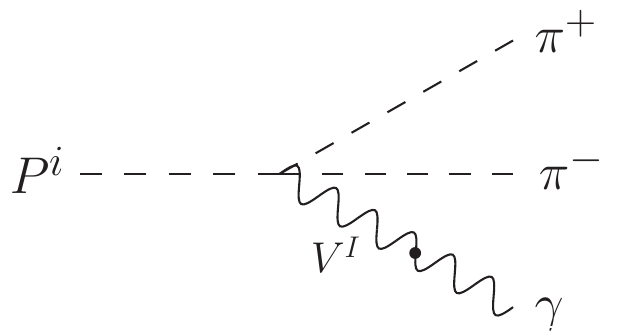}
   \label{Pto2pigamma2}}
  \end{minipage}
    \begin{minipage}[b]{\subfigwidth}
    \subfigure{\includegraphics[width=4.1cm]{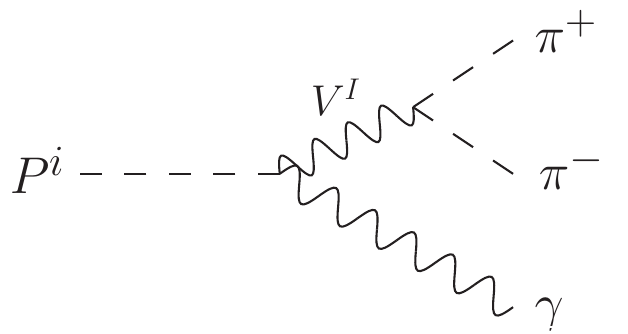}
   \label{Pto2pigamma3}}
  \end{minipage}
    \begin{minipage}[b]{\subfigwidth}
    \subfigure{\includegraphics[width=4.1cm]{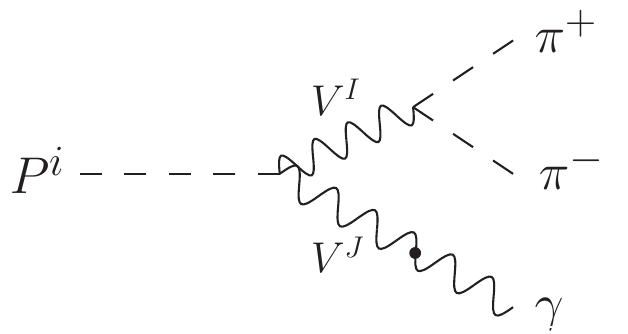}
   \label{Pto2pigamma4}}
  \end{minipage}\\
  \begin{minipage}[b]{\subfigwidth}
    \subfigure{\includegraphics[width=4.1cm]{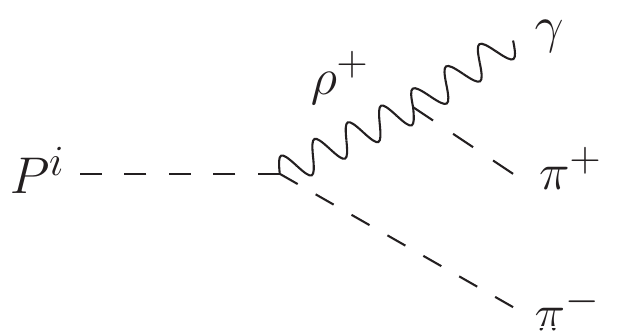}
   \label{Pto2pigamma5}}
  \end{minipage}
    \begin{minipage}[b]{\subfigwidth}
    \subfigure{\includegraphics[width=4.1cm]{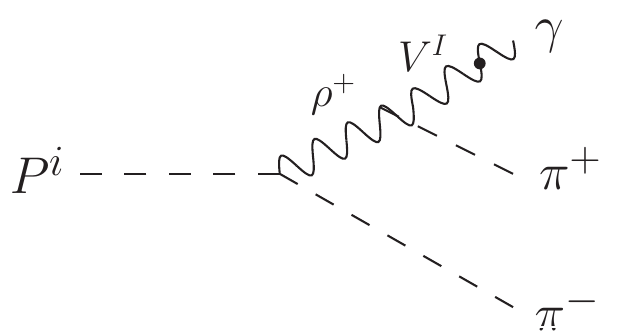}
   \label{Pto2pigamma6}}
  \end{minipage}
    \begin{minipage}[b]{\subfigwidth}
    \subfigure{\includegraphics[width=4.1cm]{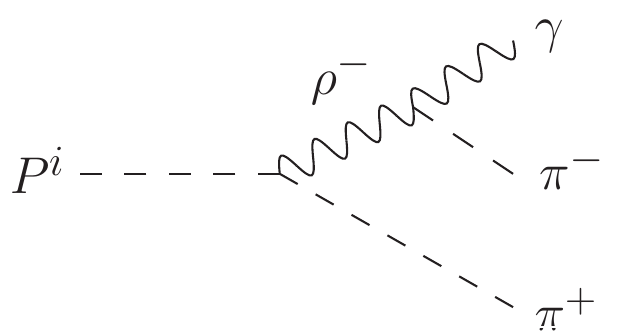}
   \label{Pto2pigamma7}}
  \end{minipage}
    \begin{minipage}[b]{\subfigwidth}
    \subfigure{\includegraphics[width=4.1cm]{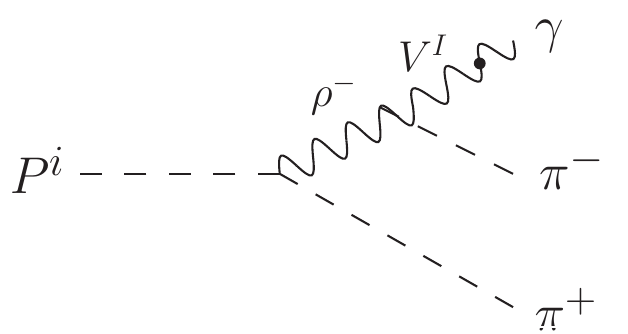}
   \label{Pto2pigamma8}}
  \end{minipage}
        \caption{Diagrams contributing to 
the decay width for $P^i\to\pi^+\pi^-\gamma$.}
\label{fig:Pto2pigamma}
\end{figure}

In this subsection, differential decay widths for $P\rightarrow \pi^+\pi^-\gamma$ are calculated.
The diagrams contributing to this process
are given in Fig.\ \ref{fig:Pto2pigamma}.
%%%%%%%%%%%%%%%%
%%% End of Figure %%
%%%%%%%%%%%%%%%%
The transition amplitude for the process
$P^i\rightarrow \pi^+\pi^-\gamma\ (i=2, 3)$ is,
\bea
\mathcal{M}_{P^i\rightarrow \pi^+\pi^-\gamma}
&=&Y^{\gamma}_i\epsilon^{\mu\nu\rho\sigma}
\epsilon_{\mu}^{\gamma*}p^-_\nu p^+_\rho p^\gamma_\sigma,\label{ygamma}\\
Y_i^\gamma&=&
-\frac{e}{f^3_{\pi}}\left[\bar{A}_{i}+\bar{B}_{iI}D^I(s)
+\bar{C}_{i}
(D_{+}(s_{+0})+D_{-}(s_{-0}))\right]
\label{Ppipigamma},\\
\bar{A}_{i}&=&\left(\frac{1}{4\pi^2}+2c_{34}^+\right)\left(O_{1i}+\frac{1}{\sqrt{3}}\sqrt{\frac{Z_2^\pi}{Z_1^\pi}}O_{2i}\right),\nn\\
\bar{B}_{iI}&=&-2g_{\rho\pi\pi}f^2_{\pi}\bar{\chi}_{iI}\Pi^I,\quad
\bar{C}_{i}=-\frac{4g_{\rho\pi\pi}f^2_{\pi}}{3}gc_{34}^+O_{1i}\label{Eq:ABC}
.\eea
Using the above equations, one can obtain the differential decay width,
\bea
\displaystyle\frac
{\mathrm{d}^2\Gamma[P^i\rightarrow\pi^+\pi^-\gamma]}{
\mathrm{d}s\mathrm{d}\cos\theta}
=\displaystyle\frac{1}{8192\pi^3M_{P^i}^3}|Y_i^\gamma|^2
\sin^2\theta s^4\beta^3_{\pi^+}\left(1-\frac{M_{P^i}^2}{s}\right)^3,
\label{ingamma}
\eea
where $s$ denotes the squared invariant mass in $\pi^+\pi^-$
system and $\theta$ implies the angle between
$\pi^+$ and $\gamma$ in the rest frame of $\pi^+\pi^-$.
%%%%%%%%%%%%
%%%   Fitting %%%
%%%%%%%%%%%%
\section{Numerical analysis}
\label{sec4}
In this section, phenomenological analysis is carried out
in the model.
In the following subsection,
we perform $\chi^2$ fittings
in order to estimate the parameters in the model.
As input data in the fittings,
the following data are utilized:
(1) the spectral function of $\tau$ decay,
(2) the masses of vector mesons and
(3) the IP violating decay widths, the masses
of pseudoscalars and the TFFs of $V\to Pl^+l^-$.
Subsequently, using the parameters estimated from
the aforementioned observables,
we give the prediction of the model.
Specifically, the results are presented
for Dalitz distributions and partial decay widths of IP 
violating modes.
\par
In order to carry out the analysis,
the following points are addressed:
\begin{itemize}
\item For the parameter $c$ defined in Eq.\ (\ref{Eq:cexp}),
we take $f=f_\pi$.
\item For $\mu_P$ given in Eq.\ (\ref{Eq:Defmu}),
$f=f_\pi$ is also taken.
\item 
In the expression of $(g_{\rho\pi\pi})_{\mathrm{tree}}$
in Eqs.\ (\ref{Eq:DeltagKKP}, \ref{Eq:DeltaRPP},
\ref{Eq:PhiPP}-\ref{Eq:GOPP}, \ref{Eq:LVPP}, \ref{Eq:PII}),
we use tree-level decay constant $f$,
which is a free parameter.
\end{itemize}
\subsection{Parameter fit}
%%%%%%%%%%%%%%%%%%%%%%%%%%%%%%%%%%%%%%%%%%%%%%%%%%%%%%%%
%%%%%%%%%%%%%%%%%%%%%%%%%%%%%%%%%%%%%%%%%%%%%%%%%%%%%%%%%%%%%%%%%%%%%%%%%
\label{Sec:para}
\subsubsection{$\tau^- \to K_s \pi^- \nu$}
%%%%%%%%%%%%%%%%%%%%%%%%%%%%%%%%%%%%%%%%%%%%%%%%%%%%%%%%%%%%%%%%%%%%%%%%%
In this subsection, we estimate
parameters in the model with the decay distribution
for $\tau^- \to K_s \pi^- \nu$. To evaluate the decay distribution, we use
the procedure similar to the method in Ref.\ \cite{Kimura:2014wsa}.
Throughout the analysis, we take isospin limit in the decay distribution.
\par
The differential branching fraction for $KP\nu\ (P=\pi, \eta)$ is given as,
\bea
\frac{\mathrm{d}{\rm Br}[\tau \to KP\nu]}{\mathrm{d}\sqrt{Q^2}}
&=& \frac{1}{\Gamma_\tau} \frac{G_F^2 |V_{us}|^2}{2^5 \pi^3} 
   \frac{(m_{\tau}^2-Q^2)^2}{m_{\tau}^3}p_{K} \nn \\
&&\times \left[ \left(\frac{2 m_{\tau}^2}{3 Q^2}+\frac{4}{3} \right)  
{p_{K}}^2  |F_V^{KP}(Q^2)|^2 
+\frac{m_{\tau}^2}{2} |F_S^{KP}(Q^2)|^2 \right] ,
\label{bran}
\eea
where $p_K$ is the momentum of $K$ in the
hadronic center of mass (CM) frame.
The vector and scalar form factors are written in
App.\ \ref{Formfactor}.
In order to compare the model prediction
with the Belle data,
we use the method in Ref.\ \cite{%Boito:2008fq, 
Kimura:2014wsa}.
Including the overall normalization,
the differential width in Eq.\ (\ref{bran}) is rewritten as,
\bea
\frac{N_{\mathrm{tot}}}{\mathrm{Br}^{\mathrm{Belle}}[\tau^-
\to K_s\pi^-\nu]}\times
11.5\ \mathrm{MeV}\times \frac{\mathrm{d} {\rm Br}[\tau^-\to K_s\pi^-\nu]}{\mathrm{d}\sqrt{Q^2}},
\label{inclOverall}
\eea
where $N_{\mathrm{tot}}$ denotes
the observed number
of events for $\tau$ decay
while $11.5\ \mathrm{MeV}$
indicates the width of bins in the Belle experiment.
We carry out the $\chi^2$ fitting based on
Eq.\ (\ref{inclOverall}), which represents
the expected number of events in the model.
\par
In this paper, we take the tree-level pion 
decay constant, $f$, as a parameter.
Since the effect of the $K^*$ resonance 
is important in the decay mode,
we choose the mass and the decay width of $K^*$ meson
as fitting parameters.
Additionally, the octet vector meson mass and
the finite parts of 1-loop ordered
coefficients, $K_1^r+K_2^r$ and $L_9^r$,
are also free parameters.
To summarize, $(M_V,\ K_1^r+K_2^r,\ L_9^r,\ M_{K^*}, \ \Gamma_{K^*},\ f)$
are the relevant fitting parameters in this mode.
These six parameters are estimated from
90 bins of the data in the region
$M_K+M_\pi \leq
\sqrt{Q^2} \leq 1665$ MeV.
As a result of fitting,
the parameters are determined as,
\begin{eqnarray}
M_V &=& 851 \pm 100 {\rm MeV}, 
\ \   
K_1^r+K_2^r = 0.0268 \pm 0.0091,\ \
L_9^r = (2.06 \pm 1.89)\times 10^{-3},\quad \quad \label{Taupara}\\
M_{K^*} &=& 895.6 \pm 0.3 {\rm MeV}, \ \ \quad\;\;
\Gamma_{K^*} = 48.4 \pm 0.6 {\rm MeV},\ \ \;\;\;
f =136\pm 19 {\rm MeV},\quad\quad
\nonumber
\end{eqnarray}
where the obtained
$\chi^2_\mathrm{min}/{\rm n.d.f.}$ is $147.1 /84$.
The correlation matrix of
$(M_V,\ K_1^r+K_2^r,\ L_9^r,\ M_{K^*}, \ \Gamma_{K^*},\ f)$ is,
\bea
\begin{pmatrix}
1 & \hspace{5.2mm}0.28 &\hspace{4.2mm} 0.26 &\hspace{2.mm} 0.49 &\hspace{2.mm} 0.29 &\hspace{3.5mm} -0.64\\
* & \hspace{5.2mm}1      &\hspace{4.2mm} 1.0   &\hspace{2.mm} -0.071&\hspace{2.mm} 0.41 &\hspace{3.5mm} -0.92\\
* & \hspace{5.2mm}*      & \hspace{4.2mm} 1      &\hspace{2.mm} -0.067&\hspace{2.mm} 0.45 &\hspace{3.5mm} -0.91\\
* & \hspace{5.2mm}*      &\hspace{4.2mm}  *      &\hspace{2.mm}  1         &\hspace{2.mm} 0.26&\hspace{3.5mm} -0.15\\
* & \hspace{5.2mm}*      &\hspace{4.2mm}  *      &\hspace{2.mm}  *         &\hspace{2.mm}   1    &\hspace{3.5mm} -0.44\\
*  & \hspace{5.2mm}*     & \hspace{4.2mm} *      &\hspace{2.mm}  *         &\hspace{2.mm}  *      &\hspace{3.5mm}  1
\end{pmatrix}.
\label{Taucorr}\eea
\par
Since the tree-level $K^*K\pi$ coupling is
proportional to $M_V^2/f^2$,
tree-level expressions of the form factors in Eq. (\ref{bran})
does not depend on $M_V$ and $f$ solely,
but on the ratio.
Due to this fact, $M_V$ and $f$ are determined through
1-loop correction in the form factors
so that large errors arise from the fitting,
as shown in Eq.\ (\ref{Taupara}).
\par
\begin{figure}[!t]
  \begin{center}
    \includegraphics[width=10cm]{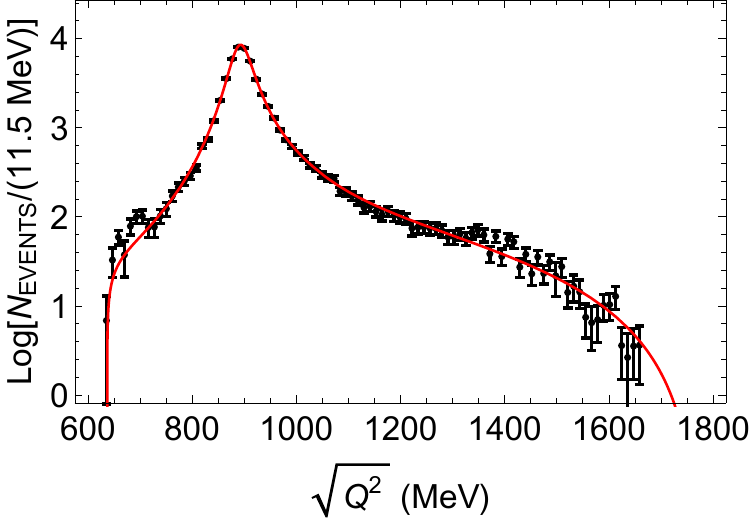}
  \end{center}
  \vspace{-0.5cm}
  \caption{The fitting result of the decay distribution for $\tau^- \to K_s\pi^-\nu$.
The red line corresponds to the prediction of our model. The closed circles with the 
error bars are experimental data \cite{Epifanov:2007rf}.}
  \label{dir_Kpi}
\end{figure}

The result of the decay distribution is shown in Fig.\ \ref{dir_Kpi}. In this plot, one can find that the resonance
of $K^{*}$ is seen
around $\sqrt{Q^2}\simeq 900$ MeV.
%%%
The prediction for the branching fraction is
%%%%%%%%%%%%%%%%%%%%%%%%%%%%%%%%%%%%%%%%
% Median \pm 34.1% $0.364^{+0.066}_{-0.097} \%$%%
%%%%%%%%%%%%%%%%%%%%%%%%%%%%%%%%%%%%%%%%
$0.403\pm0.069 \%$
(the experimental value is
($0.404\pm0.002\pm0.013$)$\%$
\cite{Epifanov:2007rf}).
\par
%%%
%%%
%%%%%%
\begin{table} [!h]
\begin{center}
 \caption{Numerical values of the parameters in the model.}
 \label{fit_para0}
%\begin{ruledtabular}
 \begin{tabular}{cccccc} \hline
  $g$ & $6.68\pm1.56$ & $C_2^r$ & $-0.415\pm0.331$ & 
  \ \ 
  $g_{\rho\pi\pi}$ & $6.37\pm0.04$
  \\  
  $Z_V^r$ & $0.819\pm0.002$ &\ \ $C_3^r-4C_4^r$ & $-0.149^{+0.080}_{-0.086}$ 
& \ \ 
$(g_{\rho\pi\pi})_\mathrm{tree}$ & $2.9^{+1.1}_{-0.7}$
\\ 
$C_1^r$ & $0.275\pm0.007 $ & $C_5^r$ & $(9.92^{+18.62}_{-8.88})\times 10^{-4}$ 
    & $c$ & $-0.91^{+0.37}_{-0.53}$\\ 
 $\sqrt{Z_1^\pi}$ & $1.49^{+0.27}_{-0.24}$ &
 $L_4^r$& $(-1.6^{+1.0}_{-1.1})\times10^{-3}$&
 \\
 $\sqrt{Z_2^\pi}$ & $0.96^{+0.17}_{-0.14}$ &
 $L_5^r$ &
 $(4.6^{+2.1}_{-7.3})\times10^{-3}$ &
  \\
    \hline
 \end{tabular}
\end{center}
\end{table}
In Table \ref{fit_para0}, we show other parameters
which are also determined through
Eqs.\ (\ref{Taupara}, \ref{Taucorr}).
In the following, we clarify how the parameters
in Table \ref{fit_para0} are determined.
In order to obtain the $\rho\pi\pi$ coupling,
we note that
the decay width of $K^*$ is given by the imaginary part of the self-energy \cite{Kimura:2014wsa},
\bea
\Gamma_{K^*}=\frac{1}{16\pi M_{K^*}}
\frac{\nu_{K\pi}^3(M_{K^*}^2)}{M_{K^*}^4}
\left( \frac{g_{\rho\pi\pi}}{2} \right)^2,
\label{eq:GKstar}
\eea
where $\nu_{P\pi}(Q^2)$ is defined in Eq.\ (\ref{Eq:nuDef}).
Solving Eq.\ (\ref{eq:GKstar}) with respect to $g_{\rho\pi\pi}$,
one can fix the $\rho \pi \pi$ coupling since $(M_{K^*}, \Gamma_{K^*})$ are determined from the fitting
result in Eq.\ (\ref{Taupara}).
Moreover, $g$ is also obtained from
the definition of the $\rho\pi\pi$ coupling in Eq.\ (\ref{Defgpp}).
Since the large error of $M_V$ propagates to $g$, the error of $g$ increases.
In Table\ \ref{fit_para0}
we also give the value of $(g_{\rho\pi\pi})_\mathrm{tree}$ in Eq.\ (\ref{eq:gpptree}).
One can find that $(g_{\rho\pi\pi})_\mathrm{tree}$
and $g_{\rho\pi\pi}$ are deviated from each other.
This is because the tree-level decay constant
denoted as $f$ given in Eq.\ (\ref{Taupara})
is deviated from PDG value, $f_\pi=92.2\ \mathrm{MeV}$.
In order to calculate $L_4^r$ and $L_5^r$, we use the following pion and kaon decay constants
\cite{Gasser:1984ux} with obtained $f$,
\begin{eqnarray}
f_\pi &=& f \left\{ 1 - c (2\mu_\pi+\mu_K)
 +4\left(\frac{M_\pi^2+2M_K^2}{f^2}L_4^r + \frac{M_\pi^2}{f^2}L_5^r
\right) \right\},
\label{fpi} \\
f_K &=& f \left\{ 1 - \frac{3c}{4} (\mu_\pi+2\mu_K+\mu_{\eta_8})
 + 4 \left(\frac{M_\pi^2+2M_K^2}{f^2}L_4^r + \frac{M_K^2}{f^2}L_5^r
\right) \right\} ,
\label{fK}
\end{eqnarray}
where in the above expressions,
$f$ represents the tree-level parameter
given in Eq.\ (\ref{Taupara}).
The coefficients of 1-loop ordered interaction,
$(C_3^r-4C_4^r,\ C_5^r)$,
are also determined from the procedure
similar to one of Ref.\ \cite{Kimura:2014wsa}.
Wavefunction renormalizations for
$\pi_3$ and $\eta_8$
are calculated from Eq.\ (\ref{WFR1}) and Eq.\ (\ref{WFR2}), respectively.
If one fixes the parameters as the best fit values in
Eq.\ (\ref{Taupara}), the wavefunction renormalizations
are,
\bea
\sqrt{Z_1^\pi}=1.52,\quad
\sqrt{Z_2^\pi}=0.763.
\label{WAVEbest}
\eea
Hereafter, the values in Eq.\ (\ref{WAVEbest})
are referred to as {\it best fit values} of the wavefunction
renormalizations for pseudoscalars.
\par
For the ratio of decay constants of pseudoscalars,
we verify whether the model prediction of
$f_{K^-}/f_{\pi^-}$ is consistent with
the experimental data if one uses $f_\pi$
instead of the tree-level parameter in Eq.\ (\ref{rafKfpi}).
The results in the model and the experimental data extracted
from the PDG data \cite{PDG} are,
\bea
f_{K^-}/f_{\pi^-}=
\begin{cases}
1.40^{+0.18}_{-0.11}\qquad (68.3\%~ \mathrm{C.L.} ~\mathrm{in}~\mathrm{the}~\mathrm{model})\\
1.40^{+0.96}_{-0.24}\qquad (99.7\%~ \mathrm{C.L.} ~\mathrm{in}~\mathrm{the}~\mathrm{model})\\
1.197\pm0.006\qquad (1\sigma ~\mathrm{in}~\mathrm{the}~\mathrm{PDG})
\end{cases}.
\eea
In the above result,
one can find that the model prediction is slightly
deviated from the case of
the tree-level $\rho\pi\pi$ coupling in Eq.~(\ref{rafKfpi}).
This is because the estimated value of $f$ is
deviated from the experimental value of $f_\pi$.
However, up to the $99.7\%$ confidence interval of the model prediction,
it is shown that the central value
of the PDG data \cite{PDG} is included.
%%%%%%%%%%%%%%%%%%%%%%%%%%%%%%%%%%%%%
\subsubsection{Mass and width of vector mesons}
%%%%%%%%%%%%%%%%%%%%%%%%%%%%%%%%%%%%%%%%%%%%%%%%%%%%%%%%%%%%%%%%%%%%%%%%%
In this subsection, we explain how the parameters, $C_1^r,
C_2^r, Z_V^r, \hat{g}_{1V}$ and $M_{0V}$ are fixed, 
and evaluate the vector meson mass,
the renormalization constant and the decay width.

At first, we consider the off-diagonal elements of $V_\mu$, {\it i.e.}, $\rho^+$, $K^{*+}$ and 
$K^{*0}$ to obtain the parameters, $C_1^r, C_2^r$ and $Z_V^r$. We define the masses of
$\rho^+$ and $K^{*+}$ mesons as the momentum-squared for which real parts of inverse
propagators vanish,
\begin{eqnarray}
M_V^2 + {\rm Re}[\delta A_{\rho^{+}} (Q^2=m_{\rho^{+}}^2; C_1^r, C_2^r) ] &=& 0 , \\
M_V^2 + {\rm Re}[\delta A_{K^{*+}} (Q^2=m_{K^{*+}}^2; C_1^r, C_2^r) ] &=& 0 ,
\end{eqnarray}
where $\delta A_{\rho^{+}}$ and $\delta A_{K^{*+}}$ are shown in Eqs.\ (\ref{deltaAB_Kp}) 
and (\ref{deltaAB_rho}), respectively.
Solving the above equations, we have $C_1^r$ and $C_2^r$,
\begin{eqnarray}
C_1^r &=& \frac1{\Delta_{K^+ \pi}} \left\{ Z_V^r \Delta_{K^{*+}\rho^+} 
 - {\rm Re}[\Delta A_{K^{*+}} (m_{K^{*+}}^2) ] + {\rm Re}[\Delta A_{\rho^{+}} (m_{\rho^{+}}^2) ]
 \right\} , \label{C1r} \\
C_2^r &=& -\frac1{2\bar{M}_K^2 + M_\pi^2} \left\{ M_V^2 - Z_V^r m_{K^{*+}}^2
 + {\rm Re}[\Delta A_{K^{*+}} (m_{K^{*+}}^2) ] + C_1^r M_{K^+}^2
 \right\} . \label{C2r}
\end{eqnarray}
Imposing the condition for the residue of the vector meson propagator, 
${\rm Res}D(Q^2 = m_{K^{*+}}^2)=1$, we have
\begin{eqnarray}
Z_V^r= 1 + \left. \frac{d{\rm Re}[\Delta A_{K^{*+}} (Q^2) ]}
 {d Q^2} \right|_{Q^2=m_{K^{*+}}^2} .
\end{eqnarray}
Since $\Delta A_V$ only depends on $g_{\rho \pi \pi}$, $C_1^r$ and $Z_V^r$ can be fixed 
by $g_{\rho \pi \pi}$.
On the other hand, $C_2^r$ is related to two parameters, $g_{\rho \pi \pi}$ 
and $M_V$.

The decay widths are given
by the imaginary part of the inverse propagators,
\begin{eqnarray}
m_I \Gamma_I &=& - Z_I {\rm Im}[\delta A_I (m_I^2)],
\end{eqnarray}
where $I=\rho^+, K^{*+}, K^{*0}$,
\bea
(Z_I)^{-1}
= Z_V^r - \left. \frac{d{\rm Re}[\Delta A_I (Q^2) ]}
 {d Q^2} \right|_{Q^2=m_I^2} , 
\eea
where $\Delta A_I$ is defined in Eqs.\ (\ref{DeltaAB_Kp}, \ref{deltaAB_K0}, 
\ref{DeltaAB_rhop}).
The renormalization constants $Z_i$ are estimated as follows,
\bea
Z_{\rho^+} &=& 1.0461 \pm 0.0006, \quad Z_{K^{*0}} = 1.00244 \pm 0.00003.
\eea
In the following, we determine the parameters $\hat{g}_{1V}$ and 
$M_{0V}$ with the $\chi^2$ fitting 
of the neutral vector meson mass.
The masses in Eq.\ (\ref{poleMv})
are written in terms of the mass
eigenvalues of the mass matrix
and the mixing angles of vector 
mesons.
In Ref. \cite{Pilaftsis:1999qt},
the authors introduced 
a method to express
mixing angles and the eigenvalues
in terms of the elements of the
mass matrix.
With varying the parameters
in the elements of the mass matrix,
one can conduct $\chi^2$
fitting with respect to physical
masses.
The fitted results of the masses are shown in Table \ref{tab_mv},
where the
obtained $\chi^2$/n.d.f is 0.386/1.
The parameters $\hat{g}_{1V}$ and $M_{0V}$ are
fixed as,
\bea
\hat{g}_{1V} =3.185 \pm 0.001,
\quad M_{0V} =871.8 \pm 0.1 {\rm MeV}.
\eea
\begin{table}[h!]
\caption{\label{tab_mv}
    The results of the neutral vector meson mass.}
\begin{center}
\begin{tabular}{ccc}
\hline\hline
Mass & \hspace{8mm} Theory(MeV) \hspace{8mm}& PDG (MeV) \\
\hline
$m_{\rho^0}$ & $775.42\pm0.01$ & $775.26 \pm 0.25$ \\ 
$m_\omega$ & $782.65^{+0.18}_{-0.15}$ & $782.65 \pm 0.12$ \\  
$m_\phi$  & $1019.46\pm0.04$ & $1019.461 \pm 0.019$ \\
\hline\hline
\end{tabular}
\end{center}
\end{table}
Using the above parameters, we have the orthogonal matrix $O_V$ which diagonalizes 
vector meson mass matrix,
\begin{eqnarray}
O_V=\begin{pmatrix}
0.9979\pm0.0001 & 0.0644\pm0.001 & -0.00375\pm0.00004 \\
-0.0447\pm0.0006 & 0.6475\pm0.0006 & -0.7608\pm0.0005 \\
-0.0466\pm0.0009 & 0.7594\pm0.0004 & 0.6490\pm0.0005 
\end{pmatrix},
\label{MixingOV}
\end{eqnarray}
where $\omega-\phi$ mixing angle is $(40.59\pm0.04)^\circ$.
The wave function renormalization of the neutral vector meson and the eigenvalues for 
the mass matrix are obtained as follows,
\begin{eqnarray}
&Z_{\rho^0} = 1.0462\pm0.0005, \quad Z_\omega = 1.0281\pm0.0004, \quad 
 Z_\phi = 0.9167\pm0.0008,&\nn \\
&\cM_1 = 791.8\pm0.2 {\rm MeV}, \quad \cM_2 = 763.2\pm0.3 {\rm MeV}, \quad 
\cM_3 = 1001.9\pm0.2 {\rm MeV}.&
\end{eqnarray}
%%%%%%%%%%%%%%%%%%%%%%%%%%%%%%%%%%%%%%%%%%%%%%%%%%%%
\subsubsection{Intrinsic parity violating decays}
%%%%%%%%%%%%%%%%%%%%%%%%%%%%%%%%%%%%%%%%%%%%%%%%%%%%
In this subsection, we estimate model parameters
by using the IP violating observables for light hadrons. 
As input data of $\chi^2$ fittings,
experimental data of decay widths
and Dalitz distributions are used.
We also utilize experimental values of masses
for pseudoscalars to
estimate parameters in the mass matrix.
\par
The widths of radiative decays,
$\Gamma[\rho^{+}\to\pi^+\gamma]$,
$\Gamma[K^{*0}\to K^0\gamma]$
and $\Gamma[K^{*+}\to K^+\gamma]$,
are proportional to the IP violating parameter
$(gc_{34}^+)^2$.
In order to estimate this parameter,
we consider the following statistic,
\bea
\chi^2=\left(\frac{\Gamma[\rho^+\to\pi^+\gamma]-
\Gamma^\mathrm{PDG}[\rho^+\to\pi^+\gamma]}{\delta
\Gamma^{\mathrm{PDG}}[\rho^+\to\pi^+\gamma]}
\right)^2
+\left(\frac{\Gamma[K^{0*}\to K^0\gamma]-
\Gamma^{\mathrm{PDG}}[K^{0*}\to K^0\gamma]}{\delta
\Gamma^{\mathrm{PDG}}[K^{0*}\to K^0\gamma]}\right)^2,
\quad\quad\quad
\label{Radchisq}\eea
where $\delta\Gamma^{\mathrm{PDG}}$ denotes
experimental errors of the widths.
As a result of the fitting,
we find that the minimum of Eq.\ (\ref{Radchisq}) is $\chi^2_{\mathrm{min}}/\mathrm{d.o.f.}=1.08/1$,
which results in the estimated parameter
as,
\bea
g|c_{34}^+|=0.102\pm0.05.\label{gc34}
\eea
In this fitting,
the sign of $c_{34}^+$ is not fixed since the widths
in Eqs.\ (\ref{RhopWid}-\ref{Kstraddecay})
depend on square of this parameter.
In Table\ \ref{Tabcharged}, the widths calculated in the model
are compared with the PDG values \cite{PDG}.
The model prediction for $\Gamma[K^{*+}\to K^+\gamma]$
is also given in Table\ \ref{Tabcharged}.
For the PDG value \cite{PDG} of $\Gamma[K^{*+}\to K^+\gamma]$,
we adopt the full width of $K^{*+}$
obtained from tau decays.
One finds $3.4\sigma$ discrepancy between
the model prediction and the experimental value
of the width for $K^{*+}\to K^+\gamma$.
Since the $K^{*0}K^0\gamma$ coupling
in Eq.\ (\ref{XKstver})
is two times larger than one for
$K^{*+}K^+\gamma$ in Eq.\ (\ref{Xrho+}),
the widths are related as
$\Gamma[K^{*0}\to K^0\gamma]\sim
4\Gamma[K^{*+}\to K^+\gamma]$.
However, this relation is not valid for
the present PDG values \cite{PDG}
so that the deviation arises.
%%%%%%%%%%%%%%%%%%%%%%
%%% Radiative Decays   %%%%%
%%%%%%%%%%%%%%%%%%%%%%
\begin{table}[h!]
\caption{Partial widths of radiative decays.
For $\rho^+\to\pi^+\gamma$ and 
$K^{*0}\to K^{0}\gamma$, the fitting result
for $\chi^2_{\mathrm{min}}/\mathrm{d.o.f.}=1.08/1$ is shown.
The model prediction is given for $K^{*+}\to K^{+}\gamma$.
For comparison, the PDG data \cite{PDG} are also written.
} 
\label{Tabcharged}
\begin{tabular}{ccc}\hline\hline
Decay mode &\hspace{5mm} Model [MeV] \hspace{5mm} & PDG [MeV]\\
\hline
$\Gamma[\rho^+\rightarrow\pi^+\gamma]$ & $(7.3\pm0.7)\times 10^{-2}$ & $(6.7\pm 0.7)\times 10^{-2}$
\\
$\Gamma[K^{*0}\rightarrow K^0\gamma]$ & $0.11\pm0.01$ & $0.12\pm0.01$
\\
\hline
$\Gamma[K^{*+}\rightarrow K^+\gamma]$ & $(2.8\pm0.3)\times 10^{-2}$ & $(4.6\pm 0.4)\times 10^{-2}$ \\
\hline\hline
\end{tabular}
\end{table}
\par
For parameter estimation,
we use observables for pseudoscalars.
In particular,
the PDG data \cite{PDG} for masses of $\pi^0, \eta^\prime$
and decay widths of
$P\to2\gamma\ (P=\pi^0, \eta, \eta^\prime)$
are adopted.
In order to constrain parameters in the model, we consider the following system of equations,
\bea
M_{\pi^0}(\hat{g}_{2p}, \Delta_{\mathrm{EM}}, M_{88}^{\prime2}, M_{80}^{\prime2})&=&M_{\pi^0}^{\mathrm{PDG}},\label{multi1}\\
M_{\eta^\prime}(\hat{g}_{2p}, \Delta_{\mathrm{EM}}, M_{88}^{\prime2}, M_{80}^{\prime2})&=&M_{\eta^\prime}^{\mathrm{PDG}},
\label{multi2}\\
\Gamma[\pi^0\to2\gamma]
(c_{6-9-10}, \hat{g}_{2p}, \Delta_{\mathrm{EM}}, M_{88}^{\prime2}, M_{80}^{\prime2})&=&
\Gamma^{\mathrm{PDG}}[\pi^0\to2\gamma],
\label{multi3}\\
\Gamma[\eta\to2\gamma]
(c_{6-9-10}, \hat{g}_{2p}, \Delta_{\mathrm{EM}}, M_{88}^{\prime2}, M_{80}^{\prime2})&=&
\Gamma^{\mathrm{PDG}}[\eta\to2\gamma],\label{multi4}\\
\Gamma[\eta^\prime\to2\gamma]
(c_{6-9-10}, \hat{g}_{2p}, \Delta_{\mathrm{EM}}, M_{88}^{\prime2}, M_{80}^{\prime2})&=&
\Gamma^{\mathrm{PDG}}[\eta^\prime\to2\gamma]\label{multi5},
\eea
where the left-handed sides in these equations denote
the model expressions.
Solution to Eqs.\ (\ref{multi1}-\ref{multi5})
leads to estimated values for
the parameters given as
$(c_{6-9-10}, \hat{g}_{2p}, \Delta_{\mathrm{EM}}, M_{88}^{\prime2}, M_{80}^{\prime2})$.
This procedure of solving the equations
is carried out in the following way:
provided that the PDG data \cite{PDG} obey Gaussian distributions,
the right-handed sides in Eqs.\ (\ref{multi1}-\ref{multi5})
are generated as Gaussian data.
For $(\sqrt{Z_1^\pi}, \sqrt{Z_2^\pi})$,
we use the parameter list obtained from the fitting
of tau decays, which is summarized in Table \ref{fit_para0}.
In order to determine model values of the masses
in Eqs.\ (\ref{multi1}-\ref{multi2}),
we use formalism in App.\ \ref{APPneutral}
which incorporates 1-loop correction
to the mass matrix.
One can numerically calculate the model values for pseudoscalar masses,
which are eigenvalues of the mass matrix
in Eq.\ (\ref{MASSMATRIX}).
For Eq.\ (\ref{multi3}-\ref{multi5}),
the widths in the model are calculated on the basis of
Eq.\ (\ref{pi2gamma}).
Since $\Gamma[P\to2\gamma]$ depends on the
pseudoscalar mixing matrix elements,
we adopt a method \cite{Pilaftsis:1999qt}
to write a mixing matrix in terms of
mass matrix elements.
Using $10^{4}$ data samples, we solve the system of
Eqs.\ (\ref{multi1}-\ref{multi5}) to obtain the parameters
$(c_{6-9-10}, \hat{g}_{2p}, \Delta_{\mathrm{EM}}, M_{88}^{\prime2}, M_{80}^{\prime2})$.
Confidence intervals of the parameters are estimated
from a list of the solutions to Eqs.\ (\ref{multi1}-\ref{multi5}).
In Table\ \ref{Tabparams}, we show confidence intervals
of the model parameters which are determined in
this procedure.
Since the parameters in the mass matrix are estimated,
a mixing angle for pseudoscalars is also determined.
$\theta_{1}$ is obtained as $\theta_1=\arccos(O_{33})$,
where $O_{33}$ is the mixing matrix element in
Eq.\ (\ref{mixO}).
The numerical value of this angle is,
\bea
\theta_1=
\begin{cases}
-28^{+2}_{-5} [\mathrm{degree}]\quad (68.3\% \ \mathrm{C.L.})\\
-28^{+5}_{-43} [\mathrm{degree}]\quad (99.7\% \ \mathrm{C.L.})
\end{cases}.
\eea
%%%%%%%%%%%%%
%%%%%%%%%%%%%
\begin{table}[h!]
\caption{Confidence intervals
of the model parameters estimated from the
data \cite{PDG} of widths and masses for pseudoscalars.
See the text for a detailed explanation
of parameter estimation.}
\label{Tabparams}
\begin{tabular}{cccccc}\hline\hline
Parameter & $c_{6-9-10}\times10^2$ & $\hat{g}_{2p}$ &
$\Delta_{\mathrm{EM}}\ [\mathrm{MeV}^2]$& $\sqrt{M_{88}^{\prime2}}\ [\mathrm{MeV}]$
&$\sqrt{|M_{80}^{\prime2}|}\ [\mathrm{MeV}]$\\
\hline
$68.3\%\ \mathrm{C.L.}$ & $1.1^{+0.2}_{-0.2}$ &
$-1.0^{+0.7}_{-0.7}$
&$1220^{+20}_{-60}$&$660^{+30}_{-20}$ & $510\pm20$\\
$99.7\%\ \mathrm{C.L.}$ & $1.1_{-0.5}^{+0.7}$ & $-1.0^{+2.1}_{-2.3}$
&$1220^{+30}_{-310}$& $660^{+260}_{-60}$ & $510\pm80$
\\
\hline
\end{tabular}
\end{table}
\par
In Eqs.\ (\ref{multi1}-\ref{multi5}),
if one adopts the best fit model parameters
in Eq.\ (\ref{Taupara}) on left-handed sides
and central values of the PDG data \cite{PDG}
on right-handed sides, solution is obtained as,
\bea
&c_{6-9-10}=1.1\times10^{-2},\quad
\hat{g}_{2p}=-1.0,&\nn\\
&\Delta_{\mathrm{EM}}=1220\ [\mathrm{MeV^2}],\quad
M_{88}^\prime=662\ [\mathrm{MeV}],\quad
M_{80}^\prime=507\ [\mathrm{MeV}].&\label{c6910}
\eea
Using the above values, the mixing matrix
and the wavefunction renormalizations of
pseudoscalars are calculated as,
\bea
&O=\begin{pmatrix}
0.99998 & -3.3\times10^{-4} & -6.8\times10^{-3}\\
3.5\times10^{-3} & 0.88 & 0.47\\
5.8\times10^{-3} & -0.47 & 0.88\\
\end{pmatrix},&\label{mixingbest}
\eea
In the following analysis, the parameter values
in Eq.\ (\ref{mixingbest})
are referred to as {\it best fit values} for
the mixing matrix elements.
\par
Here, we discuss a case in which singlet-induced contribution is absent.
If one takes the limit $c_{6-9-10}\to 0$,
the partial width of $\eta^\prime$ becomes
$\Gamma[\eta^\prime\to2\gamma]=7\times 10^{-5}$MeV.
This value is much smaller than the experimental data,
$\Gamma^{\mathrm{PDG}}[\eta^\prime\to2\gamma]=(4.4\pm 0.3)\times 10^{-3}$MeV.
Hence, one notices that the presence of singlet-induced
IP violation is necessary in the framework of
the singlet$+$octet scheme.
\par
For parameter estimation of the IP violating parameters,
the ratio of
the effective coupling for
$VP\gamma$ to one for $\rho^+\pi^+\gamma$ in Eq.\ (\ref{7-2}) are
compared with experimental values.
Model parameters are estimated
from the following statistic,
\bea
\chi^2&=&\displaystyle\sum_{(i, I)}^{(1, 2), (2, 2),
(3, 2), (3, 3)
}\left(\frac{|X_{iI}/X_{\rho^+}|-|X_{iI}/X_{\rho^+}|^{\mathrm{PDG}}}
{\delta|X_{iI}/X_{\rho^+}|^{\mathrm{PDG}}}\right)^2.
\label{CHISTAT2}
\eea
The experimental data used in the above $\chi^2$
are extracted from PDG \cite{PDG} through
r.h.s in Eq.\ (\ref{7-1}).
In Eq.\ (\ref{7-2}),
the wavefunction renormalizations
and the mixing matrices for mesons
are set as the best fit values obtained
in Eqs.\ (\ref{WAVEbest}, \ref{mixingbest}).
(for vector meson mixing, Eq.\ number should
be referred.)
In the procedure to minimize the statistic in
Eq.\ (\ref{CHISTAT2}),
one can vary model parameters, $c_{69}/g^2c_{34}^+$
and $c_8^{\mathrm{IP}}/gc_{34}^+$.
The fitting results are shown in Table \ref{Tabfour}.
The parameter ranges
estimated from this fitting are,
\bea
c_{69}/g^2c_{34}^+=-0.91\pm0.04,
\qquad c_8^{\mathrm{IP}}/gc_{34}^+=0.85\pm0.05,\label{c69c34}
\eea
where the correlation coefficient of these parameters
is $0.12$.
Predictions for effective coupling ratios of
$\Gamma[V^I\to P^i\gamma]$ to $\Gamma[\rho^+
\to\pi^+\gamma]$ for $(i, I)=(1, 1), (1, 3), (2, 1), (2, 3)$
are given in Table \ref{Tabfour}.
Furthermore, the prediction for the decay widths
of $V^I\to P^i\gamma$ are shown in Table \ref{dataf}.
%%%%%%%%%%%%%%
%%% pi 0 to gamma
%%%%%%%%%%%%%%
\begin{table}[h!]
\caption{Fitting result and model prediction of the ratio of effective coupling for $V^I\to P^i\gamma$ to one for $\rho^+\to \pi^+\gamma$.
For $(i, I)=(1, 2), (2, 2), (3, 2), (3, 3)$,
the fitting result for $\chi^2_{\mathrm{min}}
/\mathrm{d.o.f.}=1.12/2$ is shown
while the model predictions are given
for $(i, I)=(1, 1), (1, 3), (2, 1), (2, 3)$.
For comparison, the experimental data extracted
from the PDG data \cite{PDG} are also shown.
In the fourth column, the model prediction in the
isospin limit is displayed with wave function
renormalizations set as unity.
Mixing angles for vector meson are defined as
$\cos\theta_{V}^{08}=O_{V22}\sim O_{V33}$
and $\sin\theta_{V}^{08}=O_{V23}\sim -O_{V32}$.
} \label{Tabfour}
\begin{center}
\begin{tabular}{cccc} 
\hline\hline
\hspace{1mm} Ratio \hspace{1mm} & Model \hspace{5mm}& PDG \hspace{5mm}&
\hspace{5mm} Model in the isospin limit \hspace{5mm}
\\ \hline \vspace{1mm}
$|X_{12}/X_{\rho^+}|$&
$3.1\pm0.1$&$3.2\pm0.2$
&$\sqrt{3}|\cos\theta_V^{08}
-\frac{3c_8^{\IP}}{gc_{34}^+}\sin\theta_V^{08}|$
\\  \vspace{1mm}
$|X_{22}/X_{\rho^+}|$&$0.71^{+0.09}_{-0.08}$
& $0.62\pm0.04$
&$|\cos\theta_V^{08}||\cos\theta_1+\sqrt{3} (\frac{c_{69}}{g^2c_{34}})\sin\theta_1
-\frac{\sqrt{3}c_8^\IP}{gc_{34}^+}\cos\theta_1\tan\theta_V^{08}|$
\\  \vspace{1mm}
$|X_{32}/X_{\rho^+}|$&
$0.53^{+0.16}_{-0.13}$& $0.60\pm0.04$
&$|\cos\theta_V^{08}||\sin\theta_1-\sqrt{3} (\frac{c_{69}}{g^2c_{34}})\cos\theta_1-
\frac{\sqrt{3}c_8^\IP}{gc_{34}^+}\sin\theta_1\cot\theta_V^{08}|$
\\  \vspace{1mm}
$|X_{33}/X_{\rho^+}|$&
$1.15^{+0.15}_{-0.13}$&$0.99\pm 0.06$
&$|\sin\theta_V^{08}||\sin\theta_1-\sqrt{3} (\frac{c_{69}}{g^2c_{34}})\cos\theta_1
+\frac{\sqrt{3}c_8^\IP}{gc_{34}^+}\sin\theta_1\cot\theta_V^{08}|$\hspace{10mm}
\\  \hline
\vspace{1mm}
$|X_{11}/X_{\rho^+}|$&
$0.80\pm0.02$&$1.15\pm0.10$
&1
\\ \hspace{1.2mm}\vspace{1mm}
$|X_{13}/X_{\rho^+}|$&
$0.31\pm0.09$&$0.18\pm0.01$
&$\sqrt{3}|\sin\theta_{V}^{08}
+\frac{3c_8^{\IP}}{gc_{34}^+}\cos\theta_V^{08}|$
\\  \vspace{1mm}
$|X_{21}/X_{\rho^+}|$&
$1.8\pm0.2$&  $2.2\pm0.1$
&$\sqrt{3}|\cos\theta_1-(\frac{c_{69}}{g^2c_{34}})\tan\theta_1|$
\\  \vspace{1mm}
$|X_{23}/X_{\rho^+}|$&
$0.6^{+0.1}_{-0.2}$& $0.96\pm0.05$
&$
|\sin\theta_V^{08}||\cos\theta_1+\sqrt{3} (\frac{c_{69}}{g^2c_{34}})\sin\theta_1
+\frac{\sqrt{3}c_8^\IP}{gc_{34}^+}\cos\theta_1\cot\theta_V^{08}|$
\\ \hline\hline
\end{tabular}
\end{center}
\end{table}
\begin{table}[h!]
\caption{Partial widths of the radiative
decays for vector mesons.
For comparison, the PDG data \cite{PDG} are also shown.
}\label{dataf}
\centering
\begin{tabular}{ccc}
\hline\hline
& \hspace{5mm} Model [MeV] \hspace{5mm} & PDG [MeV] \\ 
\hline
$\Gamma[\omega\rightarrow\pi^0\gamma]$ &
$0.71\pm0.09$ & $0.70\pm0.02$ \\
$\Gamma[\omega\rightarrow\eta\gamma]$ & $
(5.5^{+1.6}_{-1.3})\times10^{-3}$ & $(3.9\pm0.3)\times10^{-3}$ \\
$\Gamma[\eta^\prime\rightarrow\omega\gamma]$ & $
(4.6^{+3.3}_{-2.0})\times10^{-3}$ & $(5.4\pm0.5)\times10^{-3}$ \\
$\Gamma[\phi\rightarrow\eta^\prime\gamma]$ & $
(3.9^{+1.2}_{-0.9})\times10^{-4}$ & $(2.67\pm0.09)\times10^{-4}$ \\\hline
$\Gamma[\rho^0\rightarrow\pi^0\gamma]$ &
$(4.6\pm0.5)\times10^{-2}$ & $(9\pm1)\times10^{-2}$ \\
$\Gamma[\phi\rightarrow\pi^0\gamma]$ &
$(17^{+12}_{-9})\times10^{-3}$ & $(5.4\pm0.3)\times10^{-3}$ \\
$\Gamma[\rho\rightarrow\eta\gamma]$ &
$(3.3^{+0.8}_{-0.9})\times10^{-2}$ & $(4.5\pm0.3)\times10^{-2}$ \\
$\Gamma[\phi\rightarrow\eta\gamma]$ & $
(2.2^{+0.9}_{-1.2})\times10^{-2}$ & $(5.6\pm0.1)\times10^{-2}$ \\
\hline\hline
\end{tabular}
\end{table}
\newpage
\par
In the following, TFFs for
Dalitz decay of vector mesons are analyzed.
In particular, we fit $|F_{V^I P^i}|^2$
for $(i, I)=(1, 2), (1, 3)$ and $(2, 3)$, in each bin
for di-lepton invariant mass.
In order to minimize the statistic,
\bea
\chi^2=\displaystyle\sum_{\mathrm{Available}\ \mathrm{data}}
\displaystyle\sum_{(i, I)}^{(1, 2), (1, 3), (2, 3)}
\left(\frac{|F_{V^I P^i}|^2-(|F_{V^I P^i}|^2)^{\mathrm{Exp.}}}
{\delta(|F_{V^I P^i}|^2)^{\mathrm{Exp.}}}\right)^2,
\label{chisq3}
\eea
we vary the IP violating parameters:
$(c_{3}^{\mathrm{IP}}, c_{5}^{\mathrm{IP}}, c_{6}^{\mathrm{IP}}, c_{7}^{\mathrm{IP}})$.
For the expression of $|F_{V^IP^i}|^2$ in Eq.\ (\ref{TFFVECTORM}),
the mixing matrices and
wavefunction renormalizations of mesons 
are set as the best fit values
in Eqs.\ (\ref{WAVEbest}, \ref{MixingOV},
\ref{mixingbest}).
In Eq.\ (\ref{chisq3})
the experimental data extracted from Refs.\ \cite{Arnaldi:2016pzu,
Achasov:2008zz,
Dzhelyadin:1980tj,
Akhmetshin:2005vy,
Babusci:2014ldz,
Achasov:2000ne,
::2016hdx} are adopted for
parameter estimation.
In the fitting procedure, two cases: $c_{34}^+<0$ and $c_{34}^+>0$ are considered.
For these cases,
one can find that the goodness-of-fit
is comparable with each other.
We find that the minimum of Eq.\ (\ref{chisq3}) is
$\chi^2_{\mathrm{min}}/\mathrm{d.o.f.}=211.8/151\; (215.4/151)$
for $c_{34}^+<0\; (>0)$.
As an alternative analysis, 
we also fit $\chi^2$ in the case without the Lepton-G data \cite{Dzhelyadin:1980tj}.
This fitting analysis leads to $\chi^2_{\mathrm{min}}/\mathrm{d.o.f.}=170.1/144\; (173.7/144)$
for $c_{34}^+<0\; (>0)$, which is a slightly improved
result.
In this case,
we find that the best fit values
and the errors of $(c_3^{\IP}, c_5^{\IP}, c_6^{\IP}, c_7^{\IP})$
are almost identical to ones
in the case with the Lepton-G data.
For each fitting,
$\chi^2_{\mathrm{min}}/\mathrm{d.o.f.}$,
corresponding p-values
and the estimated parameters are summarized
in Table\ \ref{FITTFF}.
%%%%%%%%%%%%%%%%
\begin{table}[h!]  
\caption{Fitting results of the TFFs for vector meson decays.
For the four cases of fitting, the-goodness-of fit is shown.
Estimated 1$\sigma$ ranges for the IP violating parameters are also given.}
\label{FITTFF}
\centering
\begin{tabular}{lccrrcc}
\hline\hline
& \hspace{2mm}$\chi^2_{\mathrm{min}}/\mathrm{d.o.f.}
\hspace{2mm}$ & \hspace{2mm} p-value \hspace{2mm}
& \hspace{2mm}$c_3^{\mathrm{IP}}\times 10^2$ \hspace{2mm}
& \hspace{2mm}$c_5^{\mathrm{IP}}\times 10^2$ 
& \hspace{2mm}$c_6^{\mathrm{IP}}$ \hspace{2mm}
& \hspace{2mm}$c_7^{\mathrm{IP}}$ \hspace{2mm}
\\\hline
$c_{34}^+<0$ without Lepton-G&
170.1/144&
$0.068$&
$1.12\pm0.05$&
$6.5\pm0.2$&
$-1.3\pm0.7$&
$-1.8\pm1.3$
\\
$c_{34}^+<0$ with Lepton-G&
211.8/151&
$8.1\times10^{-4}$&
$1.12\pm0.05$&
$6.5\pm0.2$&
$-1.3\pm0.7$&
$-1.8\pm1.3$
\\
$c_{34}^+>0$ without Lepton-G&
173.7/144&
0.046&
$-1.12\pm0.05$&
$-6.5\pm0.2$&
$-0.3\pm0.8$&
$-1.1\pm1.4$
\\
$c_{34}^+>0$ with Lepton-G&
215.4/151&
$4.5\times10^{-4}$&
$-1.12\pm0.05$&
$-6.5\pm0.2$&
$-0.3\pm0.8$&
$-1.1\pm1.4$
\\
\hline\hline
\end{tabular}
\end{table}
As a result of the fittings without the Lepton-G data,
the correlation matrices for
$(c_3^{\IP}, c_5^{\IP}, c_6^{\IP}, c_7^{\IP})$ are,
\bea
\begin{pmatrix}
1 & \hspace{3mm}0.75 & -0.12 & -0.082\\
*   & \hspace{3mm}1      & -0.14 & -0.11 \\
*   & \hspace{3mm} *       &     1    &  1.0   \\
*   &  \hspace{3mm}*       &     *      &      1
\end{pmatrix}
(c_{34}^+<0),\qquad 
\begin{pmatrix}
   1&\hspace{3mm} 0.75 & -0.12 & -0.086\\
   *&\hspace{3mm} 1      & -0.14 & -0.11 \\
   *&\hspace{3mm}  *       &     1    &  1.0 \\
   *&\hspace{3mm} *        &    *       &      1
\end{pmatrix}
(c_{34}^+>0).
\eea
\par
In Table\ \ref{FITTFF},
one can find that the errors of
$c_6^\IP$ and $c_7^\IP$ are large.
This is because the contributions of
$c_6^\IP$ and $c_7^\IP$ are suppressed by
either isospin breaking or the $\eta-\eta^\prime$
mixing angle for $|F_{\omega\pi^0}|^2$,
$|F_{\phi\pi^0}|^2$ and $|F_{\phi\eta}|^2$
in Eq.\ (\ref{TFFVECTORM}).
To improve the precisions of $c_6^\IP$
and $c_7^\IP$, the experimental errors
of the TFFs should be reduced,
especially for $|F_{\phi\eta}|^2$.
\par
\begin{figure}[h!]
\begin{center}
  \setlength{\subfigwidth}{.5\linewidth}
  \addtolength{\subfigwidth}{-.5\subfigcolsep}
  \begin{minipage}[h]{\subfigwidth}
    \subfigure[]{\includegraphics{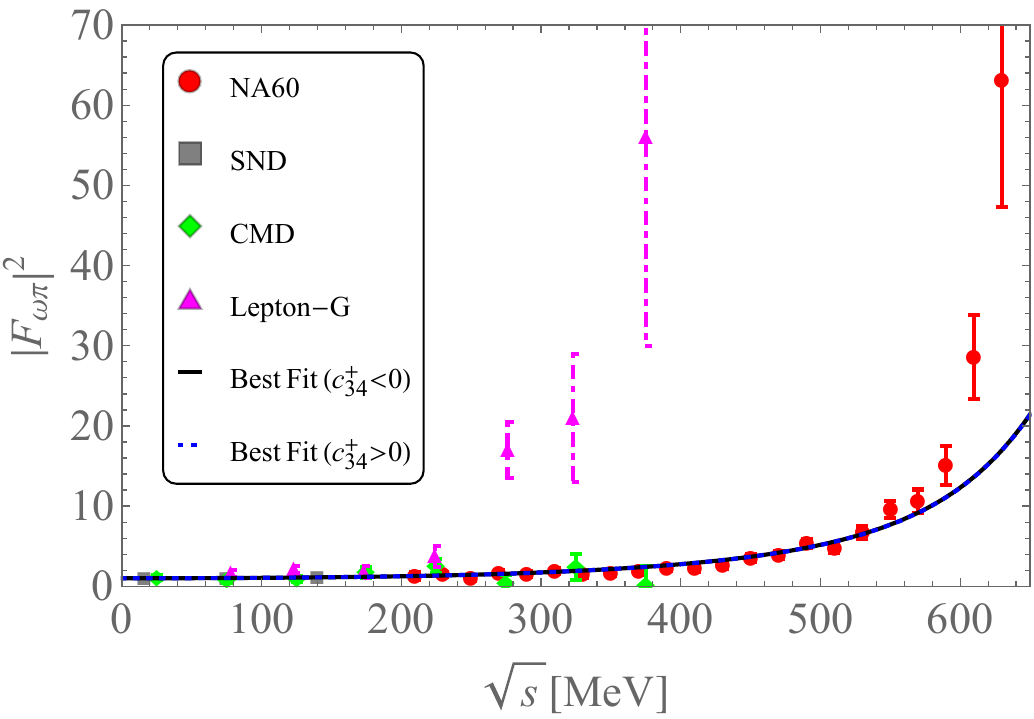}
   \label{Data1}}
  \end{minipage}\\
    \begin{minipage}[h]{\subfigwidth}
    \subfigure[]{\includegraphics{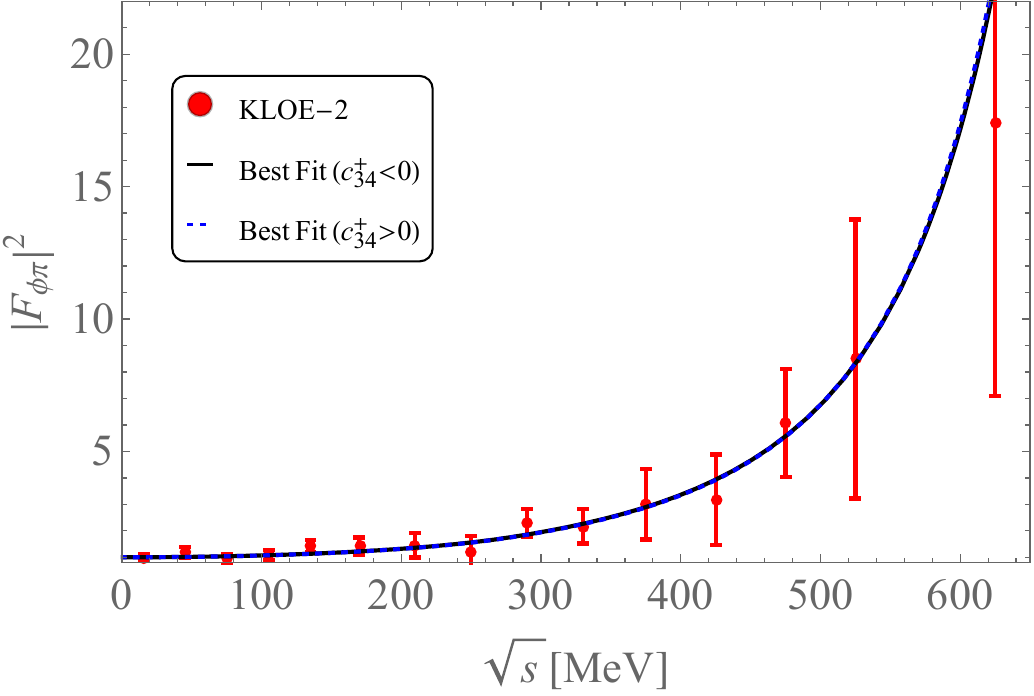}
   \label{Data5}}
  \end{minipage}\\
  \begin{minipage}[h]{\subfigwidth}
    \subfigure[]{\includegraphics{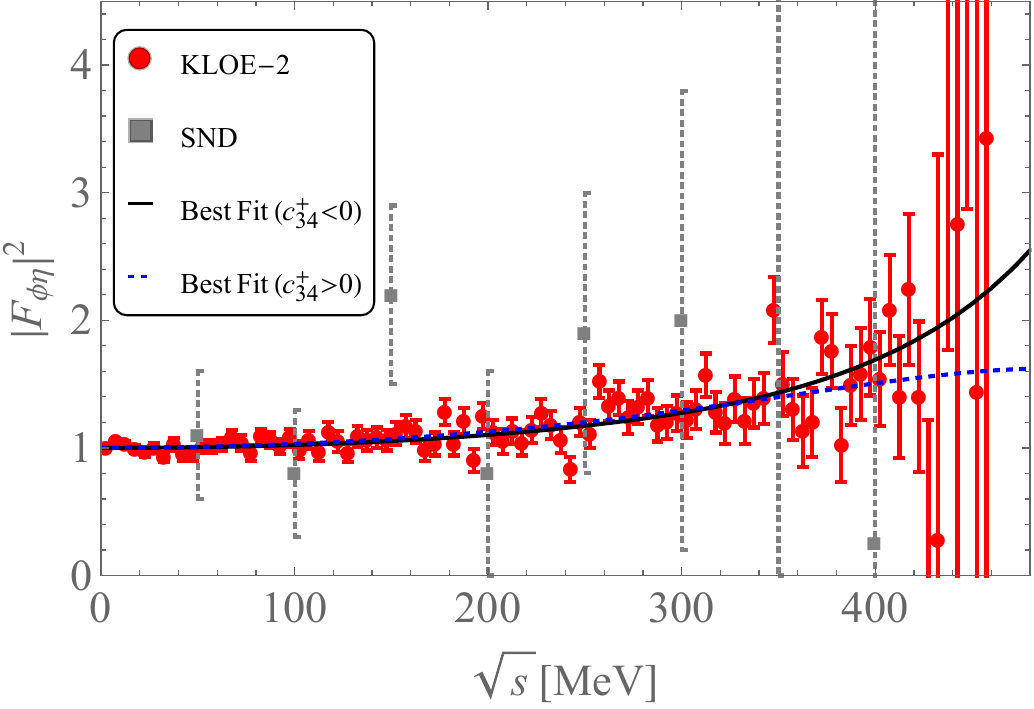}
   \label{Data6}}
  \end{minipage}\\
    \caption{Transition form factors versus 
    di-lepton invariant mass:
    (a) $\omega\to\pi^0l^+l^-$,
    (b) $\phi\to\pi^0 l^+l^-$
    and (c) $\phi\to \eta l^+l^-$.
    Black solid lines indicate best fit curves for
    $c_{34}^+<0$ while blue dotted lines
    imply ones for $c_{34}^{+}>0$.
    For comparison, experimental data are shown for
    (a) NA60 \cite{Arnaldi:2016pzu}, SND \cite{Achasov:2008zz},
    Lepton-G \cite{Dzhelyadin:1980tj} and     
    CMD-2 \cite{Akhmetshin:2005vy},
    (b) KLOE-2 \cite{Babusci:2014ldz} and
    (c) SND \cite{Achasov:2000ne} and KLOE-2 \cite{::2016hdx}.}
    \label{EXP2}
    \end{center}
\end{figure}

In the following analysis in this paper,
we adopt parameter sets which are estimated
from the case without the Lepton-G data.
The TFFs obtained in the model,
which result from the case without the Lepton-G data,
are shown in Fig.\ \ref{EXP2}.
One can see that best fit curves for $c_{34}^{+}<0$
and $c_{34}^{+}>0$ are slightly deviated
from one another
in $\phi\to\eta\l^+l^-$ whereas the two predictions
mostly overlap with each other
for $\omega\to\pi^0l^+l^-$ and
$\phi\to\pi^0l^+l^-$.
\par
We determine the IP violating parameters,
$(c_4^{\mathrm{IP}}, c_8^{\mathrm{IP}},
c_9^{\mathrm{IP}}, c_{10}^{\mathrm{IP}})$, from
Eqs.\ (\ref{gc34}, \ref{c6910}, \ref{c69c34})
and Table\ \ref{FITTFF}.
The result is shown in Table \ref{rez} for  two cases, $c_{34}^+<0$ and $c_{34}^+>0$, separately.
%%%%%%%%%%%%%%%%%%%%%
% V->Pl^+l^- distribution  %
%%%%%%%%%%%%%%%%%%%%%
\begin{table}[h!]
\caption{Intrinsic parity violating
parameters estimated in the fittings.
For $c_{34}^+<0$ and $c_{34}^+>0$, the confidence
intervals are shown, respectively.}
\label{rez}
\begin{tabular}{ccccc}\hline\hline
$(c_{34}^+<0)$\hspace{0mm}
& \hspace{5mm}$c_4^{\mathrm{IP}}\times10^{2}$ \hspace{5mm}& \hspace{5mm}$c_8^{\mathrm{IP}}\times10^{2}$ \hspace{5mm}
& \hspace{5mm} $c_9^{\mathrm{IP}}$
\hspace{5mm}
& \hspace{5mm}$c_{10}^{\mathrm{IP}}$\hspace{5mm}
 \\\hline
68.3\% C.L. & \hspace{5mm}$-2.7^{+0.3}_{-0.4}$&
\hspace{-4mm}$-8.6^{+0.7}_{-0.7}$&
$0.47^{+0.26}_{-0.22}$&\hspace{5mm}$-0.11^{+0.05}_{-0.06}$
\\\vspace{1.5mm}
99.7\% C.L. & \hspace{5mm}$-2.7^{+0.7}_{-2.1}$&
\hspace{-4mm}$-8.6^{+2.0}_{-2.2}$&
$0.47^{+1.11}_{-0.66}$&\hspace{5mm}
$-0.11^{+0.15}_{-0.24}$
\\\hline\hline
$(c_{34}^+>0)$ & \hspace{5mm}$c_4^{\mathrm{IP}}\times10^{2}$ & $c_8^{\mathrm{IP}}\times10^{2}$ &
$c_9^{\mathrm{IP}}$ & \hspace{5mm}
$c_{10}^{\mathrm{IP}}$\\\hline
68.3\% C.L. & \hspace{5mm}$0.33^{+0.43}_{-0.31}$&
$8.6^{+0.7}_{-0.7}$&
$0.29^{+0.26}_{-0.22}$&\hspace{5mm}
$-0.06^{+0.05}_{-0.06}$
\\
99.7\% C.L. & \hspace{5mm}$0.33^{+2.11}_{-0.74}$&
$8.6^{+2.2}_{-2.0}$&
$0.29^{+1.09}_{-0.67}$&\hspace{5mm}$-0.06^{+0.15}_{-0.23}$
\\
\hline\hline
\end{tabular}
\end{table}
%%%%%%%%%%%%%%%%%%%
%%%  Model Prediction %%
%%%%%%%%%%%%%%%%%%%
\subsection{Model prediction}
In this subsection, predictions of the model
are given for the TFFs of Dalitz decays,
partial widths and differential decay widths
of IP violating modes.
We utilize the parameter set obtained in the previous
subsection.
\par
\begin{figure}[ht]
  \setlength{\subfigwidth}{.5\linewidth}
  \addtolength{\subfigwidth}{-.5\subfigcolsep}
  \begin{minipage}[b]{\subfigwidth}
    \subfigure[]{\includegraphics[width=8.2cm]{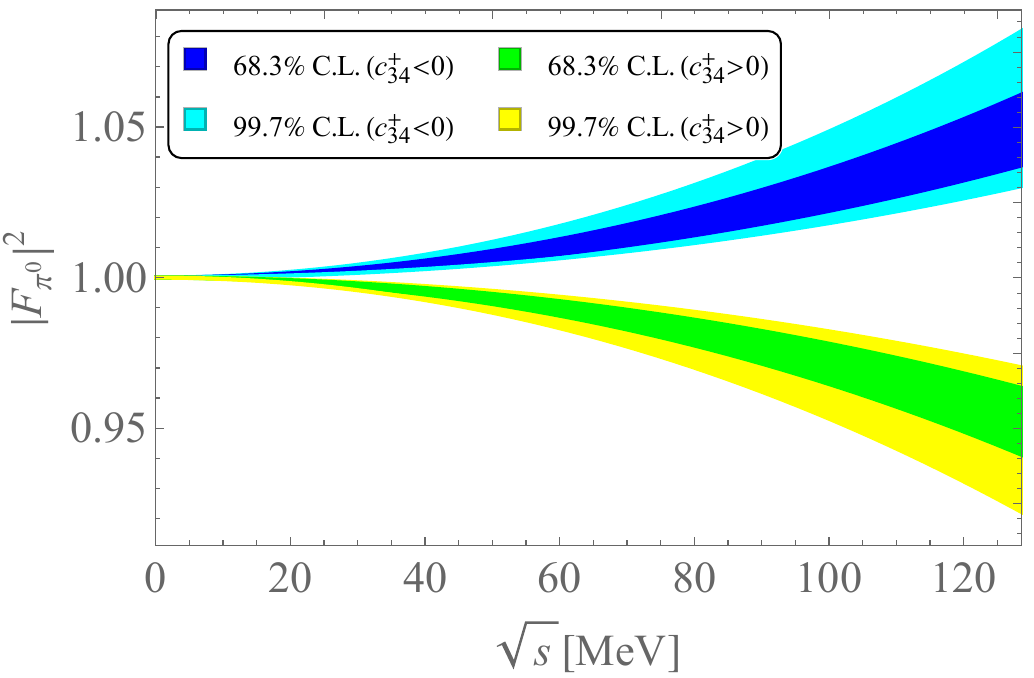}
   \label{Data1}}
  \end{minipage}\\
  \begin{minipage}[b]{\subfigwidth}
    \subfigure[]{\includegraphics[width=8.2cm]{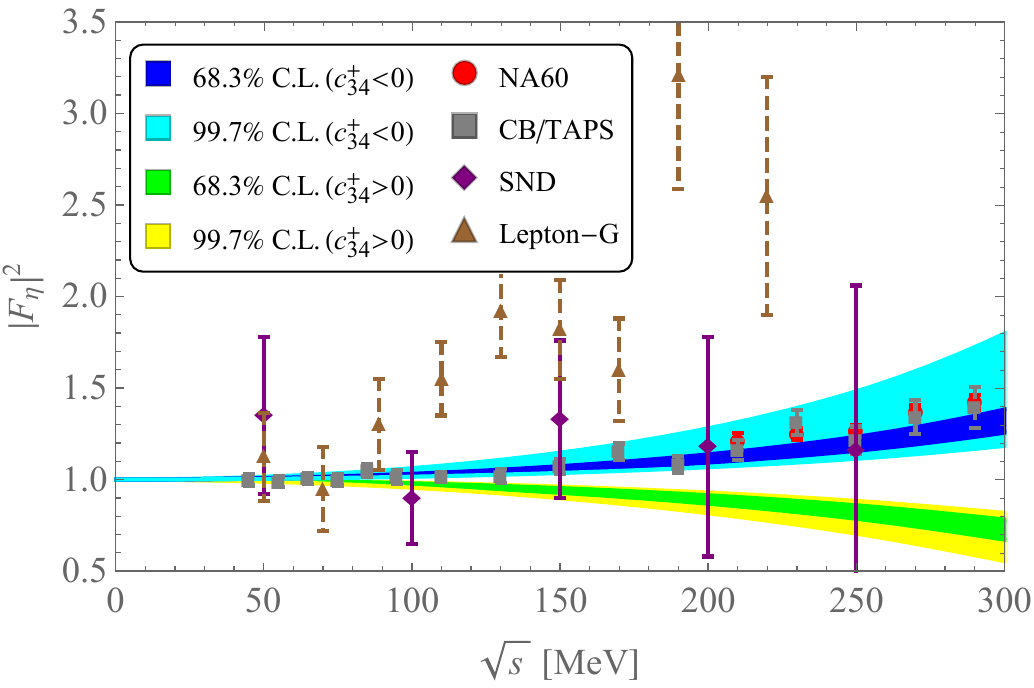}
   \label{Data2}}
  \end{minipage}
  \begin{minipage}[b]{\subfigwidth}
    \subfigure[]{\includegraphics[width=8.2cm]{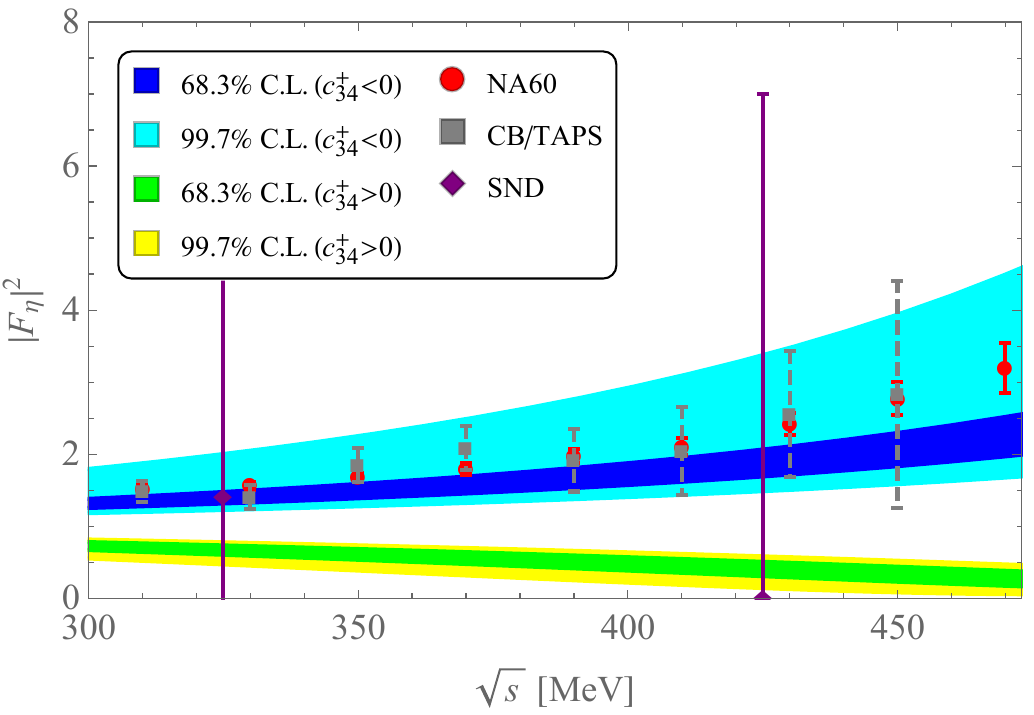}
   \label{Data3}}
  \end{minipage}\\
  \begin{minipage}[b]{\subfigwidth}
    \subfigure[]{\includegraphics[width=8.2cm]{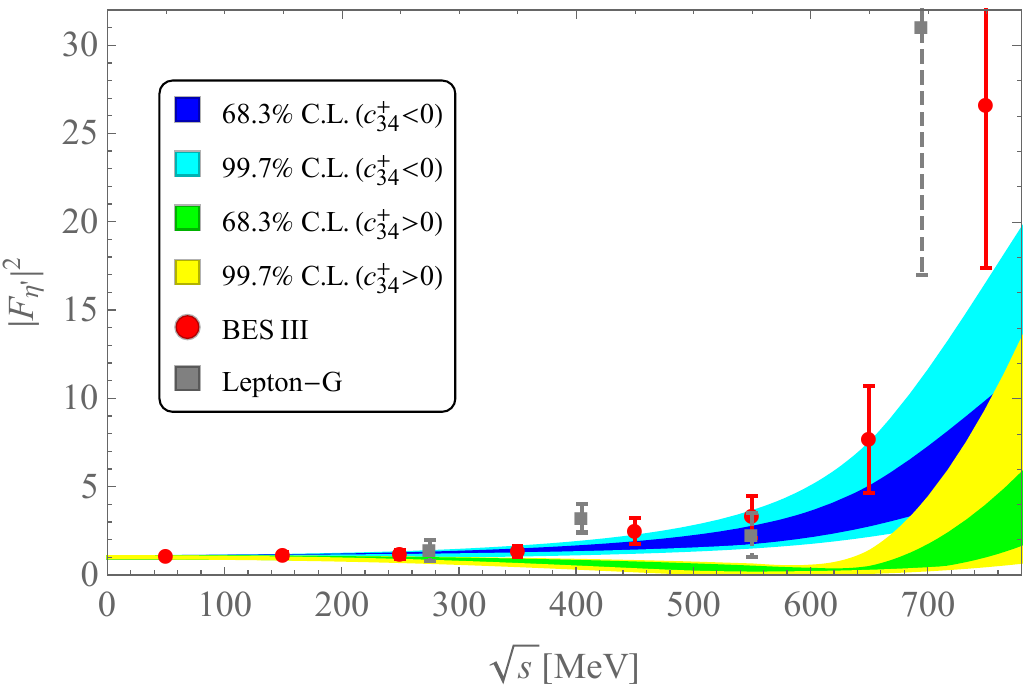}
   \label{Data4}}
  \end{minipage}
  \caption{Transition form factors versus
  di-lepton invariant mass:
  (a) $\pi^0\to\gamma l^+l^-$,
  (b)-(c) $\eta\to\gamma l^+l^-$ in the mass range
  for $[2m_e, 300] \mathrm{MeV}$ and for $[300, 470]
 \mathrm{MeV}$ respectively
and (d) $\eta^\prime\to\gamma l^+l^-$.
For $c_{34}^+<0$,
blue (cyan) bands indicate theoretical predictions
in $68.3\%$ $(99.7\%)$ C.L. 
while for $c_{34}^+>0$, green (yellow) bands
represent ones in $68.3\%$ $(99.7\%)$ C.L. 
For comparison, the experimental data obtained by
(b)-(c) NA60 \cite{Arnaldi:2016pzu}, Lepton-G \cite{Dzhelyadin:1980kh}, CB/TAPS \cite{Aguar-Bartolome:2013vpw} and
SND \cite{Achasov:2000ne}
and (d) BES III \cite{Ablikim:2015wnx}, Lepton-G \cite{Dzhelyadin:1980ki} are shown.}
  \label{EXP}
  \end{figure}

In Fig.\ \ref{EXP}, the model predictions for $P^i\to\gamma l^+l^-
\ (i=1, 2, 3)$ are given. We show the result for the two cases,
$c_{34}^+<0$ and $c_{34}^+>0$, respectively.
For $c_{34}^+>0$,
one can find a discrepancy between
the model prediction and
the precise data obtained by the NA60 collaboration
\cite{Arnaldi:2016pzu}.
Thus, we do not give a further result of analysis
for the case of $c_{34}^+>0$ since this case is disfavored.
\par
In Table \ref{resIP},
the model predictions for widths
of IP violating decays are exhibited.
Within $99.7\%\ \mathrm{C.L.}$ of the model predictions,
one can find no disagreement with experimental data.
The substantial error of $\phi\to\omega\pi^0$
comes from $g\bar{\theta}_{123}$ in Eq.\ (\ref{phiomegapi}),
which is proportional to a $\phi\omega\pi^0$ coupling.
Using the best-fit mixing matrix for mesons,
one can obtain the coupling,
\bea
g\bar{\theta}_{123}=0.11 c_3^\IP g^2-0.15 c_5^\IP g
+0.011c_6^\IP-0.023c_7^\IP,\label{Eq:123}
\eea
where each term has a comparable contribution to
$g\bar{\theta}_{123}$.
In Eq.\ (\ref{Eq:123}), the errors of $g, c_6^\IP$ and $c_7^\IP$
in Tables \ref{fit_para0} and \ref{FITTFF} are large, and give rise to uncertainty
of $\Gamma[\phi\to\omega\pi^0]$ in Eq.\ (\ref{Eq:phiomagapi0}).
Likewise, for $\Gamma[\phi\to\pi^0l^+l^-]$,
the substantial error arises  since the width includes
the VVP coupling in Eq.\ (\ref{TFFVECTORM}).
For $\eta\to\pi^+\pi^-\gamma$
the coupling associated with $\eta\pi\pi\gamma$
is given in Eq.\ (\ref{Ppipigamma}).
To determine $c_{34}^+$ in Eq.\ (\ref{Eq:ABC}),
we used the relation $c_{34}^+=gc_{34}^+/g$.
With Eq.\ (\ref{gc34}), Table \ref{fit_para0}
and $c_{34}^+<0$, one can obtain $c_{34}^+=(-1.5\pm0.4)\times10^{-2}$, which leads to uncertainty of $\Gamma[\eta\to\pi^+\pi^-\gamma]$.
\par
\begin{figure}[ht]
  \setlength{\subfigwidth}{.5\linewidth}
  \addtolength{\subfigwidth}{-.5\subfigcolsep}
  \begin{minipage}[b]{\subfigwidth}
    \subfigure[]{\includegraphics[width=8.2cm]{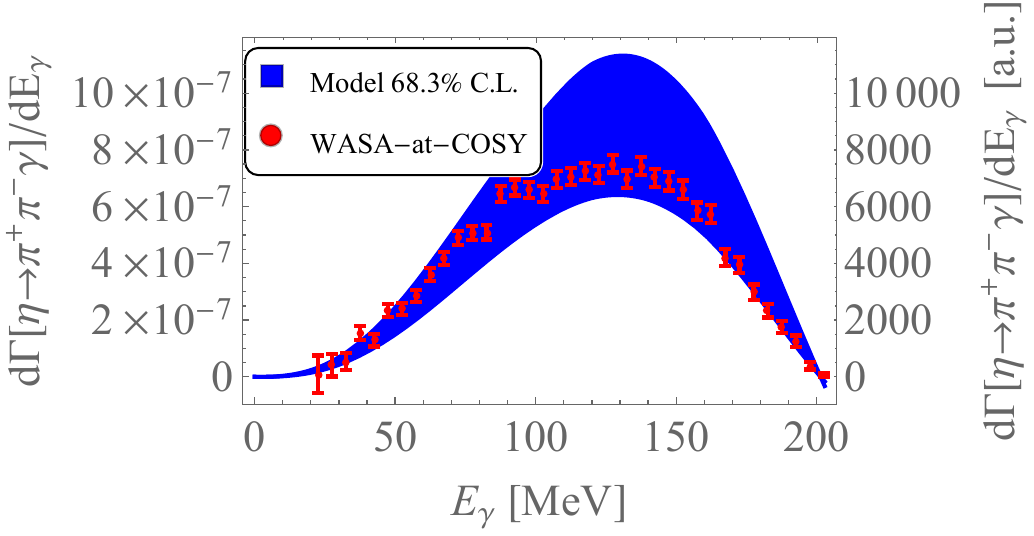}
   \label{Data10}}
  \end{minipage}
    \begin{minipage}[b]{\subfigwidth}
    \subfigure[]{\includegraphics[width=8.2cm]{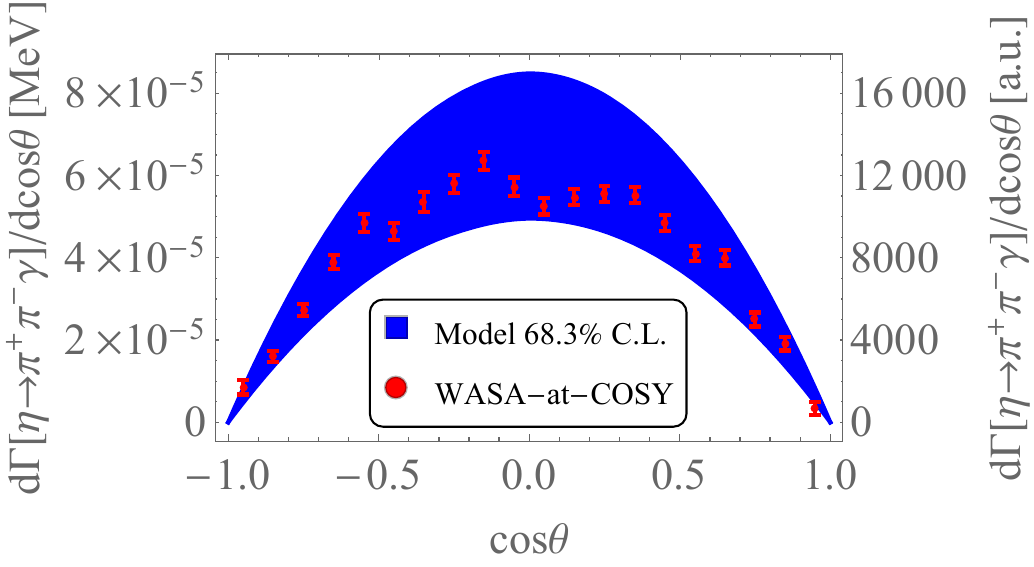}
   \label{Data6}}
  \end{minipage}\\
    \begin{minipage}[b]{\subfigwidth}
    \subfigure[]{\includegraphics[width=6.4cm]{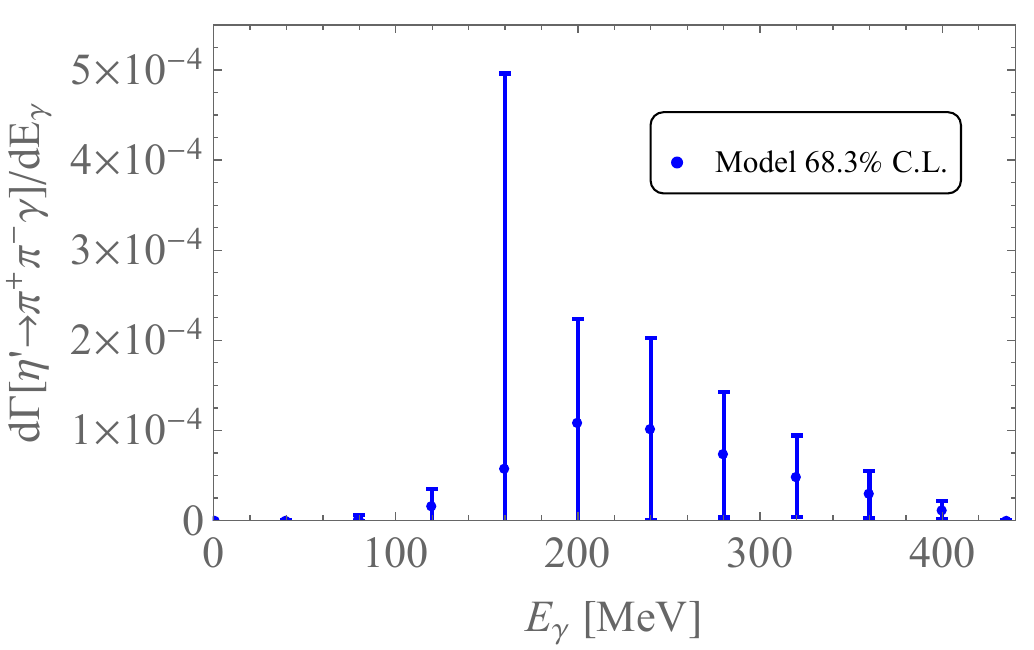}
  \label{Data11}}
  \end{minipage}
    \begin{minipage}[b]{\subfigwidth}
    \subfigure[]{\includegraphics[width=6.4cm]{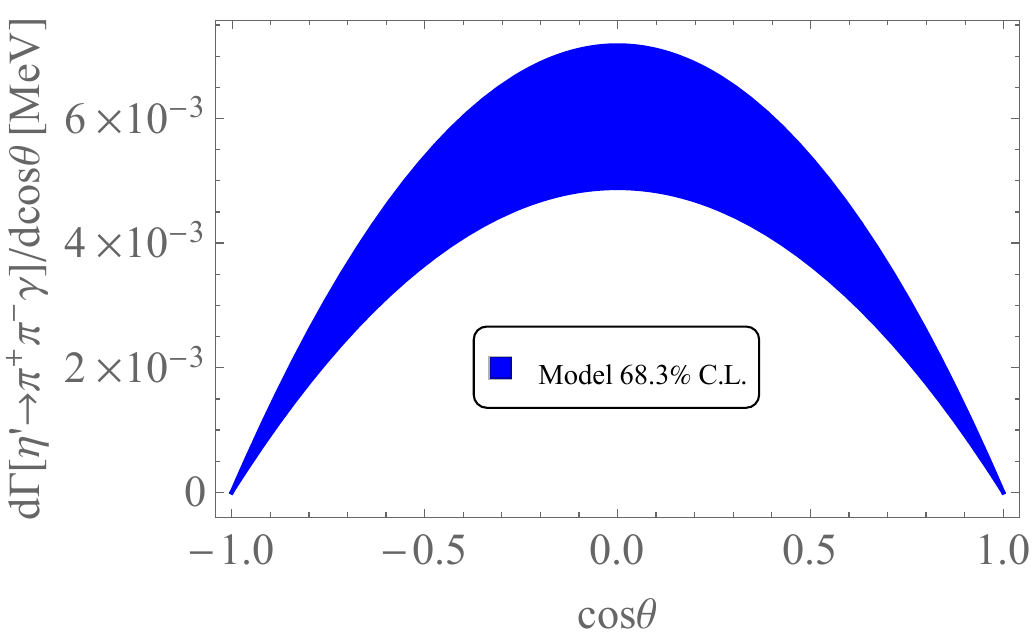}
   \label{Data8}}
  \end{minipage}
  \caption{
Plots of differential decay width of $P\to\pi^+\pi^-\gamma$:
(a), (c) the distributions of photon energy in the rest frame of 
$\eta$ and $\eta^{\prime}$,
(b), (d) the distributions of cosine of the angle between
$\pi^+$ and $\gamma$ in the rest frame of $\pi^+\pi^-$
for decays of $\eta$ and $\eta^\prime$, respectively.
For comparison,
the data measured by the WASA-at-COSY collaboration \cite{Adlarson:2011xb}
are shown as red circles in (a) and (b).
For both (a) and (b),
the vertical axis on the left side denotes
the physical differential width
while one on the right side shows arbitrary unit.
See the text for a detailed explanation of
units in which the differential widths
of $\eta\to\pi^+\pi^-\gamma$ are calculated.
}
\label{etada}
\end{figure}

In Fig.\ \ref{etada},
the differential decay widths for
$P^i\to\pi^+\pi^-\gamma\ (i=2, 3)$
are displayed.
For comparison, the data measured by
the WASA-at-COSY collaboration \cite{Adlarson:2011xb},
which are originally given in arbitrary unit,
are also shown for $\eta\to\pi^+\pi^-\gamma$.
For $\eta\to\pi^+\pi^-\gamma$ in (a) and (b),
the widths are given in two units:
one is physical unit which is based on
the calculation of decay width,
while another is arbitrary unit.
In order to compare the model values
in physical unit with
the experimental data,
we multiplied WASA-at-COSY data
(including central values and 1$\sigma$ errors)
by (a) $10^{-10}$ and (b) $5\times10^{-9}$,
respectively.
Likewise, in arbitrary unit,
our data are rescaled by the same factors.
We find that our numerical result agrees
with the experimental data if one chooses
the appropriate rescaling factors
for comparison.
In (c), one can find a resonance region around
$E_\gamma \sim 160 \mathrm{MeV}$.
This is because the photon energy in the rest frame of
$\eta^\prime$ is related to $\pi^+\pi^-$ invariant
mass as $E_\gamma=
(M_{\eta^\prime}^2-s_{+-})/2M_{\eta^\prime}$,
which indicates
that $E_\gamma=164.9 \mathrm{MeV}
(159.1 \mathrm{MeV})$
corresponds to the pole which arises from intermediate
$\rho \ (\omega)$.
\par
\begin{figure}[ht]
  \setlength{\subfigwidth}{.5\linewidth}
  \addtolength{\subfigwidth}{-.5\subfigcolsep}
     \begin{minipage}[b]{\subfigwidth}
    \subfigure[]{\includegraphics[width=8.2cm]{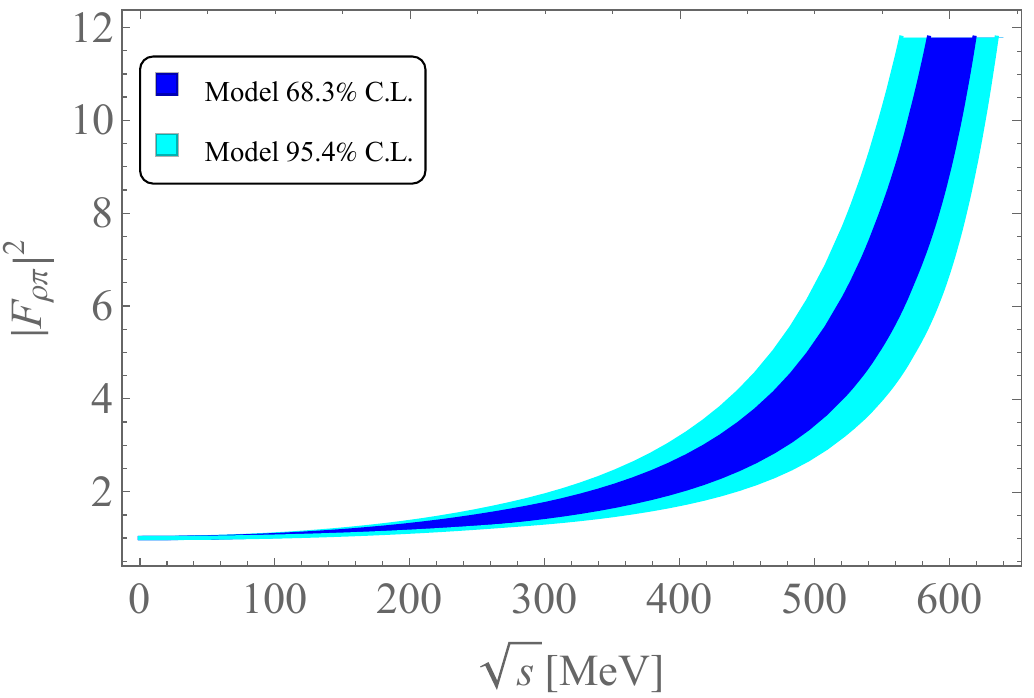}}
  \end{minipage}
  \begin{minipage}[b]{\subfigwidth}
    \subfigure[]{\includegraphics[width=8.2cm]{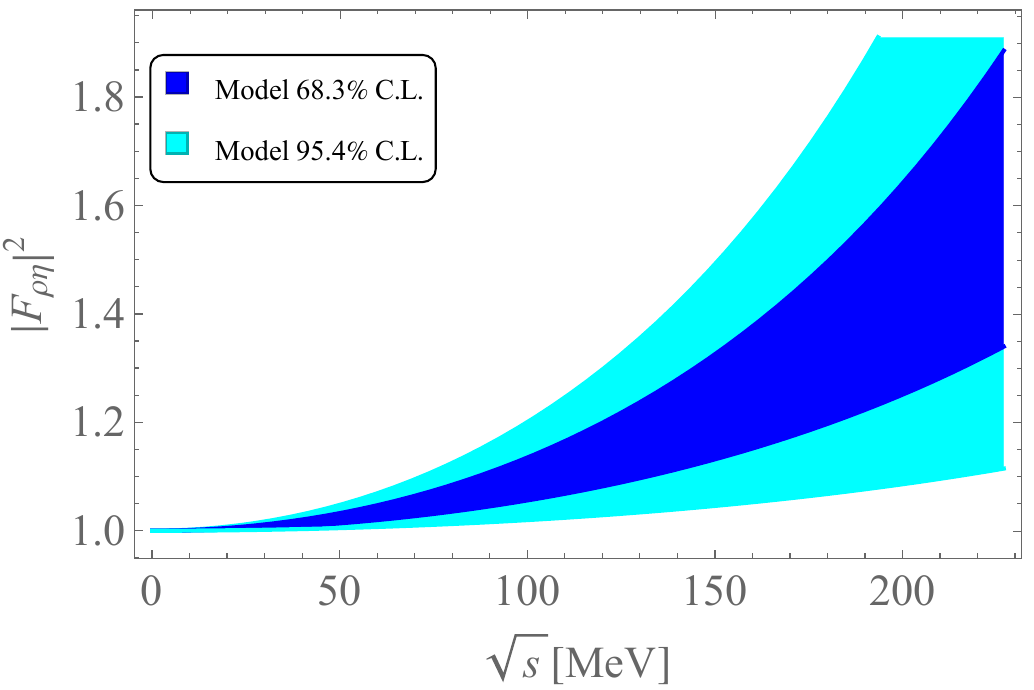}
   \label{Data9}}
  \end{minipage}\\
  \begin{minipage}[b]{\subfigwidth}
    \subfigure[]{\includegraphics[width=8.2cm]{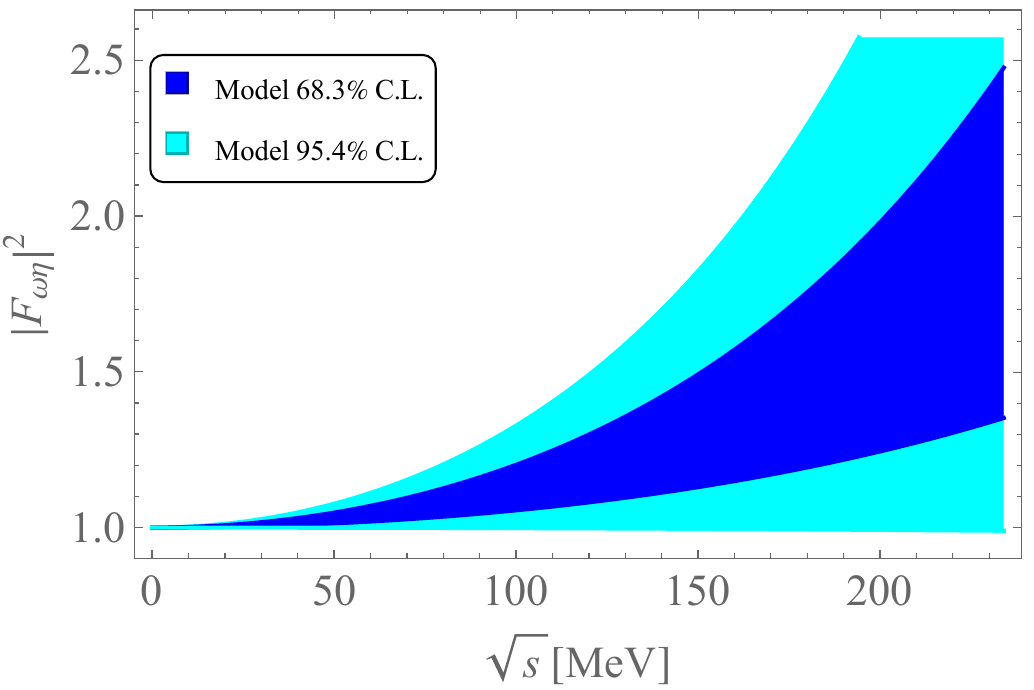}}
  \end{minipage}
  \begin{minipage}[b]{\subfigwidth}
    \subfigure[]{\includegraphics[width=8.2cm]{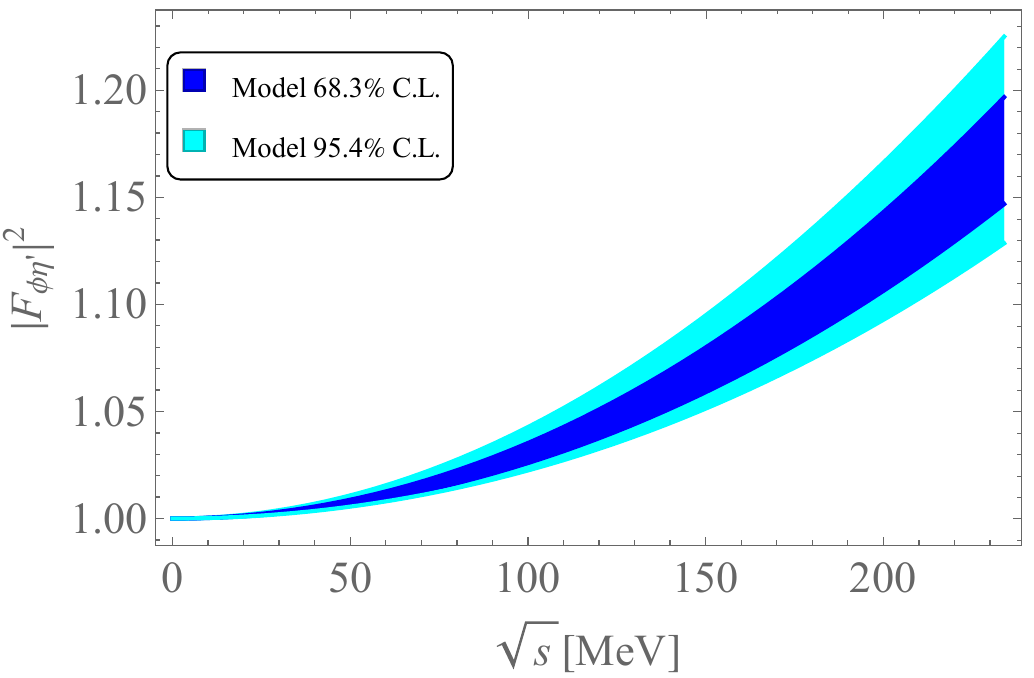}
   \label{Data4}}
  \end{minipage}
  \caption{Prediction of the model for TFFs:
  (a) $\rho^0\to\pi^0l^+l^-$, (b) $\rho^0\to\eta l^+l^-$,
  (c) $\omega\to\eta l^+l^-$ and
  (d) $\phi\to\eta^\prime l^+l^-$.
  Blue (cyan) bands imply model prediction
  in $68.3\%$ $(95.4\%)$ C.L.}
  \label{VPpred}
  \end{figure}

In Fig.\ \ref{VPpred},
we present the numerical result for the Dalitz distributions of $V^I\to P^il^+l^-$ for $(i, I)=(1, 1), (2, 1), (2, 2), (3, 3)$.
Since these modes are not measured yet,
it is expected that one can test the validity of
the model via future experiments.
\par
\begin{figure}[ht]
  \centering
  \setlength{\subfigwidth}{.5\linewidth}
  \addtolength{\subfigwidth}{-.5\subfigcolsep}
  \hspace{1.5mm}
  \begin{minipage}[b]{\subfigwidth}
    \subfigure[]{\includegraphics[width=8.5cm]{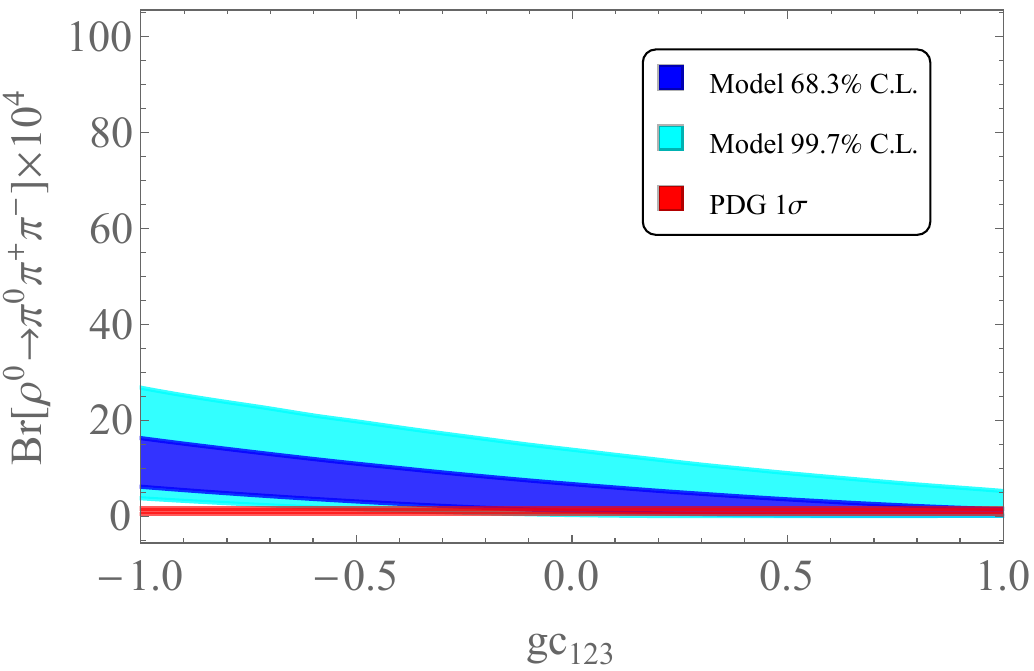}
   \label{3pi1}}
  \end{minipage}\\
  \hspace{2mm}
    \begin{minipage}[b]{\subfigwidth}
    \subfigure[]{\includegraphics[width=8.5cm]{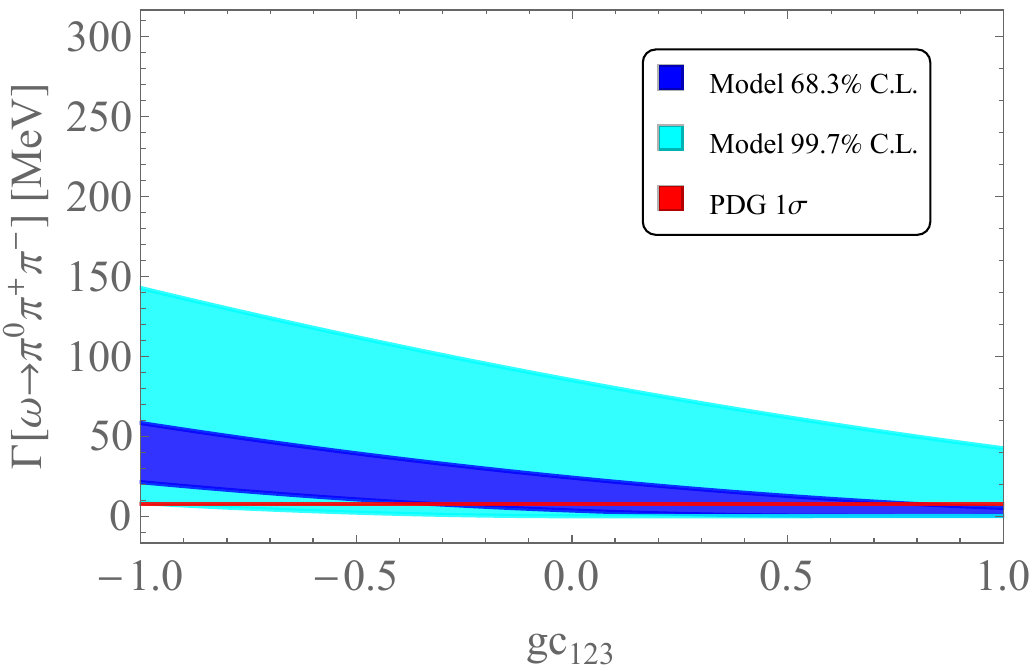}
   \label{3pi2}}
  \end{minipage}\\
  \hspace{-3mm}
    \begin{minipage}[b]{\subfigwidth}
    \subfigure[]{\includegraphics[width=8.5cm]{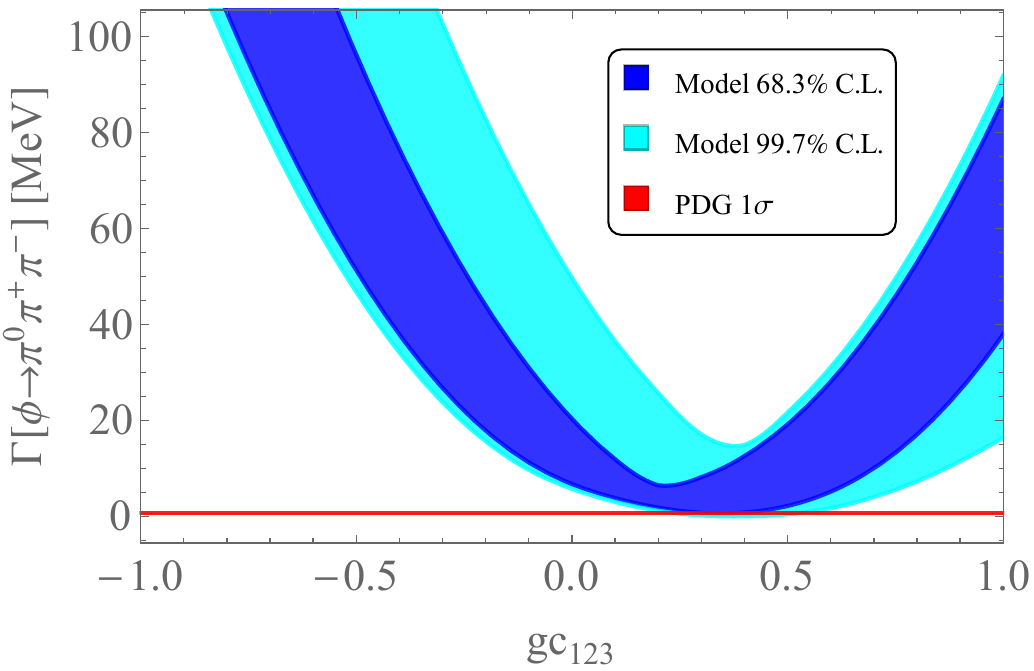}
   \label{3pi3}}
  \end{minipage}\\
  \caption{Plots of model prediction:
  (a) branching ratio of $\rho^0\to
  \pi^0\pi^+\pi^-$, (b) decay width of $\omega\to
  \pi^0\pi^+\pi^-$ and (c)
  decay width of $\phi\to \pi^0\pi^+\pi^-$.
  In these plots, blue (cyan) bands represent
  $68.3\%$ (99.7$\%$)
  confidence intervals of the model predictions
  while red bands indicate 1$\sigma$ ranges of the PDG  
  data \cite{PDG}.}
    \label{V3pigc123}
  \end{figure}

In Fig.\ \ref{V3pigc123}, predictions for a
branching ratio for $\rho^0\to \pi^0\pi^+\pi^-$
and decay widths of $V^I\to \pi^0\pi^+\pi^-\ (I=2, 3)$
are shown.
Varying the value of $gc_{123}$, we estimate
error bands of the model prediction.
For simplicity, we do not account uncertainty
which arises from parameters in the
vector meson propagators in Eq.\ (\ref{propVdiag}).
We find that if one fixes $gc_{123}\sim 0.35$,
the predictions for $V^I\to \pi^0\pi^+\pi^-\ (I=1, 2, 3)$
are consistent with
the PDG data \cite{PDG}.
\par
\begin{figure}[ht]
  \centering
  \setlength{\subfigwidth}{.5\linewidth}
  \addtolength{\subfigwidth}{-.5\subfigcolsep}
  \begin{minipage}[b]{\subfigwidth}
    \subfigure[]{\includegraphics{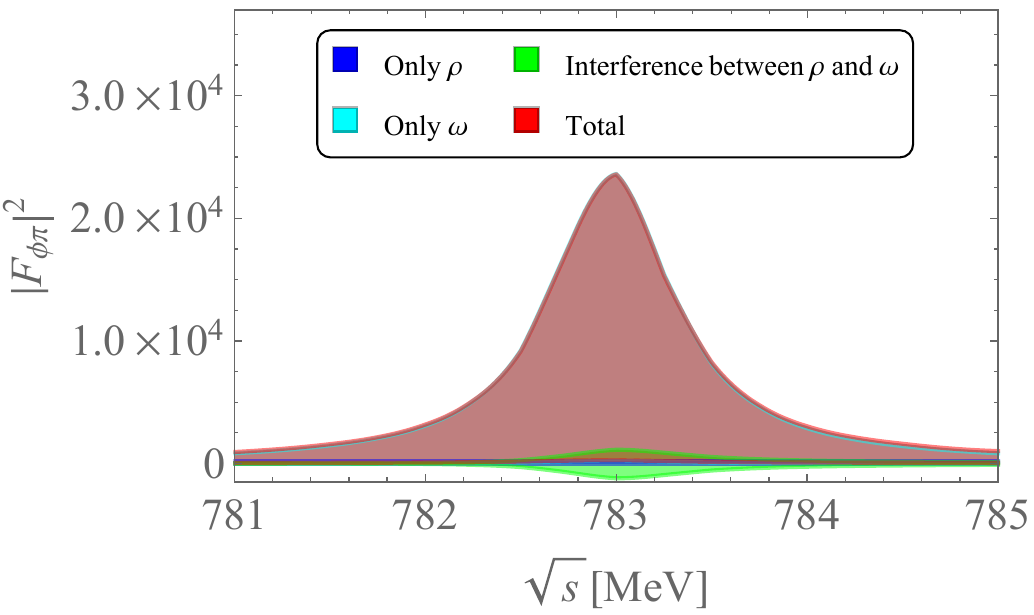}
   \label{resophi}}
  \end{minipage}
    \begin{minipage}[b]{\subfigwidth}
    \subfigure[]{\includegraphics{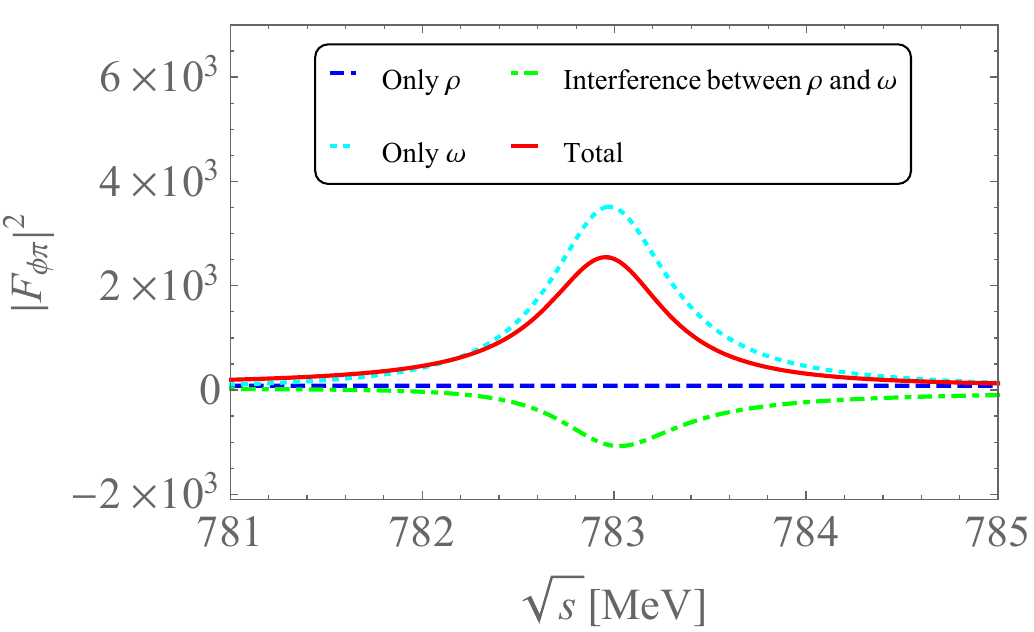}
   \label{resophi}}
  \end{minipage}\\
    \begin{minipage}[b]{\subfigwidth}
    \subfigure[]{\includegraphics{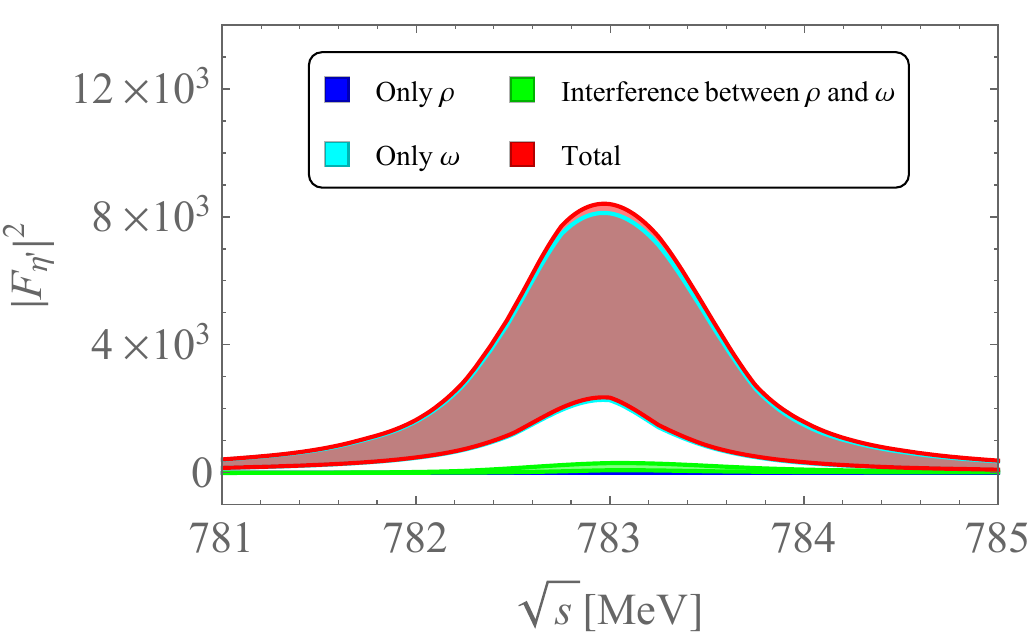}
   \label{resoetap}}
  \end{minipage}
      \begin{minipage}[b]{\subfigwidth}
    \subfigure[]{\includegraphics{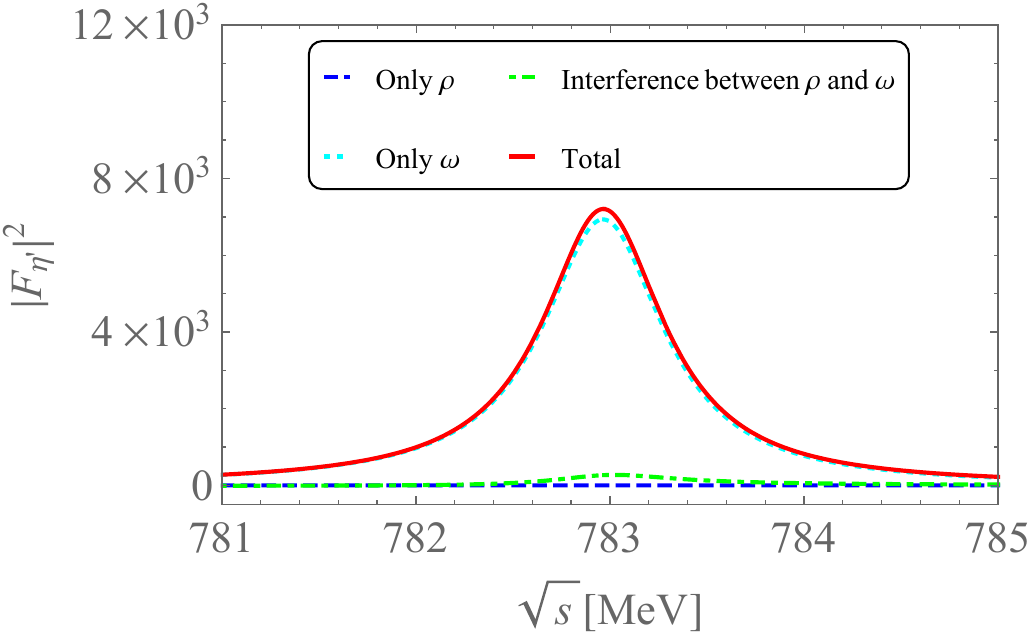}
   \label{resoetap}}
  \end{minipage}
  \caption{Transition form factors versus di-lepton invariant mass in the vicinity of the resonance regions:
(a) $\phi\to\pi^0 l^+l^-$ with 68.3$\%$ C.L. error bands,
(b) $\phi\to\pi^0 l^+l^-$ for the best fit prediction,
(c) $\eta^\prime\to\gamma l^+l^-$ with
68.3$\%$ C.L. error bands and (d) $\eta^\prime\to\gamma l^+l^-$ for the best fit prediction.
For each figure,
partial contributions from
$\rho^0$, $\omega$ and interference between
$\rho^0$ and $\omega$ are shown, respectively.}
    \label{Vpeak}
  \end{figure}

In the vicinity of the peak region, plots of the TFFs are exhibited for Dalitz decays in Fig.\ \ref{Vpeak}.
The partial contributions from $\rho, \omega$,
and interference between $\rho$ and $\omega$
are also indicated.
In (a) and (c), the predictions in 68.3$\%$ C.L.
are shown for TFFs of $\phi\to\pi^0 l^+l^-$
and $\eta^\prime\to\gamma l^+l^-$, respectively.
In (b) and (d), the best fit predictions,
in which the model parameters are fixed,
are given for the two modes.
For both $\phi\to\pi^0l^+l^-$ and
$\eta^\prime\to\gamma l^+l^-$,
we find that the contribution from $\omega$ pole
is dominant around the region of resonance.
It is shown that the partial contribution of
interference between $\rho$ and
$\omega$ is not negligible.
In particular, for (b), one can see that
the contribution of the interference is sizable.
\par
Using Eq.\ (\ref{vpp}), we obtain the decay widths for $\omega \to \pi\pi, 
\phi \to K^+K^-$ and $\phi \to K^0\bar{K}^0$
which are shown in Table \ref{tabVPP2}.
We should note that the leading contribution
of the decay is a one-loop level and  isospin breaking amplitude. 
About the $\phi \to K^+K^-$ and $\phi \to K^0\bar{K}^0$, they are smaller than the 
experimental values. However the discrepancy depends on the choice of 
$f_\pi/f_K$ and its  deviation from unity leads to two loop order effect.
If the ratio is modified properly, one can obtain theoretical
predictions which are in good agreement with the experimental results. We also note that the ratio of the decay widths of
 $\phi \to K^+K^-$ and $\phi \to K^0\bar{K}^0$ deviates from unity for both theoretical prediction and experimental
result. This implies the presence of the isospin breaking contribution. We note the ratio of the two decay widths
is in good agreement  between
theory and experiment:
\bea
(\Gamma[\phi\to K^+K^-]/\Gamma[\phi\to K^0\bar{K^0}])_{\mathrm{Th}.}&=&
1.53^{+0.22}_{-0.15},\\
(\Gamma[\phi\to K^+K^-]/\Gamma[\phi\to K^0\bar{K^0}])_{\mathrm{PDG}}&=&
1.430\pm0.026.
\eea
\par
With tree-level formulae, one may not explain the width of  $K^*$ and $\rho$ simultaneously, while the
1-loop formulae in Eqs.\ (\ref{Eq:rhopipi1loop}, \ref{Eq:KstKpi1loop})
can reproduce both of them within errors.
In Table \ref{tab_VPP}, the model predictions with
1-loop correction
for $\Gamma[\rho\to\pi\pi]$ and $\Gamma[K^{*\pm}\to (K\pi)^\pm]$
are shown.
The 1-loop corrected formulae
include the parameters in Table \ref{fit_para0} so
that these parameters
lead to the sizable errors in the 1-loop prediction for
the widths which are given in Table \ref{tab_VPP}.
%The fitted results of the widths for $\rho\to\pi\pi,
%K^{*\pm}\to(K\pi)^\pm$ are shown in
%Table \ref{tab_VPP}.
%%%%%%%%%%%%%%%
%%%%  Figure  %%%%%
%%%%%%%%%%%%%%%
%%%%%%%%%%%%%%%
%%%%  Table  %%%%%
%%%%%%%%%%%%%%%
\begin{table}[h!]
\caption{Partial decay widths of IPV decay modes.
As model predictions, we give
the estimated ranges of $68.3\%$ C.L. and 
ones for $99.7\%$ C.L., respectively.
For comparison, the data obtained by the BES
III collaboration \cite{Ablikim:2015wnx}
is shown for $\Gamma[\eta^\prime\to\gamma e^+e^-]$
and the PDG data \cite{PDG} are given for other decay modes.
For $\rho^0\to\pi^0 e^+e^-$
and $\phi\to\eta \mu^+\mu^-$, the
$90\%$ C.L. upper bounds
are written while $1\sigma$ errors are shown
for the other experimental values.
}
\label{resIP}
\centering
\begin{center}
\begin{tabular}{cccc}\hline\hline
Decay mode & \hspace{2mm} Model $(68.3\% \mathrm{C.L.})$ [MeV] \hspace{2mm}
&  \hspace{2mm} Model $(99.7\% \mathrm{C.L.})$ [MeV] \hspace{2mm}
& \hspace{2mm} Exp. [MeV] \hspace{2mm} 
\\
\hline
$\Gamma[\pi^0\to\gamma e^+e^-]$ &$(9.05^{+0.01}_{-0.01})\times10^{-8}$ &$(9.05^{+1.44}_{-0.48})\times10^{-8}$
&$(9.1\pm0.3)\times10^{-8}$\\
$\Gamma[\eta\to\gamma e^+e^-]$ &$(8.54^{+0.05}_{-0.04})\times10^{-6}$ & $(8.54^{+0.41}_{-0.08})\times10^{-6}$
&$(9.0\pm0.6)\times10^{-6}$\\
$\Gamma[\eta\to\gamma \mu^+\mu^-]$ &
$(3.7^{+0.3}_{-0.2})\times10^{-7}$ &
$(3.7^{+2.2}_{-0.4})\times10^{-7}$
&$(4.1\pm0.6)\times 10^{-7}$\\
$\Gamma[\eta^\prime\to\gamma e^+e^-]$ &
$(8.7^{+0.5}_{-0.3})\times10^{-5}$ &
$(8.7^{+38.2}_{-0.8})\times10^{-5}$
&$(9.28\pm0.95)\times10^{-5}$\\
$\Gamma[\eta^\prime\to\gamma \mu^+\mu^-]$ &
$(1.6^{+0.5}_{-0.3})\times10^{-5}$ &
$(1.6^{+38.5}_{-0.7})\times10^{-5}$
&$(2.1\pm0.6)\times10^{-5}$\\
$\Gamma[\eta\to \pi^+\pi^-\gamma]$ &
$(8.7^{+2.4}_{-2.2})\times10^{-5}$ &
$(8.7^{+8.4}_{-7.2})\times10^{-5}$
&$(5.5\pm0.2)\times10^{-5}$\\
$\Gamma[\eta^\prime\to \pi^+\pi^-\gamma]$ &
$(6.2^{+1.2}_{-1.0})\times10^{-2}$ &
$(6.2^{+4.3}_{-2.6})\times10^{-2}$&
$(5.8\pm0.3)\times10^{-2}$\\
$\Gamma[\phi\to\omega\pi^0]$ &$(48^{+314}_{-44})\times10^{-4}$&
$(48^{+4845}_{-48})\times10^{-4}$
&$(2.0\pm0.2)\times10^{-4}$\\
$\Gamma[\rho^0\to\pi^0e^+e^-]$ &$(0.43^{+0.05}_{-0.05})\times10^{-3}$ &
$(0.43^{+0.16}_{-0.13})\times10^{-3}$
&$<6.0\times10^{-3}$\\
$\Gamma[\rho^0\to\pi^0\mu^+\mu^-]$ &
$(5.0^{+0.9}_{-0.7})\times10^{-5}$&
$(5.0^{+3.3}_{-2.2})\times10^{-5}$
&$-$  \\
$\Gamma[\rho^0\to\eta e^+e^-]$ &
$(2.5^{+0.7}_{-0.5})\times10^{-4}$ &
$(2.5^{+2.4}_{-2.2})\times10^{-4}$
&$-$ \\
$\Gamma[\rho^0\to\eta \mu^+\mu^-]$ &
$(3.3^{+1.1}_{-0.8})\times10^{-8}$ &
$(3.3^{+4.4}_{-2.8})\times10^{-8}$
&$-$  \\
$\Gamma[\omega\to\pi^0 e^+e^-]$ &
$(6.8^{+0.9}_{-0.8})\times10^{-3}$  &
$(6.8^{+3.3}_{-2.3})\times10^{-3}$
&$(6.5\pm0.5)\times10^{-3}$\\
$\Gamma[\omega\to\pi^0 \mu^+\mu^-]$ &
$(0.89^{+0.15}_{-0.13})\times 10^{-3}$ &
$(0.89^{+0.49}_{-0.30})\times 10^{-3}$
&$(1.1\pm0.3)\times10^{-3}$\\
$\Gamma[\omega\to\eta e^+e^-]$
&$(4.2^{+1.3}_{-1.0})\times10^{-5}$ & 
$(4.2^{+4.6}_{-3.4})\times10^{-5}$
&$-$  \\
$\Gamma[\omega\to\eta \mu^+\mu^-]$ &
$(1.7^{+0.7}_{-0.5})\times10^{-8}$ &
$(1.7^{+4.3}_{-1.3})\times10^{-8}$
&$-$  \\
$\Gamma[\phi\to\pi^0 e^+e^-]$ &
$(23^{+22}_{-13})\times10^{-5}$ &
$(23^{+424}_{-23})\times10^{-5}$
&$(4.8\pm1.2)\times 10^{-5}$\\
$\Gamma[\phi\to\pi^0 \mu^+\mu^-]$ &
$(6.7^{+16.7}_{-4.7})\times10^{-5}$ &
$(6.7^{+388.3}_{-6.6})\times10^{-5}$
&$-$  \\
$\Gamma[\phi\to\eta e^+e^-]$ &
$(1.9^{+0.8}_{-1.0})\times10^{-4}$ &
$(1.9^{+71.0}_{-1.9})\times10^{-4}$
&$(4.9\pm0.4)\times10^{-4}$\\
$\Gamma[\phi\to\eta \mu^+\mu^-]$ &
$(1.1^{+0.5}_{-0.6})\times10^{-5}$ &
$(1.1^{+105.0}_{-1.1})\times10^{-5}$
&$<4.0\times10^{-5}$\\
$\Gamma[\phi\to\eta^\prime e^+e^-]$ &
$(2.0^{+0.7}_{-0.5})\times10^{-6}$ &
$(2.0^{+2.5}_{-1.7})\times10^{-6}$
&$-$ \\
\hline\hline 
\end{tabular}
\end{center}
\end{table}
%%%%%%%%%%%%%%%
%%%%  Table  %%%%%
%%%%%%%%%%%%%%%
%%%%%%%%%%%%%%%
%%%%  Table  %%%%%
%%%%%%%%%%%%%%%
%%%%%%%%%%%%%%%%%%%%
%%  V-> 3pi
%%%%%%%%%%%%%%%%%%%%
  %%%%%%%%%%%%%%%%%%%
    \begin{table}[h]
\caption{The results of the decay widths for $V \to PP$. They are obtained 
based on the one-loop corrected formulae in Eq.(\ref{Eq:rhopipi1loop}) and Eq.(\ref{Eq:KstKpi1loop}). }
\label{tabVPP2}
\begin{center}
\begin{tabular}{ccc}
\hline\hline
Decay mode & \hspace{5mm} Theory (MeV) \hspace{5mm} & PDG (MeV) \\
\hline
 $\Gamma[\omega \to \pi \pi]$ & $0.114^{+0.03}_{-0.02}$ &  $0.130 \pm 0.016$ \\
 $\Gamma[\phi \to K^+K^-]$ & $1.43^{+0.15}_{-0.10}$ & $2.086 \pm 0.026$ \\ 
$\Gamma[\phi \to K^0 \bar{K}^0]$ & $0.935^{+0.09}_{-0.06}$ & $1.459 \pm 0.020$ \\
\hline\hline
\end{tabular}
\end{center}
\end{table}
\begin{table}[h!]
\caption{\label{tab_VPP}
    The results of the decay widths for $V \to PP$.}
\label{VPPcharge}
\begin{center}
\begin{tabular}{ccc}
\hline\hline 
 Decay mode &\hspace{5mm} Theory (MeV) \hspace{5mm}
& PDG (MeV) \\ \hline
 $\Gamma[\rho \to \pi\pi]$ & $157^{+66}_{-47}$ & $149.1 \pm 0.8$ \\ 
 $\Gamma[K^{*\pm} \to (K\pi)^\pm]$ & $45.4^{+20.6}_{-14.6}$ & $46.2 \pm 1.3$ \\   
\hline\hline
\end{tabular}
\end{center}
\end{table}
  \clearpage
%%%%%%%%%%%%%%%%%%%%%%%
% SU(3) breaking section added by Morozumi Oct.24th
%%%%%%%%%%%%%%%%%%%%%%
\section{SU(3) breaking effect in IPV interaction}
\label{secV}
In the previous sections, we include the SU(3) breaking effects from the intrinsic parity conserving part.
These effects are order of $O(p^4 m_\pi^2)$.
The full SU(3) breaking  effect for IPV processes up to this order comes from one loop diagrams
and also from  SU(3) 
breaking IPV vertex. The latter interactions are studied in  \cite{Hashimoto:1996ny}. Below we study the 
 SU(3) breaking effect from the IPV vertex. We  focus on the processes $\rho \to \pi \gamma$
and $K^\ast \to K \gamma $ 
and show how these terms improve the predictions compared with those without SU(3) breaking terms.
One loop corrections  and the renormalization of the divergence is beyond the scope of this paper.
%\ref{}%  
We consider the following SU(3) breaking IPV interaction terms  
\begin{eqnarray}
  \Delta L_3 & = & c_{31 b}^{\text{IP}} \epsilon^{\mu \nu \rho \sigma}
  \text{Tr} [g \{ F_{V \mu \nu}, \hat{\epsilon}^{} \} \{ \alpha_{L \rho}
  \alpha_{R \sigma} - \alpha_{R \rho} \alpha_{L \sigma} \}] , \nonumber \\
  \Delta L_3' & = & c_{32 b}^{\text{IP}} \epsilon^{\mu \nu \rho \sigma}
  \text{Tr} [g F_{V \mu \nu} \{ \alpha_{L \rho} \hat{\epsilon} \alpha_{R
  \sigma} - \alpha_{R \rho} \hat{\epsilon} \alpha_{L \sigma} \}], \nonumber \\
 \Delta L_3^{''} & = & c_{33 b}^{\text{IP}} \epsilon^{\mu \nu \rho \sigma}
  \text{Tr} [g \{ F_{V \mu \nu}, \hat{\epsilon}^{} \}] \text{Tr}\{ \alpha_{L \rho}
 \alpha_{R \sigma} - \alpha_{R \rho} \alpha_{L \sigma} \}, \nonumber  \\
  \Delta L_4 & = & c_{41 b}^{\text{IP}} \epsilon^{\mu \nu \rho \sigma}
  \text{Tr} [ \{ (\hat{F}_L^{} + \hat{F}_R)_{\mu \nu}, \hat{\epsilon} \} \{
  \alpha_{L \rho} \alpha_{R \sigma} - \alpha_{R \rho} \alpha_{L \sigma} \}], \nonumber \\
 \Delta L_4' & = & c_{42 b}^{\text{IP}} \epsilon^{\mu \nu \rho \sigma}
  \text{Tr} [ (\hat{F}_L^{} + \hat{F}_R)_{\mu \nu}  \{
  \alpha_{L \rho}\hat{\epsilon}  \alpha_{R \sigma} - \alpha_{R \rho} \hat{\epsilon} \alpha_{L \sigma} \}], \nonumber \\
\Delta L_4^{''} & = & c_{43 b}^{\text{IP}} \epsilon^{\mu \nu \rho \sigma}
  \text{Tr} [ \{ (\hat{F}_L^{} + \hat{F}_R)_{\mu \nu}, \hat{\epsilon} \}] 
\text{Tr} \{ \alpha_{L \rho} \alpha_{R \sigma} - \alpha_{R \rho} \alpha_{L \sigma} \},
\end{eqnarray}
where $\hat{\epsilon}$ are spurion field for SU(3) breaking,
\begin{eqnarray}
\hat{\epsilon}&=&\xi \epsilon \xi + \xi^\dagger \epsilon \xi^\dagger,
\epsilon=\text{diag.} (\epsilon_1, \epsilon_2, \epsilon_3).
\end{eqnarray}
$\Delta L_3^{''}$ and $\Delta L_4^{''}$ are newly introduced and the others are studied in \cite{Hashimoto:1996ny}.
Below we assume that $\epsilon$ is proportional to current quark mass matrix $\text{diag.}
(m_u, m_d,m_s)$.
Using the Lagrangian, one can compute the effective interactions for $\rho \rightarrow \pi \gamma$ and $K^\ast \rightarrow K \gamma$.
To obtain the interactions, not only the direct interaction of  $V \to P \gamma$ but also  the contribution
 from  $V \to V^0 P \to \gamma P$ is included.
\begin{eqnarray}
  L_{eff}& = & - \frac{4 e}{f} \epsilon^{\mu \nu \rho \sigma} A_{\rho}
  \left[ \left\{ - \frac{1}{6} g \hat{c}_{34}^{+} - \frac{2 - \widehat{m_d}}{3} a-c 
  \right\} \partial_{\mu} \rho^0_{\nu} \partial_{\sigma} \pi^0_{^{}} \right.\nonumber \\
  & + & \left\{ - \frac{1}{6} g \hat{c}_{3 4}^{+} - \frac{2 - \widehat{m_d}}{3}
  a + \frac{1 - \widehat{m_d}}{2} b -c 
\right\} \partial_{\mu} \rho^+_{\nu}
  \partial_{\sigma} \pi^- \nonumber \\
  & - & \left\{ - \frac{1}{3} g \hat{c}_{3 4}^+ - \frac{\widehat{m_d} +
  \widehat{m_s}}{3}(a+3 c) \right\} \partial_{\mu} K^{\ast 0}_{\nu}
  \partial_{\sigma} \bar{K}^0 \nonumber \\
  & + & \left. \left\{ - \frac{1}{6} g \hat{c}_{3 4}^{+} - \frac{2 -
  \widehat{m_s}}{3} a + \frac{1 - \widehat{m_s}}{2} b -c\right\}
  \partial_{\mu} K^{\ast +}_{\nu} \partial_{\sigma} K^- \right] ,
\end{eqnarray}
where  $\widehat{m_s}=\frac{m_s}{m_u}$ and  $\widehat{m_d}=\frac{m_d}{m_u}$. 
$g \hat{c}_{34}^{+}$, $a$, $b$ and $c$ are defined as,
\begin{eqnarray}
g \hat{c}_{34}^{+}&=&g c_{34}^+-2(1+\widehat{m_s}+\widehat{m_d})c, \nonumber \\
 a & = & g \epsilon_1 \left( c_{31 b}^{\text{IP}} + 2 c_{41 b}^{\text{IP}} +
  \frac{1}{2} (c_{32 b}^{\text{IP}} + 2 c_{42 b}^{\text{IP}}) \right), \nonumber \\
 b & = & g \epsilon_1 (c_{32 b}^{\text{IP}} + 2 c_{42 b}^{\text{IP}}),\nonumber \\
 c & = & g \epsilon_1 (c_{33 b}^{\text{IP}} + 2 c_{43 b}^{\text{IP}}).
\end{eqnarray}
We note that $c$ contributes to all four modes in the same manner and the strength of the contribution is proportional to $\text{Tr}{(Q M)}=\frac{2 m_u-(m_d+m_s)}{3}$. 

In our numerical calculation, we fit all four modes and  determine the parameters.
\begin{eqnarray}
(g \hat{c}_{34}^{+},a,b,c)=(-0.10, 0.031, 0.022,-0.01).
\end{eqnarray}
With the determined parameters, we reproduce the central values of  PDG 
 decay widths for $K^\ast \to K \gamma $ and $\rho \to \pi \gamma$ 
shown in Table III and IV.
%%%%%%%%%%%%%%%
%%  Conclusion  %%%
%%%%%%%%%%%%%%%
\section{Summary and discussion}
\label{sec6}
The IP violating phenomena of light hadrons are investigated
in the model of chiral Lagrangian including vector mesons.
We introduced the suitable tree-level interaction terms 
which include singlet fields of vector meson and pseudoscalar.
Power counting of superficial degree of
divergence enables us to specify
the 1-loop order interaction Lagrangian
under the presence of the tree-level part.
With introduced interactions,
1-loop correction to the self-energies of vector mesons
is analyzed.
Using the 1-loop corrected mass matrix, we obtained the
expressions of physical masses and the mixing matrix
of $\rho, \omega$ and $\phi$.
Including the kinetic mixing effect,
the model expressions of the width for $V\to PP$ decay are calculated.
We also analyzed the mixing between photon and
neutral vector mesons, which gives important
contribution to processes such as $V\to PV^*\to
P\gamma$.
\par
For pseusoscalars,
we took account of 1-loop correction to the mass matrix.
The physical states of $\pi^0, \eta, \eta^\prime$
are written in terms of SU(3) eigenstates
through wavefunction renormalizations
and an orthogonal matrix which diagonalizes
the 1-loop corrected mass matrix.
\par
On the basis of the framework incorporating octet
and singlet fields, the IP violating operators are introduced
within SU(3) invariance.
We constructed $\mathcal{L}_i (i=5-10)$,
which includes the SU(3) singlet fields
of a pseudoscalar and a vector meson
in addition to ones introduced in
Refs.\ \cite{Fujiwara:1984mp, Bando:1987br, Hashimoto:1996ny}.
In order to realize the experimental
data in the framework including the singlets and octets,
we found that the singlet-induced operators play an important
role; if $\mathcal{L}_i (i=5-10)$ were absent in the model,
$\Gamma[\eta^\prime\to2\gamma]$ would become much
smaller than the observed value in the experiments.
\par
Using the introduced IP violating operators,
we obtained the analytic formulae for the IP violating (differential) decay widths.
In particular, the widths of $P\to V\gamma, V\to P\gamma,
\phi\to\omega\gamma, P\to2\gamma, V\to P\pi^+\pi^-$
are given.
Moreover, the electromagnetic TFFs of 
$P\to\gamma l^+l^-$ and $V\to Pl^+l^-$
are also obtained.
Additionally, the formula of the differential width for
$P\to\pi^+\pi^-\gamma$ is also shown.
\par
For parameter estimation, we used
precise data of spectrum function
for $\tau^-\to K_S\pi^-\nu$ measured by
the Belle collaboration\ \cite{Epifanov:2007rf}.
Furthermore, the PDG data \cite{PDG} of physical masses of charged vector mesons,
$\rho^+$ and $K^{*+}$ are 
used for parameter estimation of the coefficient of
1-loop order interaction terms.
We also estimated the model parameters which appear in
the mass matrix of neutral vector mesons
by using the PDG data \cite{PDG} of $m_\rho, m_\omega$
and $m_\phi$.
Since the masses of vector mesons are precisely
measured in the experiment,
the model parameters in the mass matrix
are estimated with smaller uncertainty.
\par
The numerical analyses of IP violating decay widths,
the TFFs for electromagnetic decays,
are carried out in the model.
In order to estimate the IP violating parameters,
we utilized the PDG data \cite{PDG} of widths for radiative decays.
Specifically, the experimental data
of $\Gamma[K^*\to K\gamma]$ and 
the effective coupling ratios of $V^0V^0P^0$
to $\rho^+\pi^+\gamma$ are used.
We also considered constraints on a mass matrix and a mixing matrix of pseudoscalars.
To obtain a parameter region which
is consistent with the masses and
$\Gamma[P\to2\gamma]$,
we solved the system of equations to realize
the PDG data \cite{PDG}.
Furthermore, $\chi^2$ fitting for the TFFs measured in the experiments \cite{Arnaldi:2016pzu,
Achasov:2008zz,
Dzhelyadin:1980tj,
Akhmetshin:2005vy,
Babusci:2014ldz,
Achasov:2000ne,
::2016hdx}
is carried out.
We found that the goodness-of-fit is improved if one
does not use the input data measured by
the Lepton-G experiment.
Hence, we adopted the parameter set estimated
without their data. 
\par
Using the estimated model parameters,
we gave the model predictions for IP violating decays.
In particular, we found that the electromagnetic TFFs
of $\eta\to \gamma l^+l^-, \eta^\prime\to\gamma l^+l^-$
are consistent with the experimental data
for $c_{34}^+<0$.
The partial widths of
$P\to \gamma l^+l^-$,
$P\to\pi^+\pi^-\gamma$,
$\phi \to \omega\pi^0$ and
$V\to Pl^+l^-$
are calculated, none of which result in significant
deviation from the experimental data up to
$99.7\%$ C.L.
For the differential widths
of $\eta\to\pi^+\pi^-\gamma$
and $\eta^\prime\to\pi^+\pi^-\gamma$,
the model predictions are given.
The differential width of $\eta\to\pi^+\pi^-\gamma$
is compared with the data measured by
the WASA-at-COSY collaboration \cite{Adlarson:2011xb}.
Here, no significant deviation is found in this result.
The predictions are also obtained for 
the TFFs of $\rho\to\pi^0 l^+l^-, \rho\to\eta l^+l^-,
\omega\to\eta l^+l^-$ and $\phi\to\eta^\prime l^+l^-$,
which are expected be observed in future experiments.
The model predictions for
$\mathrm{Br}[\rho^0\to\pi^0\pi^+\pi^-]$,
$\Gamma[\omega\to\pi^0\pi^+\pi^-]$
and $\Gamma[\phi\to\pi^0\pi^+\pi^-]$
are also presented.
We found that these IP violating observables
are consistent with the PDG values \cite{PDG}.
In the vicinity of resonance region, the TFFs for
$\phi\to\pi^0l^+l^-$ and one for
$\eta^\prime\to\gamma l^+l^-$ are analyzed.
It is shown that the $\omega$ pole is dominant
in the peak region for both TFFs.
We also found that the contribution of the interference
between $\rho$ and $\omega$
is non-negligible in the peak region.
It is shown that 
the theoretical prediction for
$\Gamma[\phi\to K^+K^-]/
\Gamma[\phi\to K^0\bar{K^0}]$
agrees with the experimental value,
although $\Gamma[\phi\to K^+K^- (K^0\bar{K^0})]$
depends on two-loop ordered uncertainty.
Our framework, which includes $O(p^4)$ contribution both in IP conserving part and in IPV  part, can not explain simultaneously  decay widths of $K^{\ast+} \to K^{+} \gamma$ and $K^{\ast 0} \to K^0 \gamma$. 
As the possible solutions of the problem, we study the $O(p^4 m_\pi^2)$ contribution. 
In addition to  SU(3) breaking interactions for  IPV part of  Ref.\ \cite{Hashimoto:1996ny}
, we include two new terms.  With these terms, we can explain 
the decay widths for  
all four modes of $\rho \to \pi \gamma$ and $K^\ast \to K \gamma$.  
In contrast to the treatment of Ref.\ \cite{Hashimoto:1996ny}, we assume that SU(3) breaking 
is proportional to current quark mass.
\section*{Acknowledgement}
We thank H. Tagawa for helpful discussion.
This work is partially supported by Scientific Grants by the Ministry of Education, Culture, Sports, Science and Technology of Japan (Nos.\ 24540272 (HU), 26247038 (HU), 15H01037 (HU), 16H00871 (HU), 16H02189 (HU)) and by JSPS KAKENHI Grant Numbers JP16H03993 (TM), and JP17K05418 (TM). 
%%%%%%%%%%%%%
%% appendix %%%
%%%%%%%%%%%%%
\appendix
%%%%%%%%%%%%%%%%%%%
%%%  Counter terms  %%%%
%%%%%%%%%%%%%%%%%%%
\section{Counter terms}
\label{counterterms}
%%%%%%%%here%%%%%%%%%%%%%%
The counter terms are computed
with 1-loop correction of
SU(3) singlet pseudoscalar
in Ref.\ \cite{Kimura:2014wsa}.
In this work, we only consider the corrections due to SU(3) octet pseudoscalars.
The effect of SU(3)$_R$ external gauge boson is included.
The counter terms in 1-loop order are,
\bea
\mathcal{L}_{c}=&&L_1
\left({\rm Tr}(D_\mu U (D^\mu U)^\dagger)\right)^2+L_2{\rm Tr}(D_\mu U (D_\nu U)^\dagger) {\rm Tr}(D^\mu U (D^\nu U)^\dagger) \nn \\
&+&L_3 \aTr \{D^\mu U (D_{\mu} U)^\dagger D^\nu U 
(D_{\nu} U)^\dagger\} \nn \\
&+&\frac{4B}{f^2}L_4 {\rm Tr}\{D_\mu U(D^\mu U)^\dagger\} 
{\rm Tr}\{M(U+U^\dagger)\} \nn \\
&+&\frac{4B}{f^2}L_5{\rm Tr}\{D_\mu U(D^\mu U)^\dagger(U M+ M U^\dagger)\}
\nn
\\
&+&\frac{16 B^2}{f^4} L_6\{{\rm Tr}(M (U+U^\dagger))\}^2 \nn \\
&+&\frac{16 B^2}{f^4} L_7 \{{\rm Tr}(M(U-U^\dagger))\}^2 \nn \\
&+&\frac{16 B^2}{f^4} L_8{\rm Tr}(M U M U+ M U^\dagger M U^\dagger) \nn \\
&+&i L_9  {\rm Tr} \{
F_{L \mu \nu}(D^{\mu} U)(D^{\nu} U)^\dagger+
F_{R \mu \nu} (D^\mu U)^\dagger D^\nu U \} \nn \\
&+&L_{10}\mathrm{Tr} (F_{L \mu \nu} U {F_R}^{\mu \nu} U^\dagger) \nn\\
&+&
H_1 {\rm Tr}(F_{L \mu \nu} {F_L}^{\mu \nu}
+F_{R \mu \nu} {F_R}^{\mu \nu})
\nn \\
&+& H_2 \left(\frac{4B}{f^2}\right)^2 {\rm Tr}(M^2) \nn \\
&+& i\frac{K_1}{2} {\rm Tr}(\xi^\dagger D^\mu U (D^\nu U)^\dagger \xi)
(D_\mu v_\nu-D_\nu v_\mu+ i[v_\mu, v_\nu]) \nn \\
&-&\frac{1}{2} \left(K_2 {\rm Tr}
(\xi^\dagger F_{L\mu \nu} \xi+\xi F_{R \mu \nu} \xi^\dagger)(D^\mu v^\nu-D^\nu 
v^\mu+ i[v^\mu, v^\nu]) \right. \nn \\
&+& \left.K_3  {\rm Tr}(D_\mu v_\nu-D_\nu v_\mu+ i[v_\mu, v_\nu])
 (D^\mu v^\nu-D^\nu 
v^\mu+ i[v^\mu, v^\nu]) \right) \nn \\
&+&\frac{4B}{f^2}
\left(K_4 {\rm Tr}\{(\xi M \xi+ \xi^\dagger M \xi^\dagger) v^2\} +K_5
{\rm Tr} \{M(U+U^\dagger)\} {\rm Tr}(v^2) \right) \nn \\
&+&K_6 {\rm Tr}(v_\rho \alpha_\perp^\mu) {\rm Tr}(v^\rho \alpha_{\perp \mu})+ 
K_7{\rm Tr}(v^2 \alpha_{\perp \mu} \alpha_{\perp}^\mu)+K_8
{\rm Tr}(\alpha_{\perp}^2) {\rm Tr}(v^2) \nn \\
&+& 
K_9\{{\rm Tr}(v^2) \}^2+K_{10} {\rm Tr}(v^4) \nn \\
&+&i    \frac{g_{2p}}{f^2}T_1 \eta_0
{\rm Tr}\{(\xi M \xi- \xi^\dagger M \xi^\dagger) v^2 \} \nn \\
&+& i   \frac{g_{2p}}{f^2}T_2
\eta_0{\rm Tr}\{M(U-U^\dagger)\} {\rm Tr}(v^2) \nn \\
&+& T_3 i\frac{g_{2p}}{f^2}\frac{4B}{f^2} \eta_0
{\rm Tr} M(U+U^\dagger) {\rm Tr} M(U-U^\dagger) \nn \\
&+& T_4 \left(\frac{g_{2p}}{f^2} \right)^2 {\eta_0}^2 
\left(\aTr M(U-U^\dagger) \right)^2 +i T_5 \frac{4 B}{f^2}
\frac{g_{2p}}{f^2}\eta_0 {\rm Tr}(MUMU-MU^\dagger M U^\dagger)\nn \\
&+&
T_6 \left(\frac{g_{2p}}{f^2} \right)^2
\eta_0^2 \aTr(MUMU+MU^\dagger M U^\dagger-2 M^2) \nn \\
&+&i \frac{g_{2p}}{f^2}
\eta_0\bigl{[}T_7 {\rm Tr}\{M (D_{\mu}U(D^\mu U)^\dagger U-U^\dagger
D_{\mu}U(D^\mu U)^\dagger)\} \nn \\
&+&T_8
{\rm Tr}(M(U-U^\dagger)) 
{\rm Tr}(D_{\mu}U(D_{}^\mu U)^\dagger) \bigr{]}, 
\label{Count1}\eea
\bea
v_\mu&=&g_{\rho\pi\pi}
\left(V_\mu-\frac{\alpha_\mu}{g}\right),\label{Count2}\\
L_i&=&\lambda \Gamma_i+L_i^r (i=1-10),\label{Count3}\\
K_i&=&\lambda k_i + {K_i}^r (i=1-10),\label{Count4}\\
H_i&=&\lambda \Delta_i+H_i^r (i=1-2),\label{Count5}\\
T_i&=& \lambda t_i +T_i^r (i=1-8)\label{Count6},\\
\lambda&=&-\frac{1}{32\pi^2}(1+C_{UV}-\ln\mu^2),
\label{uvdiv}\\
C_{UV}&=&\displaystyle\frac{1}{2-\frac{d}{2}}-\gamma+\ln 4\pi.
\eea
In Eq.\ (\ref{Count1}), 
the contribution from singlet pseudoscalar is
omitted in the coefficients of
$\Gamma_6, \Gamma_8$ and $\Delta_2$.
We have also corrected the sign of $k_9$ and
$k_{10}$ in Ref. \cite{Kimura:2014wsa}.
\begin{table}[h!]
\begin{center}
\caption{The coefficients of the counter terms:
$k_i, \Gamma_i$ and $\Delta_i$.}
\begin{tabular}{cccc} \hline\hline
$k_1=1$ \hspace{3mm}& $t_1=-6$ & $\Gamma_1=\frac{2c^2+1}{32}$ & $\Delta_1=-\frac{1}{8}$
\\  \hline
$k_2=1$& $t_2=-2$ & $\Gamma_2=\frac{1+2c^2}{16}$ & 
$\Delta_2=\frac{5}{24}$
\\  \hline
$k_3=1$& $t_3=-\frac{11}{18}$& $\Gamma_3=\frac{3(c^2-1)}{16}$ &  \\ \hline
$k_4=\frac{3}{2}$& $t_4=-\frac{11}{9}$ & $\Gamma_4=\frac{c}{8}$ & \\ \hline 
$k_5=\frac{1}{2}$& $t_5=-\frac{5}{6}$ &  $ \Gamma_5=\frac{3c}{8}$ & \\ \hline
$k_6=4c$& $t_6=-\frac{5}{3}$ &   $\Gamma_6=\frac{11}{144}$ & \\ \hline
$k_7=6c$& $t_7=-\frac{3c}{2}$  &   $\Gamma_7=0$ & \\ \hline
$k_8=2c$& $t_8=-\frac{c}{2}$ &   
$\Gamma_8=\frac{5}{48}$ & \\ \hline 
$k_9=3$& &  $\Gamma_9=\frac{1}{4}$ &\\ \hline
$k_{10}=3$ & &$\Gamma_{10}=-\frac{1}{4}$  &\\ 
\hline\hline 
\end{tabular}
\end{center}
\label{table:2}
\end{table}
%%%%%%%%%%%%%%%%%%
%% Power Counting %%%
% Replace this section Oct.12th by Morozumi
%%%%%%%%%%%%%%%%%%
\section{Power counting with SU(3) breaking
and singlets}
\label{Power}
In this appendix, we show the power counting rule which is used to classify 
the interaction Lagrangian in Eq.\ (\ref{chiLag}) and counter terms in 
Eq.\ (\ref{Count1}).
Since we treat the electromagnetic correction due to the term proportional to $C$ in Eq.\ (\ref{chiLag}) only within tree level, in the following power counting, we do not take this term
into account.   
Since we employ the loop expansion due to pseudoscalar octet, the Lagrangian is organized as follows,
\bea
\sum_{n=0}^\infty {\cal L}^{(n)} {\hbar}^{n-1},
\eea
where we denote ${\cal L}^{(n)} $ as $n$ loop contribution.
We first evaluate  the superficial degree of divergence of the $n$ loop diagram of Nambu-Goldstone bosons using the interaction part of the tree level Lagrangian,
\bea
{\cal L}^{(0)}_{\rm int}=\frac{f^2}{4} {\rm Tr}(D_\mu U D^\mu U^\dagger)+ \frac{M_V^2}{g^2}{\rm Tr}(\alpha_\mu \alpha^\mu) 
-\frac{2{M_V}^2}{g} {\rm Tr}(V_\mu \alpha^\mu) \nn \\
+B {\rm Tr}[M(U+U^\dagger)] -i g_{2p} \eta_0 {\rm Tr}[M(U-U^\dagger)].
\label{eq:treeint}
\eea
The first two terms of Eq.(\ref{eq:treeint}) denote the interaction with the second derivatives among the Nambu-Goldstone  bosons.
The third term with the first derivative is the interaction between SU(3) octet vector mesons and SU(3) octet pseudoscalars. 
The other terms are the chiral breaking term which is proportional to the coefficient $B$ and the interaction term between SU(3) singlet $\eta_0$ and SU(3) octets. 
We compute the superficial degree of divergence $\omega$ for $N_L$  loop with $N_\chi$ insertions of the chiral breaking term and with $N_{\eta^0}$ $(N_{V_8})$ external pseudoscalar singlets
(vector meson octets) lines. It is given as follows,
\bea
\omega=4 N_L+2N_2+N_{V_8}-2N_I,
\label{eq:super}
\eea
where $N_2$ is the number of the vertex with second derivatives and $N_I$ denotes the numbers of the
propagators of pseudoscalar octets in the internal line.
It is related to the total number of the vertex ($N_v$) and the number of loop ($N_L$) as follows,
\bea
N_I=N_L+(N_v-1),
\label{eq:NI}
\eea
where $N_v$ is 
\bea
N_v=N_{\eta^0}+N_\chi+N_{V_8}+N_2.
\label{eq:Nv}
\eea
Substituting Eq.(\ref{eq:NI}) with Eq.(\ref{eq:Nv}) into
Eq.\ (\ref{eq:super}), one obtains the following formula,
\bea
\omega&=& 2 N_L +2 -N_{V_8}-2(N_{\eta^0}+N_\chi).
\eea
The ultraviolet divergence can occur when $\omega \ge 0$ and we obtain the following 
condition which the divergent diagrams satisfy, 
\bea
2 N_L+2 \ge N_{V_8} +2 (N_{\eta^0}+N_\chi).
\eea
The counter terms which subtract the divergence also satisfy the above condition on the
number of the external lines ($N_{\eta^0}, N_{V_8}$) and the powers of $B$  which correspond to $N_\chi$.   
Let us examine the types of the counter terms which are required within one-loop calculation by setting $N_L=1$.
Then the superficial degree of divergence is
\bea
\omega=4-N_{V_8}-2(N_{\eta^0}+N_\chi).
\eea
Note that the $\omega$ is equal to the number of the derivatives $\omega_0$  included in the counter terms. 
In Table \ref{tab:super}, we show  
$\omega_0(\ge 0)$, $N_\chi$;the powers of $B$, $N_{\eta_0}$
and $N_{V_8}$
in each 1-loop counter term.
We classify each counter term in Eq.\ (\ref{Count1}) according to these numbers
and show their coefficients.

Next we study the power counting of the interaction terms for singlet vector meson.
In contrast to the octet vector mesons,
the chiral invariant interaction of the singlet vector meson to the octet pseudoscalars with the first
derivative vanishes,
\bea
\phi^{0\mu} {\rm Tr}\left(V_\mu-\frac{\alpha_\mu}{g}\right)=0.
\eea
Therefore there is no tree level interaction for the singlet vector meson.
The interaction of the singlet vector meson with chiral breaking term
\bea
g_{1V} 
\phi^{\mu 0} {\rm Tr}(\xi M \xi+\xi^\dagger M \xi^\dagger)
\left(V_\mu-\frac{\alpha_\mu}{g}\right),
\label{eq:phi0}
\eea
is classified as the one loop level interaction since this term also includes a vector meson with the first derivative and the chiral breaking $M$, which has a structure similar to the one loop effective counter terms 
in Eq.\ (\ref{eq:octetcb}) for vector meson octets given below,  
\bea
&&C_1 \frac{2B}{f^2} {\rm Tr}\left\{\left(\xi M \xi+ \xi^\dagger M \xi^\dagger\right)
\left(V_\mu-\frac{\alpha_\mu}{g}\right)
\left(V^\mu-\frac{\alpha^\mu}{g}\right)\right\} \nn \\
&+&C_2 \frac{2B}{f^2} {\rm Tr}\left(\xi M \xi+ \xi^\dagger M \xi^\dagger\right)
{\rm Tr}\left\{\left(V_\mu-\frac{\alpha_\mu}{g}\right)\left(V^\mu-\frac{\alpha^\mu}{g}\right)\right\}.
\label{eq:octetcb}
\eea
\begin{table}
\caption{$(\omega_0, N_\chi, N_{\eta_0}, N_{V_8})$ for
1-loop counter terms.}
\begin{tabular}{ccccc} \hline \hline
$\omega_0$ & $N_\chi$ &$N_{\eta_0}$ & $N_{V_8}$& The coefficients of the counter terms \\ \hline
 $4$ & $0$ & $0$ & $0$ & $L_1, L_2, L_3, L_9,L_{10}, H_1,K_1,K_2,K_3,K_6,K_7,K_8,K_9, K_{10}$ \\
 $3$ & $0$ & $0$ & $1$ & $K_1,K_2,K_3,K_6,K_7,K_8,K_9,K_{10}$  \\
 $2$ & $1$ & $0$ & $0$ & $L_4, L_5,K_4, K_5$ \\
 $0$ & $2$ & $0$ & $0$ & $L_6, L_7, L_8, H_2 $\\
 $2$ & $0$ & $0$ & $2$ & $K_1,K_2,K_3,K_6,K_7,K_8,K_9, K_{10}$ \\
 $2$ & $0$ & $1$ & $0$ & $T_1,T_2, T_7,T_8$ \\
 $1$ & $0$ & $1$ & $1$ & $T_1,T_2$ \\
 $0$ & $0$ & $1$ & $2$ & $T_1,T_2$ \\
 $0$ & $1$ & $1$ & $0$ & $T_3,T_5$ \\
 $0$ & $0$ & $2$ & $0$ & $T_4,T_6$ \\
 $0$ & $1$ & $0$ & $2$ & $K_4,K_5$ \\
 $1$ & $1$ & $0$ & $1$ & $K_4,K_5$ \\
 $1$ & $0$ & $0$ & $3$ & $K_3, K_9, K_{10}$ \\
 $0$ & $0$ & $0$ & $4$ & $K_3, K_9, K_{10}$ \\ \hline \hline
\end{tabular}
\label{tab:super}
\end{table}
%%%%%%%%%%%%%%%%%%
%% Isospin breaking %%%
%%%%%%%%%%%%%%%%%%
\section{1-loop correction to self-energy for
$K^{*+}, K^{*0}$ and $\rho^+$}
\begin{figure}[h]
\begin{center}
\includegraphics[clip,width=11.5cm]{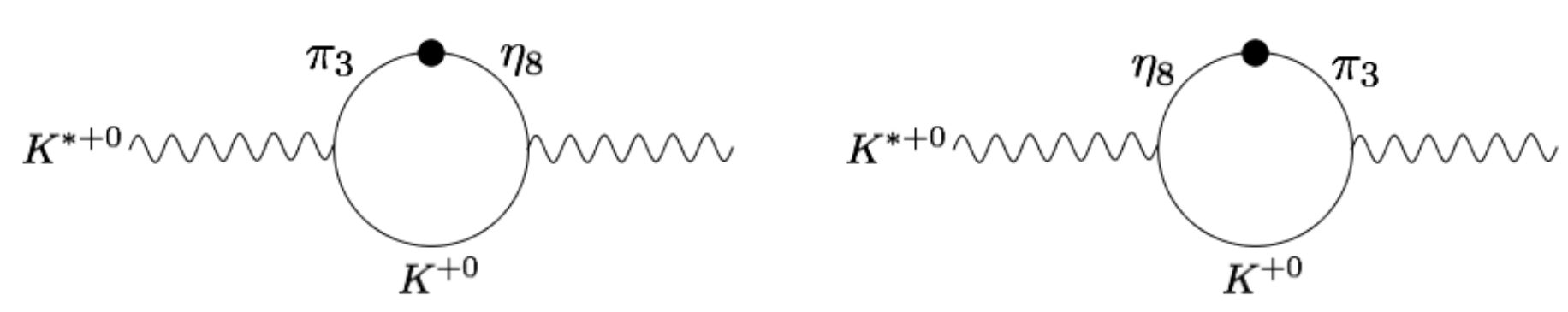}
\end{center}
\caption{Self-energy corrections to $K^{\ast + 0}$. The diagrams include the $\pi_3-\eta_8$ mixing due to the isospin breaking effect.}
\label{fig:pi0eta}
\end{figure}

In this appendix, we study self-energy corrections to  $K^{\ast + 0}$ mesons and charged $\rho$
meson taking SU(3) breaking into account.
The interaction Lagrangian for $V \to PP$ is given as,
\def\dc{\stackrel{\leftrightarrow}{\partial}}
\bea
{\mathcal L}^{VPP}&=&
-\frac{2g_{\rho\pi\pi}}{i} {\rm Tr}
(V_{\mu} [\Delta, \partial^{\mu} \Delta]) \nn \\
&=& i\frac{g_{\rho\pi\pi} }{2}
\left[
K^{\ast + \mu} \left(\hat{K}^- \dc_{\mu} \hat{\pi}_3+ 
\sqrt{3}
\hat{K}^- \dc_{\mu} \hat{\eta}_8 + \sqrt{2} \hat{\bar{K}}^0 \dc 
\hat{\pi}^- \right)
\right.   \nn \\
&&+ \left. K^{\ast 0 \mu} \left(-\hat{\bar{K}}^0 \dc_{\mu} \hat{\pi}_3+ 
\sqrt{3}
\hat{\bar{K}}^0 \dc_{\mu} \hat{\eta}_8 + \sqrt{2} \hat{K}^-  \dc_\mu 
\hat{\pi}^- \right)
\right.   \nn \\
&&+ \left. \rho^{+ \mu}
\left(2 \hat{\pi}^- \dc_\mu \hat{\pi}_3 +\sqrt{2} \hat{\bar{K}}^0
\dc_\mu \hat{K}^- \right) \right]+ h.c.,\\
\Delta&=&\frac{1}{2}\begin{pmatrix} \hat{\pi}_3+\displaystyle\frac{\hat{\eta}_8}{\sqrt{3}}
& \sqrt{2} \hat{\pi}^+  &\sqrt{2} \hat{K}^+ \\  
\sqrt{2} \hat{\pi}^- & -\hat{\pi}_3+\displaystyle\frac{\hat{\eta}_8}{\sqrt{3}} &\sqrt{2} \hat{K}^0 \\
\sqrt{2} \hat{K}^+ & \sqrt{2} \hat{\bar{K}}^0 & -2
\displaystyle\frac{\hat{\eta}_8}{\sqrt{3}} \end{pmatrix},
\eea
where $\Delta$ denotes the quantum fluctuation for the pseudoscalar octet
in the background field method \cite{Kimura:2014wsa}.
The isospin breaking leads to $\pi_3-\eta_8$ mixing and
the Feynman diagrams of the self-energy for $K^{*+0}$
are shown in Fig.\ \ref{fig:pi0eta}.
%%%%%%%%%%%%%
%%%%%%%%%%%%%
 $\pi_3-\eta_8$ mixing obtained from the chiral breaking term 
is given by 
the following Lagrangian,
\bea
{\cal L}&=&-M^2_{38}\hat{\pi}_3\hat{\eta}_8\label{mix38},
\\
M_{38}^2&=&\frac{1}{\sqrt{3}}(M^2_{K^+}-M^2_{K^0}).
\eea
We treat the mixing in Eq.\ (\ref{mix38}) as perturbation. The mixing insertion is denoted with black circles in Fig.\ \ref{fig:pi0eta}. 
Below, the amplitude corresponding to the diagrams in Fig.\ \ref{fig:pi0eta}
is shown,
\bea
&&\frac{g_{\rho\pi\pi}^2}{4} 
\int \frac{d^d k}{(2 \pi)^d i} \frac{(Q-2 k)_\mu (Q-2k)_\nu}{
((Q-k)^2-M^2_{\pi})((Q-k)^2-M^2_{\eta_8})(k^2-M_{K}^2)} 2 \sqrt{3} M^2_{38} \nn \\
&=&\frac{g_{\rho\pi\pi}^2}{2} \frac{M_{K^+}^2-M_{K^0}^2}{M^2_{\eta^8}-M^2_\pi}
(J_{\mu \nu}^{\eta_8 K}-J_{\mu \nu}^{\pi^0 K}) \nn \\
&=& 2g_{\rho\pi\pi}^2 \frac{M_{K^+}^2-M_{K^0}^2}{M^2_{\eta^8}-M^2_\pi}
 Q_{\mu\nu}(M^r_{K \eta_8}-M^r_{K\pi})
\nn \\
&& + g_{\rho\pi\pi}^2 (M_{K^+}^2-M_{K^0}^2)
g_{\mu \nu}\Bigl{(}2 \frac{L_{K \eta_8}-L_{K \pi}}{M^2_{\eta_8}-M_\pi^2}- \lambda- \frac{\mu_{\eta_8}-\mu_\pi}{M^2_{\eta_8}-M_\pi^2}\Bigr{)},\label{A4eq}
\eea
where $Q_{\mu\nu}$ is defined
in Eq.\ (\ref{Qmunudef}) and
$M_K$ denotes $M_{K^+}$ or $M_{K^0}$ and one uses the following
1-loop function,
\bea
J_{\mu \nu}^{QP}&=&\int \frac{d^d k}{(2 \pi)^d i} 
\frac{(Q-2 k)_\mu (Q-2k)_\nu}{
((Q-k)^2-M^2_{Q})(k^2-M_{P}^2)} \nn \\
&=&
Q_{\mu\nu} \left(4 M^r_{PQ}-\frac{2}{3} \lambda\right)
+g_{\mu \nu}(4 L_{PQ}-2 \lambda \Sigma_{PQ}- 2(\mu_Q+\mu_P)),
\label{eq:Jmunu}\\
\Sigma_{PQ}&=&M_P^2+M_Q^2.
\eea
In Eqs.\ (\ref{A4eq}, \ref{eq:Jmunu}),
$\lambda$ denotes the ultraviolet divergence defined
in Eq.\ (\ref{uvdiv}).
In Eq.\ (\ref{eq:Jmunu}),
$L_{PQ}$ and $M^r_{PQ}$ are functions given below,
\bea
M^r_{PQ}&=&\frac{1}{12 Q^2} \left(Q^2-2 \Sigma_{PQ}\right) \bar{J}_{PQ}
+ \frac{\Delta_{PQ}^2}{3 Q^4}\left[
\bar{J}_{PQ}-Q^2
\frac{1}{32 \pi^2} \left( \frac{\Sigma_{PQ}}{\Delta_{PQ}^2}+
2 \frac{M_P^2 M_Q^2}{\Delta_{PQ}^3} \ln \frac{M_Q^2}{M_P^2} \right)
\right] \nn \\
&&-\frac{k_{PQ}}{6}+\frac{1}{288 \pi^2}, \label{JPQ1} \\ 
L_{PQ}&=& \frac{\Delta_{PQ}^2}{4 s} \bar{J}_{PQ},\label{JPQ2}\\
k_{PQ}&=&\frac{(\mu_{P}-\mu_{Q}) f^2}{\Delta_{PQ}}.
\label{eq:ML}
\eea
In Eqs.\ (\ref{JPQ1}, \ref{JPQ2}),
$\bar{J}_{PQ}$ is a 1-loop scalar function of pseudoscalar
mesons with masses $M_P$ and $M_Q$.
Above the threshold $Q^2\geq (M_P+M_Q)^2$, it is given by,
\bea
\bar{J}_{PQ}(Q^2)&=&
\frac{1}{32 \pi^2}
\left[ 2 +\frac{\Delta_{PQ}}{Q^2} \ln 
\frac{M_Q^2}{M_P^2}-\frac{\Sigma_{PQ}}{\Delta_{PQ}} 
\ln \frac{M_Q^2}{M_P^2}\right.\nn \\
&&- \left.
\frac{\nu_{PQ}}{Q^2}
\ln
\frac{(Q^2+\nu_{PQ})^2-\Delta_{PQ}^2}{(Q^2-\nu_{PQ})^2-\Delta_{PQ}^2}
\right]+ \frac{i}{16 \pi} \frac{\nu_{PQ}}{Q^2},
\label{eq:Jbar}\\
\nu_{PQ}^2&=&Q^4-2 Q^2 \Sigma_{PQ}+ \Delta_{PQ}^2,
\label{eq:nuPQ}
\eea
while below the threshold $(M_P-M_Q)^2 \leq Q^2 \leq (M_P+M_Q)^2$,
\bea
\bar{J}_{PQ}(Q^2)&=&
\frac{1}{32 \pi^2}
\left[2 +\frac{\Delta_{PQ}}{Q^2} \ln 
\frac{M_Q^2}{M_P^2}-\frac{\Sigma_{PQ}}{\Delta_{PQ}} 
\ln \frac{M_Q^2}{M_P^2} \right. \nn \\
&&- \left. 2 \frac{\sqrt{-\nu_{PQ}^2}}{Q^2}
\left(
\arctan \frac{Q^2-\Delta_{PQ}}{\sqrt{-\nu_{PQ}^2}}
+\arctan\frac{Q^2+\Delta_{PQ}}{\sqrt{-\nu_{PQ}^2}}\right)
\right].
\label{eq:Jbar2}
\eea
We write inverse propagators of vector mesons as,
\bea
D_{V\mu\nu}^{-1}&=&(M_V^2+\delta A_{V})g_{\mu \nu} +\delta \tilde{B}_{V}  Q_{\mu} Q_{\nu},
\label{eq:invp}
\eea
where the metric part of the inverse propagator
consists of
the sum of tree-level mass $M_V$ and 
loop correction $\delta A_V$.
Using loop functions defined, we add 
the isospin breaking corrections
in Fig.\ \ref{fig:pi0eta}
to the calculation given in Ref. \cite{Kimura:2014wsa}.
We also take account of the
mass differences of $K^+-K^0$
and $\pi^+-\pi^0$, which
were not considered in the previous
study.
The self-energy
corrections to $K^{\ast +0}$ and $\rho^+$ mesons
are obtained as,
\bea
\delta \tilde{B}_{K^{\ast +}}&=& Z_V^{r}(\mu)+
g_{\rho\pi\pi}^2 \Bigr{[}2 M^r_{K^0 \pi^+}+
M^r_{K^+ \pi^0}+3 M^r_{K^+ \eta_8} +
2\frac{M^2_{K^+}-M^2_{K^0}}{M^2_{\eta_8}-M^2_{\pi}}(M^r_{K^+ \eta_8}-M^r_{K^+ \pi^0}) \Bigl{]} , \nn \\
\delta A_{K^{\ast +}}&=&\Delta A_{K^{\ast +}}+
C_1^r(\mu) M^2_{K^+} + C_2^r(\mu) (2\bar{M}^2_K+M^2_\pi)-Q^2 Z_V^r(\mu), 
\label{deltaAB_Kp} \\
\Delta A_{K^{\ast +}}&=&-Q^2g_{\rho\pi\pi}^2 \Bigr{[}2 M^r_{K^0 \pi^+}+
M^r_{K^+ \pi^0}+3 M^r_{K^+ \eta_8}+
2\frac{M^2_{K^+}-M^2_{K^0}}{M^2_{\eta_8}-M^2_{\pi}}(M^r_{K^+ \eta_8}-M^r_{K^+ \pi^0}) \Bigl{]} \nn \\
&+&g_{\rho\pi\pi}^2 
\Bigr{[} 2 \left(\frac{M^2_{K^+}-M^2_{K^0}}{M^2_{\eta_8}-M^2_\pi}
\right)
\left(L_{K^+ \eta_8}-L_{K^+\pi^0}-\frac{f^2}{2}(\mu_{\eta_8}-\mu_\pi)
\right)
+2L_{K^0 \pi^+}+L_{K^+ \pi^0} \nn \\
&+&3L_{K^+ \eta_8}
-\frac{f^2}{2}\{2(\mu_{K^0}+\mu_\pi)+\mu_{K^+}+\mu_\pi+3(\mu_{K^+}+\mu_{\eta_8})  \}\Bigr{]},\label{DeltaAB_Kp}\\
\delta\tilde{B}_{K^{\ast 0}}&=& Z_V^{r}(\mu)+
g_{\rho\pi\pi}^2 \Bigr{[}2 M^r_{K^+ \pi^-}+
M^r_{K^0 \pi^0}+3 M^r_{K^0 \eta_8}-2\frac{M^2_{K^+}-M^2_{K^0}}{M^2_{\eta_8}-M^2_{\pi}}(M^r_{K^0 \eta_8}-M^r_{K^0 \pi^0}) \Bigl{]} , \nn \\
\delta A_{K^{\ast 0}}&=&\Delta A_{K^{*0}}+
C_1^r(\mu) M^2_{K^0} + C_2^r(\mu) (2\bar{M}^2_K+M^2_\pi)-Q^2Z_V^r(\mu),
\label{DeltaAB_K0} \\
\Delta A_{K^{\ast 0}}&=&
-Q^2g_{\rho\pi\pi}^2 \Bigr{[}2 M^r_{K^+ \pi^-}+
M^r_{K^0 \pi^0}+3 M^r_{K^0 \eta_8}
-2\frac{M^2_{K^+}-M^2_{K^0}}{M^2_{\eta_8}-M^2_{\pi}}(M^r_{K^0 \eta_8}-M^r_{K^0 \pi^0}) \Bigl{]}\nn\\
&+&g_{\rho\pi\pi}^2
\Bigr{[}-2\left(\frac{M^2_{K^+}-M^2_{K^0}}{M^2_{\eta_8}-M^2_\pi}
\right)
\left(L_{K^0 \eta_8}-L_{K^0 \pi^0}-\frac{f^2}{2}(\mu_{\eta_8}-\mu_\pi)
\right)\nn\\
&+&
2 L_{K^+ \pi^-}+L_{K^0 \pi^0}+3L_{K^0 \eta_8}
-\frac{f^2}{2}\{2(\mu_{K^+}+\mu_\pi)+\mu_{K^0}+\mu_\pi+3(\mu_{K^0}+\mu_{\eta_8})  \}\Bigr{]},\qquad\quad
\label{deltaAB_K0} \\
\delta \tilde{B}_{\rho^{+}}&=&Z_V^{r}(\mu)+
g_{\rho\pi\pi}^2 (4 M^r_{\pi^+ \pi^0}+
2 M^r_{K^+ \bar{K}^0}), \nn \\
\delta A_{\rho^{+}}
&=&\Delta A_{\rho^+}+C_1^r(\mu)M_{\pi}^2+C_2^r(\mu)(2\bar{M}^2_K+M^2_\pi)-Q^2Z_V^r(\mu),
\label{deltaAB_rho}\\
\Delta A_{\rho^+}&=&
-Q^2g_{\rho\pi\pi}^2 (4 M^r_{\pi^+ \pi^0}+2 M^r_{K^+ \overline{K}^0})\nn\\
&&+
2g_{\rho\pi\pi}^2 \left( 2 L_{\pi^+ \pi^0}
+ L_{K^+ \bar{K}^0} -\frac{f^2}{2}( 4 \mu_\pi+\mu_{K^+}+\mu_{K^0}) \right)\label{DeltaAB_rhop}.
\eea
%%%%%%%%%%%%%%%%%%%%%%%%%%%%%%%%%%%%%%%%
%%  Vector Gamma Mixing
%%%%%%%%%%%%%%%%%%%%%%%%%%%%%%%%%%%%%%%%
%%%%%%%%%%%%%%%%
\section{Proof of the relation for $V-A$ mixing vertex}
\label{Proof}
%%%%%%%%%%%%%%%%
In this section, we show that
the metric tensor part of the two-point functions
for the $V-A$ mixing satisfies the relation in
Eq.\ (\ref{MassTwoP}).
Multiplying Eq.\ (\ref{tobeproven}) by $O^T_V$,
one can find,
\bea
&&\Pi^{VA}=O_V^T \Pi^{V^0 A} \nn \\
                  &&=-\frac{1}{g}
\begin{pmatrix} 
 M_\rho^2 O_{V11} + M_{V\rho 8}^2 O_{V21} + M_{V0 \rho}^2 O_{V31} +\displaystyle\frac{M_{V\rho 8}^2 O_{V11}+M_{V88}^2 O_{V21} + M_{V08}^2 O_{V31}}{\sqrt{3}}\\
 M_\rho^2 O_{V12} + M_{V\rho 8}^2 O_{V22} + M_{V0 \rho}^2 O_{V32} +\displaystyle\frac{M_{V\rho 8}^2 O_{V12}+M_{V88}^2 O_{V22} + M_{V08}^2 O_{V32}}{\sqrt{3}}\\
 M_\rho^2 O_{V13} + M_{V\rho 8}^2 O_{V23} + M_{V0 \rho}^2 O_{V33}+\displaystyle\frac{M_{V\rho 8}^2 O_{V13}+M_{V88}^2 O_{V23} + M_{V08}^2 O_{V33}}{\sqrt{3}}
\end{pmatrix}\label{ProofEq}.\quad\quad
\eea
Meanwhile, the diagonalization of the mass matrix
leads to,
\bea
\begin{pmatrix} M^2_{\rho} & M^2_{V \rho 8} & M^2_{V0\rho} \\
M^2_{V \rho 8}& M^2_{V88} & M^2_{V08}\\
M^2_{V0\rho} & M^2_{V08} & M^2_{0V}\\ 
\end{pmatrix} O_V
=O_V\begin{pmatrix} \cM_1^2 & 0 & 0 \\
                             0 & \cM_2^2 & 0 \\
                             0 & 0 & \cM_3^2 \end{pmatrix}
                             .
\eea
In the above equation, the matrix elements for $(i,j)=(1,I), (2,I)$ indicate the following relations,
\bea
M^2_{\rho} O_{V1I} + M^2_{V \rho 8} O_{V2I}+ M^2_{V0\rho} O_{V3I}&=& \cM_I^2 O_{V1I},\label{Eigen1}\\
M^2_{V \rho 8} O_{V1I} + M^2_{V 88} O_{V2I}+ 
M^2_{V08} O_{V3I}&=& \cM_I^2 O_{V2I}.\label{Eigen2}
\begin{comment}
M^2_{\rho} O_{V11} + M^2_{V \rho 8} O_{V21}+ M^2_{V0\rho} O_{V31}= M_1^2 
O_{V11} , \label{Proof1}\\
M^2_{V \rho 8} O_{V11} + M^2_{V 88} O_{V21}+ M^2_{V08} O_{V31}= M_1^2 
O_{V21} , \\
M_\rho^2 O_{V12} + M_{V\rho 8}^2 O_{V22} + M_{V0 \rho}^2 O_{V32}=M_2^2 O_{V12} , \\
M_{V\rho 8}^2 O_{V12}+ 
M_{V88}^2 O_{V22} + M_{V08}^2 O_{V32}=M_2^2 O_{V22} , \\
M_\rho^2 O_{V13} + M_{V\rho 8}^2 O_{V23} + M_{V0 \rho}^2 O_{V33}=M_3^2 O_{V13} , \\
M_{V\rho 8}^2 O_{V13}+ 
M_{V88}^2 O_{V23} + M_{V08}^2 O_{V33}=M_3^2 O_{V23}.\label{Proof6}
\end{comment}
\eea
Plugging Eqs.\ (\ref{Eigen1}, \ref{Eigen2})
into Eq.\ (\ref{ProofEq}), one can find
that the relation in Eq.\ (\ref{MassTwoP}) is satisfied.
%%%%%%%%%%%%%%%%%
%%   Inserted          %%%
%%%%%%%%%%%%%%%%%
\section{1-loop correction to self-energy
for $\pi^+, K^+$ and $K^0$}
\label{APPcharged}
In this appendix, the radiative correction to charged pseudoscalar masses is discussed.
Background field method is used to evaluate the
chiral loop correction\cite{Donoghue:1992dd, Gasser:1984gg}.
Kinetic terms and 1-loop corrected masses 
in effective Lagrangian are given as,
\b
\mathcal{L}_{\mathrm{eff}}&=&
\sum_{P}^{\pi^+, K^+, K^0}\left(\frac{1}{Z_{P}}\partial_\mu P_f\partial^\mu
\bar{P}_f-M_{P}^{2}P_f\bar{P}_f\right)
\label{C0}\\
&=&\sum_{P}^{\pi^+, K^+, K^0}(\partial_\mu P\partial^\mu 
\bar{P}-M_{P}^{\prime2}P\bar{P}).
\label{C1}\e
In Eq.\ (\ref{C0}), we denote $P_f$ as the pseudoscalar
in original flavor basis and
the coefficient of the kinetic term is,
\b
\frac{1}{Z_{P}}&=&1-Z_{P(1)},\\
Z_{\pi^+(1)}&\sim&-8\left(\frac{M_{\pi^+}^2+2\bar{M}_{K}^2}{f^2}L_4^r
+\frac{M_{\pi^+}^2}{f^2}L_5^r
\right)+2c(2\mu_{\pi^+}+\bar{\mu}_{K}),\\
Z_{K^+(1)}&\simeq&Z_{K^0(1)}\sim-8\left(\frac{M_{\pi^+}^2+2\bar{M}_{K}^2}{f^2}L_4^r
+\frac{\bar{M}_{K}^2}{f^2}L_5^r\right)+c\left(\frac{3}{2}\mu_{\pi^+}
+\frac{3}{2}\mu_{88}+3\bar{\mu}_{K}\right).
\e
In Eq.\ (\ref{C0}), normalization of the
kinetic term is slightly deviated from unity
due to 1-loop correction.
In order to canonically normalize $Z_P$
in Eq.\ (\ref{C0}), one should implement the transformation
in the following form as,
\b
\overset{\small (-)}{P_f}&=&\sqrt{Z_P}\overset{\small (-)}{P},\label{TransP}\\
\sqrt{Z_P}&\sim&1+\frac{Z_{P(1)}}{2}.
\e
Using transformation in Eq.\ (\ref{TransP}),
one obtains Lagrangian in Eq.\ (\ref{C1}).
We keep linear order of the small quantities
(we neglect  quadratic terms with respect to isospin breaking and
1-loop correction multiplied by isospin violation).
The masses in Eq.\ (\ref{C1}) are,
\b
M_{\pi^+}^{\prime 2}&\simeq&\left(M_{\pi^+}^2\right)_\mathrm{tr}
\left[1+(4c-3)\mu_{\pi^+}-\frac{1}{3}\mu_{88}+2(c-1)\bar{\mu}_{K} \right. \nn \\
&& \qquad\qquad\qquad\left. -8\frac{M_{\pi^+}^2+2\bar{M}_{K}^2}{f^2}L_{46}^r-8\frac{M_{\pi^+}^2}{f^2}L_{58}^r
\right]+\Delta_{\mathrm{EM}},
\label{isoMpi}
\\
M_{K^+}^{\prime 2}&\simeq&\left(M_{K^+}^2\right)_{\mathrm{tr}}
\left[1+\frac{3}{2}(c-1)\mu_{\pi^+}+\frac{1}{6}(9c-5)\mu_{88}+3(c-1)
\bar{\mu}_{K} \right. \nn \\
&& \qquad\qquad\qquad\left. -8\frac{M_{\pi^+}^2+2\bar{M}_{K}^2}{f^2}L_{46}^r
 -8\frac{\bar{M}_{K}^2}{f^2}L_{58}^r
\right]+\Delta_{\mathrm{EM}},\label{isoMKp}\\
M_{K^0}^{\prime 2}&\simeq&\left(M_{K^0}^2\right)_{\mathrm{tr}}
\left[1+\frac{3}{2}(c-1)\mu_{\pi^+}+\frac{1}{6}(9c-5)\mu_{88}
+3(c-1)\bar{\mu}_{K} \right. \nn \\
&& \qquad\qquad\qquad\left. -8\frac{M_{\pi^+}^2+2\bar{M}_{K}^2}{f^2}L_{46}^r
 -8\frac{\bar{M}_{K}^2}{f^2}L_{58}^r
\right], \label{isoMK}
\e
where low energy constants are denoted as,
\b
L_{46}^r = L_4^r-2L_6^r,\quad
L_{58}^r = L_5^r-2L_8^r,\quad
\Delta_{\mathrm{EM}} = \frac{2C}{9f^2}.
\e
In Eqs.\ (\ref{isoMpi}-\ref{isoMK}), $(M_{P}^2)_{\mathrm{tr}}$
denotes the tree-level mass parameters,
and in the loop corrections, pseudoscalar masses are identified with physical masses
expressed as $M_{\pi^+}^2$ and $\bar{M}_{K}^2$ defined
in Eq.\ (\ref{MKbardef})
since their difference gives rise to minor correction in
Eqs.\ (\ref{isoMpi}-\ref{isoMK}).
The tree-level mass parameters in r.h.s. of
Eqs.\ (\ref{isoMpi}-\ref{isoMK})
are given as Gell-Mann-Oakes-Renner (GMOR) relation,
\b
\left(M_{\pi^+}^2\right)_\mathrm{tr}=\frac{2B(m_u+m_d)}{f^2},\label{GMOR1}\;\;
\left(M_{K^+}^2\right)_{\mathrm{tr}}=\frac{2B(m_u+m_s)}{f^2},\;\;
\left(M_{K^0}^2\right)_{\mathrm{tr}}=\frac{2B(m_d+m_s)}{f^2}.\;\;\quad
\e
One can clarify that the 1-loop masses are renormalization scale invariant. Therefore, we find that
the following equation is satisfied,
\b
\frac{\partial M_{\pi^+}^{\prime2}}{\partial \ln \mu}
=\frac{\partial M_{K^+}^{\prime2}}{\partial \ln \mu}
=\frac{\partial M_{K^0}^{\prime2}}{\partial \ln \mu}=0.
\e
%%
%%%%%%%%%%%%%%%
%%%   Neutral Part  %
%%%%%%%%%%%%%%%
\section{1-loop correction to self-energy for
neutral pseudoscalars}
\label{APPneutral}
In this appendix, the radiative correction to pseudoscalar masses
is evaluated for neutral particles.
As analogous to the previous section, the background field method
is used to evaluate the quantum correction.
We consider the framework in which
chiral octet loop correction is taken into account.
Masses and kinetic terms of pseudoscalars
in 1-loop corrected effective Lagrangian
are written as,
\b
\mathcal{L}_{\mathrm{eff}}&=&
\frac{1}{2}(\partial_\mu\pi_3, \partial_\mu\eta_8, \partial_\mu\eta_0)\frac{1}{Z}
(\partial^\mu\pi_3, \partial^\mu\eta_8, \partial^\mu\eta_0)^T
-\frac{1}{2}(\pi_3, \eta_8, \eta_0)M^{2}(\pi_3, \eta_8, \eta_0)^T\quad \label{D0}\\
&&(\mathrm{SU}(3) \:\mathrm{eigenstate})\nn\\
&=&\frac{1}{2}\partial_\mu\pi_3^R\partial^\mu\pi_3^R
+\frac{1}{2}\partial_\mu\eta_8^R\partial^\mu\eta_8^R
+\frac{1}{2}\partial_\mu\eta_0^R\partial^\mu\eta_0^R
-\frac{1}{2}(\pi_3^R, \eta_8^R, \eta_0^R)M^{\prime2}(\pi_3^R, \eta_8^R, \eta_0^R)^T\quad 
\label{D1}\\
&&(\mathrm{kinetic}\:\mathrm{terms}\:\mathrm{rescaled})\nn\\
&=&\frac{1}{2}\partial_\mu\pi^0\partial^\mu\pi^0
+\frac{1}{2}\partial_\mu\eta\partial^\mu\eta+\frac{1}{2}\partial_\mu\eta^\prime\partial^\mu\eta^\prime
-\frac{1}{2}(\pi^0, \eta, \eta^\prime)\mathrm{diag}(M^{\prime2}_{\pi^0}, M^{\prime2}_{\eta}, 
M^{\prime2}_{\eta^\prime})(\pi^0, \eta, \eta^\prime)^T.\quad \;\;\qquad\label{D2}\\
&&(\mathrm{mass}\:\mathrm{eigenstate})\nn
\e
In Eq.\ (\ref{D0}), the coefficient of kinetic
terms is given as a $3\times 3$ matrix,
\b
\frac{1}{Z}&\simeq&
\begin{pmatrix}
1-Z_{33(1)}&0&0\\
0&1-Z_{88(1)}&0\\
0&0&1
\end{pmatrix},\label{matZ}\\
Z_{33(1)}&\sim&Z_{\pi^+(1)},\label{Z33}\\
Z_{88(1)}&=&-8\left(\displaystyle\frac{M_{\pi^+}^2+2\bar{M}_K^2}{f^2}L_4^r+\displaystyle\frac{M_{88}^2}{f^2}L_5^r\right)
+6c\mu_{\bar{K}}, \label{Z88}\\
M_{88}^2&=&\frac{2(M_{K^+}^2)_{\mathrm{tr}}+2(M_{K^0}^2)_{\mathrm{tr}}-
(M_{\pi^+}^2)_{\mathrm{tr}}}{3}.
\e
The matrix in Eq.\ (\ref{matZ}) implies that the kinetic terms in Eq.\ (\ref{D0})
are slightly deviated from unity with 1-loop correction.
The mass matrix denoted as $M^2$
in Eq.\ (\ref{D0}) indicates
the 1-loop corrected mixing mass matrix in the SU(3) basis.
To normalize the kinetic terms in Eq.\ (\ref{D0}) canonically, 
one should implement basis transformation,
\b
\begin{pmatrix}
\pi_3 \\
\eta_8\\
\eta_0
\end{pmatrix}
&=&\sqrt{Z}
\begin{pmatrix}
\pi_3^R \\
\eta^R_8\\
\eta^R_0
\end{pmatrix},
\qquad\quad\sqrt{Z}\sim\begin{pmatrix}
\sqrt{Z_1^\pi}&0&0\\
0& \sqrt{Z_2^\pi}&0\\
0&0&1\end{pmatrix}, \label{Dp}\\
\sqrt{Z^{\pi}_1}&=&1+\displaystyle\frac{Z_{33(1)}}{2}
\sim \sqrt{Z_{\pi^+}} , \label{WFR1}\\
\sqrt{Z^{\pi}_2}&=&1+\displaystyle\frac{Z_{88(1)}}{2}.
\label{WFR2}
\e
The transformation in Eq.\ (\ref{Dp}) relates the basis in
Eq.\ (\ref{D0}) to
one given in Eq.\ (\ref{D1}).
Thus, the kinetic terms are canonically normalized in
Eqs.\ (\ref{D1}-\ref{D2}).
One diagonalizes the mass matrix in Eq.\ (\ref{D1})
and obtains Lagrangian with mass eigenstates in
Eq.\ (\ref{D2}).
The mass matrix given in Eq.\ (\ref{D1}) is
expressed as,
\b
M^{\prime 2}=
\begin{pmatrix}
M_{33}^{\prime2} & M_{38}^{\prime2} & M_{30}^{\prime2} \\
* &  M_{88}^{\prime2} & M_{80}^{\prime2} \\
* &  *& M_{00}^{2}
\end{pmatrix}.\label{D3}
\e
In the above mass matrix, the 1-loop corrected masses are
denoted with primes.
We ignore quadratic terms with respect to
the small quantities
so that the 1-loop corrected masses in
Eq.\ (\ref{D3}) are simplified as,
\b
M_{33}^{\prime 2}&\simeq&
\left(M_{\pi^+}^2\right)_{\mathrm{tr}}\left[1+(4c-3)\mu_{\pi^+}
-\frac{1}{3}\mu_{88}
+2(c-1)\bar{\mu}_{K}
\right. \nn \\
&&\left. \qquad\qquad\qquad-8\frac{M_{\pi^+}^2+2\bar{M}_{K}^2}{f^2}L_{46}^r
-8\frac{M_{\pi^+}^2}{f^2}L_{58}^r\right],\label{D4}\\
M_{38}^{\prime 2}&\simeq&M_{38}^2
=\frac{(M_{K^+}^2)_{\mathrm{tr}}
-(M_{K^0}^2)_{\mathrm{tr}}}{\sqrt{3}},\label{D5}\\
M_{88}^{\prime 2}&\simeq&M_{88}^2-M_{\pi^+}^2\mu_{\pi^+}
-\left(\frac{16\bar{M}^2_K-7M_{\pi^+}^2}{9}
\right)\mu_{88}+\frac{2}{3}\left(9cM_{88}^2+3M_{\pi^+}^2-8\bar{M}_{K}^2\right)\bar{\mu}_{K}
\nn\\
&&-\frac{8M_{88}^2}{f^2}(M_{\pi^+}^2+2\bar{M}_{K}^2)L_{46}^r-
\frac{8}{f^2}M_{88}^4L_5^r
+\frac{16}{3f^2}[8(M_{\pi^+}^2-\bar{M}_{K}^2)^2L_7^r  \nn \\
&& +(M_{\pi^+}^4+2(M_{\pi^+}^2
-2\bar{M}_{K}^2)^2)L_8^r], \label{D6}\\
M_{30}^{\prime 2}&\simeq&M_{30}^2=-\hat{g}_{2p}\left[(M_{K^+}^2)_{\mathrm{tr}}
-(M_{K^0}^2)_{\mathrm{tr}}\right],\label{D7}\\
M_{80}^{\prime 2}&\simeq&M_{80}^2+\frac{\hat{g}_{2p}}{\sqrt{3}}\left[3M_{\pi^+}^2\mu_{\pi^+}+\frac{1}{3}\left(5M_{\pi^+}^2-8\bar{M}_{K}^2\right)\mu_{88}+2\left\{3c(M_{\pi^+}^2-\bar{M}_{K}^2)
\right.\right. \nn \\
&& \left.\left. +(3M_{\pi^+}^2-4\bar{M}_{K}^2) \right\} \bar{\mu}_{K}\right] 
-2M_{80}^2\left[\frac{M_{\pi^+}^2+2\bar{M}_{K}^2}{f^2}T_{34}^r-\frac{2\bar{M}_{K}^2}{f^2}T_5^r+\frac{2M_{88}^2}{f^2}L_5^r\right],\qquad
\label{D8}
\e 
where $T_{34}^r=2L_4^r-T_3^r$ and $\hat{g}_{2p}=fg_{2p}/B$. 
Since 1-loop corrected masses in Eqs.\ (\ref{D4}-\ref{D8})
are invariant under renormalization,
one can confirm that they satisfy the following relation,
\b
\frac{\partial M_{33}^{\prime2}}{\partial\ln\mu}=
\frac{\partial M_{38}^{\prime2}}{\partial\ln\mu}=
\frac{\partial M_{88}^{\prime2}}{\partial\ln\mu}=
\frac{\partial M_{30}^{\prime2}}{\partial\ln\mu}=
\frac{\partial M_{80}^{\prime2}}{\partial\ln\mu}
=0.
\e
Comparing Eqs.\ (\ref{isoMpi}-\ref{isoMK}) with
Eqs.\ (\ref{D4}, \ref{D5}, \ref{D7}),
we find that the neutral mass matrix elements are
related to charged ones as,
\b
M_{33}^{\prime2}&\sim& M^{\prime2}_{\pi^+}-\Delta_{\mathrm{EM}},
\label{D10}\\
M_{38}^{\prime2}&\sim& \frac{1}{\sqrt{3}}(M^{\prime2}_{K^+}
-M^{\prime2}_{K^0}-\Delta_{\mathrm{EM}}),\label{D11}\\
M_{30}^{\prime2}&\sim& -\hat{g}_{2p}(M^{\prime2}_{K^+}
-M^{\prime2}_{K^0}-\Delta_{\mathrm{EM}})\label{D12}.
\e
Using Eqs.\ (\ref{D10}-\ref{D12}),
one can write the mass matrix in Eq.\ (\ref{D3}) as,
\bea
M^{\prime2}=
\begin{pmatrix}
M^{\prime2}_{\pi^+}-\Delta_{\mathrm{EM}}&
\displaystyle\frac{1}{\sqrt{3}}(M^{\prime2}_{K^+}
-M^{\prime2}_{K^0}-\Delta_{\mathrm{EM}})&
-\hat{g}_{2p}(M^{\prime2}_{K^+}
-M^{\prime2}_{K^0}-\Delta_{\mathrm{EM}})\\
*&M_{88}^{\prime2}& M_{80}^{\prime2}\\
*&*& M_{\pi^0}^{\prime2}+M_{\eta}^{\prime2}
+M_{\eta^\prime}^{\prime2}-
M^{\prime2}_{\pi^+}-\Delta_{\mathrm{EM}}
-M_{88}^{\prime2}
\end{pmatrix},
\nn\\
\label{MASSMATRIX}
\eea
where we utilized the relation of trace for the mass matrix,
\bea
M_{00}^2=M_{\pi^0}^{\prime2}+M_{\eta}^{\prime2}
+M_{\eta^\prime}^{\prime2}-
M_{33}^{\prime2}
-M_{88}^{\prime2}.
\eea
Provided that physical masses, $M_{\pi^+}^{\prime2}, 
M_{K^+}^{\prime2}, M_{K^0}^{\prime2}, M_{\pi^0}^{\prime2}, M_{\eta}^{\prime2}$
and $M_{\eta^\prime}^{\prime2}$ are given as experimental values,
the mass matrix in Eq.\ (\ref{MASSMATRIX})
is written in terms of four model parameters:
$(\hat{g}_{2p}, \Delta_{\mathrm{EM}}, M_{88}^{\prime2},
M_{80}^{\prime2})$.
The mixing matrix should be determined to
diagonalize the mass matrix in Eq.\ (\ref{MASSMATRIX}) as,
\b
O^T
M^{\prime 2}
O
=\mathrm{diag}(M_{\pi^0}^{\prime2}, M_{\eta}^{\prime2}, M_{\eta^\prime}^{\prime2}).\label{mixingdetem}
\e
%%%%%%%%%%%%%%%%%
%%  Decay Constant %%
%%%%%%%%%%%%%%%%%
\section{1-loop correction to decay constants of
$\pi^+$ and $K^+$}
\label{decayC}
In this appendix, 1-loop corrected decay constants
are analyzed for charged pseudoscalars.
The decay constants are defined with parameterizing
matrix elements as,
\b
\braket{\pi^+(p)|\bar{u}\gamma_\mu\gamma_5 d|0}|_{1-\mathrm{loop}\:\mathrm{order}}
&=&i\sqrt{2}f_{\pi^+}p_\mu,\\
\braket{K^+(p)|\bar{u}\gamma_\mu\gamma_5 s|0}|_{1-\mathrm{loop}\:\mathrm{order}}
&=&i\sqrt{2}f_{K^+}p_\mu.
\e
One can find that 1-loop corrected decay constants are
related with wave function renormalization in
Eq.\ (\ref{TransP})
in the following as,
\b
f_{\pi^+}=\frac{f}{\sqrt{Z_{\pi^+}}},\qquad
f_{K^+}=\frac{f}{\sqrt{Z_{K^+}}}\label{KpDec},
\e
where one can show that the quantities in
Eq.\ (\ref{KpDec}) are renormalization scale invariant, {\it i.e.},
\b
\frac{\partial}{\partial \mathrm{ln}\mu}f_{\pi^+}
=\frac{\partial}{\partial \mathrm{ln}\mu}f_{K^+}
=0.
\e
Equation\ (\ref{KpDec}) leads to the 
relation between the decay constants of pion and one
for kaon in Eq.\ (\ref{rafKfpi}).
%%%%%%%%%%%%%%%%%%%%%%%%%%%%%%%%%%%%
\section{Wess-Zumino-Witten term}
\label{WZter}
%%%%%%%%%%%%%%%%%%%%%%%%%%%%%%%%%%%%
In this appendix, we give the expression for the WZW term.
As suggested in Ref.\ \cite{Wess:1971yu},
one can obtain
the WZW term by integrating the Bardeen form
anomaly.
Following Ref. \cite{Fujikawa:2004cx},
we can write the expression
for the WZW term,
\bea
\mathcal{L}_{\mathrm{WZ}}&=&-\frac{N_c}{16\pi^2}
\epsilon^{\mu\nu\rho\sigma}\int_0^1\mathrm{d}t
\mathrm{tr}\frac{\pi}{f}\left[V_{\mu\nu}(t)V_{\rho\sigma}(t)+\frac{1}{3}
A_{\mu\nu}(t)A_{\rho\sigma}(t)\right.\nn\\
&&\left.-\frac{8i}{3}(V_{\mu\nu}(t)
A_\rho(t)A_\sigma(t)+
A_\mu(t)V_{\nu\rho}(t)A_\sigma(t)+
A_\mu(t)A_\nu(t)V_{\rho\sigma}(t))\right.\nn\\
&&\left.-\frac{32}{3}A_{\mu}(t)A_{\nu}(t)A_{\rho}(s)A_{\sigma}(t)
\right],
\label{eq:A-1}
\eea
where $N_c=3$ indicates the color factor.
The notations in Eq.\ (\ref{eq:A-1}) are defined as,
\bea
V_\mu(t)&=&\frac{1}{2}(\xi(t)V_\mu\xi(-t)+\xi(-t)V_\mu\xi(t)
+\xi(t)A_\mu\xi(-t)-\xi(-t)A_\mu\xi(t)\nn\\
&&
-i\xi(t)\partial_\mu\xi(-t)-i\xi(-t)\partial_\mu\xi(t))
,\label{5.1}\\
A_\mu(t)&=&\frac{1}{2}(\xi(t)V_\mu\xi(-t)-\xi(-t)V_\mu\xi(t)
+\xi(t)A_\mu\xi(-t)+\xi(-t)A_\mu\xi(t)\nn\\
&&-i\xi(t)\partial_\mu\xi(-t)+i\xi(-t)\partial_\mu\xi(t)),\\
V_{\mu\nu}(t)&=&\partial_\mu V_\nu(t)-\partial_\nu V_\mu(t)
+i[V_\mu(t), V_\nu(t)]+i[A_\mu(t), A_\nu(t)],\\
A_{\mu\nu}(t)&=&\partial_\mu A_\nu(t)-\partial_\nu A_\mu(t)
+i[V_\mu(t), A_\nu(t)]+i[A_\mu(t), V_\nu(t)],\\
\xi(t)&=&e^{-i(1-t)\pi/f}\label{eq:A-6}.
\eea
The expressions given in Eqs.\ (\ref{eq:A-1}-\ref{eq:A-6})
are all defined in Minkowski space-time.
%%%%%%%%%%%%%%%%%%%%%
%%%  Wess-Zumino ends %%
%%%%%%%%%%%%%%%%%%%%%
%%%%%%%%%%%%%%%%%
%%%%%%%%%%%%%%%%%
%%%%%%%%%%%%%%%%%
\section{Form factors at $O(p^4)$ for  $\tau^- \to K_s \pi^- \nu$ decay}
\label{Formfactor}
The vector form factors for $\tau^- \to K_s \pi^- \nu $ decays 
including $\eta^0$ meson loop were computed in
Ref.\ \cite{Kimura:2014wsa}.
 In the present work, we do not include the loop contribution of the singlet meson. 
Below, we show the expression for form factors without the singlet meson loop contribution, which is used to calculate the decay spectrum of 
 $\tau^- \to K_s \pi^- \nu $.
The expression in this appendix can be obtained from
Eqs.\ (40-54) in \cite{Kimura:2014wsa}, by simply setting  the mixing angle ( $\theta_{08}$) between the
singlet meson  and the octet meson to be zero.
In the formulas shown below, the isospin breaking effect and the mixing induced CP violation of the neutral kaon system is also neglected.  By ignoring CP violation due to the mixing,
$K_s$ is CP even state,
\bea
|K_s \rangle =\frac{1}{\sqrt{2}}( |K^0 \rangle -|\bar{K}^0 \rangle), 
\eea  
where $|\bar{K}^0 \rangle =-CP|K^0 \rangle$.
Since $\Delta S=\Delta Q=-1$ rule holds,  one finds the following relation,
\bea
\langle K_s \pi^- |\bar{s} \gamma_\mu u |0 \rangle
= -\frac{1}{\sqrt{2}} \langle \bar{K}^0 \pi^- |\bar{s} \gamma_\mu u|0 \rangle.   \label{eq:ffrelation}
\eea
One defines the vector form factors  for 
$\bar{K}^0 \pi^- (K^- \pi^0)$ and its CP conjugate states,
\bea
\langle \bar{K}^0 \pi^-|\bar{s} \gamma_\mu u|0 \rangle&=&F_V^{\bar{K}^0 \pi^-}(q_\mu-Q_\mu
\frac{\Delta_{K \pi}}{Q^2}) + F_S^{\bar{K}^0 \pi^-}\frac{Q_\mu}{Q^2} , \nn \\
\langle K^0 \pi^+|\bar{u} \gamma_\mu s|0 \rangle&=&F_V^{K^0 \pi^+}(q_\mu-Q_\mu
\frac{\Delta_{K \pi}}{Q^2}) + F_S^{K^0 \pi^+}\frac{Q_\mu}{Q^2},\nn \\
\langle K^- \pi^0|\bar{s} \gamma_\mu u|0 \rangle&=&F_V^{K^- \pi^0}(q_\mu-Q_\mu
\frac{\Delta_{K \pi}}{Q^2}) + F_S^{K^- \pi^0}\frac{Q_\mu}{Q^2} ,\nn \\
\langle K^+ \pi^0|\bar{u} \gamma_\mu s |0 \rangle&=&F_V^{K^+ \pi^0}(q_\mu-Q_\mu
\frac{\Delta_{K \pi}}{Q^2}) + F_S^{K^+ \pi^0}\frac{Q_\mu}{Q^2}.
\label{eq:CP}
\eea
Since under  CP transformation, the charged currents are related to each other as follows,
\bea
CP (\bar{s} \gamma_\mu u) (CP)^{-1}=-\bar{u} \gamma^\mu s,
\eea
the following relations among the form factors are derived,
\bea
F_V^{\bar{K}^0 \pi^-}&=&-F_V^{K^0 \pi^+}, F_S^{\bar{K}^0 \pi^-}=-F_S^{K^0 \pi^+}, \nn \\
F_V^{K^- \pi^0}&=&-F_V^{K^+ \pi^0}, F_S^{K^- \pi^0}=-F_S^{K^+ \pi^0}.
\label{eq:ff}
\eea 
In the isospin limit, we also obtain the relations,
\bea
F_V^{\bar{K}^0 \pi^-}&=&\sqrt{2}F_V^{K^- \pi^0},
F_S^{\bar{K}^0 \pi^-}=\sqrt{2}F_S^{K^- \pi^0},\nn \\
F_V^{K^0 \pi^+}&=&\sqrt{2}F_V^{K^+ \pi^0},
F_S^{K^0 \pi^+}=\sqrt{2}F_S^{K^+ \pi^0}.
\label{eq:isospin}
\eea 
Using Eq.\ (\ref{eq:ffrelation}), Eq,\ (\ref{eq:CP}), Eq.\ (\ref{eq:ff}) and  Eq.\ (\ref{eq:isospin}), one can relate the form factor of $K_s \pi^-$ of Eq.\ (\ref{eq:ffrelation}) to
that of  $K^+ \pi^0$,
\bea
\langle K_s \pi^- |\bar{s} \gamma_\mu u |0 \rangle= \langle K^+ \pi^0|\bar{u} \gamma_\mu s |0 \rangle.
\eea
The contribution to the form factors is divided into two parts. One of them comes from
1 PI diagrams and the other comes from the diagrams which include the propagator of  
$K^\ast$ meson,
\bea
F_V^{K^+\pi^0}&=&F_V^{1 PI}+ F_V^{K^\ast}, 
\label{F_V}\\
F_S^{K^+\pi^0}&=&F_S^{1 PI}+ F_S^{K^\ast}.
\label{F_S}
\eea
Each contribution to form factors is given below (See also \cite{Kimura:2014wsa}),
\bea
F_V^{1 PI}&=&-\frac{1}{\sqrt{2}}(1-\frac{M_V^2}{2 g^2 f^2})
          +\frac{1}{\sqrt{2}}
%%%%%%%%%%%%%%%%%%%%%%%%%%%%%%%%%%%%%%%%
\Bigl{[}
-\frac{3 c}{2} (H_{K \pi}+H_{K \eta_8}) 
+\frac{c M_V^2}{8 g^2 f^2}
(10 \mu_K + 3 \mu_{\eta_8} + 11 \mu_\pi)\nn \\
&-&\frac{3}{8}
\left(\frac{M_V^2}{g^2 f^2}\right)^2 (H_{K \pi}+H_{K \eta_8}+\frac{2\mu_K+\mu_\pi+\mu_{\eta_8}}{2})-\frac{C_5^{r}}{2} \frac{Q^2}{f^2} \nn \\
&+& \frac{M_V^2}{2 g^2 f^2} \Bigl\{ \frac{M_V^2}{2 g^2 f^2}K_4^{r} \frac{m_K^2}{f^2}- 4 L_5^{r} \frac{\Sigma_{K \pi}}{f^2}+\frac{2 m_K^2+m_\pi^2}{f^2} (\frac{M_V^2}{2 g^2 f^2} K_5^{r}-8 L_4^{r})
\Bigr\}  \Bigr{]}, \\ 
F_V^{K^\ast}
&=&-\frac{1}{2\sqrt{2}g} \frac{M_V^2}{M_V^2+\delta A_{K^\ast}}
 \left[ 4E + \sqrt{2}\frac{G+Q^2{\mathcal H}}{f^2}
 - \frac{M_V^2}{gf^2}
\right] , 
\label{FKs_V}\\
F_S^{1 PI}&=&\frac{1}{\sqrt{2}} \frac{1}{Q^2} \Bigl{[}(1-\frac{M_V^2}{2g^2f^2})
\Bigl\{-\frac{\Delta_{K \pi} \bar{J}_{K \pi}}{8f^2}\{
5 c Q^2-(5c-3) \Sigma_{K \pi}\}
+\frac{
\Delta_{K \eta_8} \bar{J}_{K \eta_8}
}
{8f^2}
\{3c Q^2 -(3c-1) \Sigma_{K \pi}\} \Bigr\} \nn \\
&&
+\frac{3 \Delta_{K \pi}}{8f^2}(1-\frac{M_V^2}{2g^2f^2})^2 
\{\frac{\Delta_{K \pi}^2}{s} \bar{J}_{K \pi} 
+\frac{\Delta_{K \eta_8}^2}{s} 
\bar{J}_{K \eta_8} 
\}   \Bigr{]} \nn \\
&+&\frac{1}{\sqrt{2}}\frac{\Delta_{K \pi}}{Q^2}
\Bigl{[}-(1-\frac{M_V^2}{2 g^2 f^2})+\frac{c}{4} Q^2 \frac{3 \mu_{\eta_8}+2 \mu_K-5 \mu_\pi}{\Delta_{K \pi}}
+c \frac{M_V^2}{8 g^2 f^2}(10 \mu_K +3 \mu_8+11\mu_\pi)\nn \\
&-&\frac{3}{16}
\left(\frac{M_V^2}{g^2 f^2}\right)^2(2 \mu_K+\mu_\pi+\mu_{\eta_8}) -4 L_5^{r} 
\frac{Q^2}{f^2} \nn \\
&+& \frac{M_V^2}{2 g^2 f^2} 
\Bigl\{ \frac{M_V^2}{2 g^2 f^2}K_4^{r} \frac{m_K^2}{f^2}- 4 L_5^{r} \frac{\Sigma_{K \pi}}{f^2}+\frac{2 m_K^2+m_\pi^2}{f^2} (\frac{M_V^2}{2 g^2 f^2} K_5^{r}-8 L_4^{r})
\Bigr\} \Bigr{]},\\
F_S^{K^\ast}
&=& -\frac{1}{2\sqrt{2}g}\frac{\Delta_{K \pi}}{Q^2} 
\frac{M_V^2}{M_V^2+\delta A_{K^\ast}+ Q^2\delta \tilde{B}_{K^\ast}}
\left[ 4(E+Q^2{\mathcal F}) + \sqrt{2}\frac{G}{f^2}
-\frac{M_V^2}{gf^2}
\right],
\label{FKs_S}
\eea
where $G,{\mathcal H}, E$ and ${\mathcal F}$ are given as,
\bea
G&=&
\frac{1}{\sqrt{2}g}\{M_V^2+
\delta A_{K^\ast}+Q^2 \delta \tilde{B}_{K^\ast}-\frac{3 M_V^2}{2f^2}(L_{K \pi}+L_{K \eta_8})\}, \\
{\mathcal H}&=& \frac{1}{\sqrt{2}g}\{Z^r_V-2 g C^{r}_4-\delta \tilde{B}_{K^\ast}+ \frac{3 M_V^2}{2f^2}(M^{r}_{K \pi}
+M^{r}_{K \eta_8})\},\\
E&=&\frac{M_V^2}{4 gf^2}-\frac{g}{2 M_V^2}\{
(\delta A_{K^\ast}+ Q^2 \delta \tilde{B}_{K^\ast})(1-\frac{M_V^2}{2 g^2 f^2})
-C_1^{r}m_K^2-C_2^r(2 m_K^2+m_\pi^2) \} \nn \\
 &+& \frac{M_V^2}{16 gf^2}\{-3 (2 \mu_K+\mu_\pi+\mu_{\eta_8})
+c(10 \mu_K+3\mu_{\eta_8}+11 \mu_\pi)-32 L_4^r\frac{2 m_K^2+m_\pi^2}{f^2}
-16 L_5^r \frac{\Sigma_{K \pi}}{f^2} \} \nn \\
&+& \{ \frac{g}{2 M_V^2}(1-\frac{M_V^2}{2 g^2 f^2})(\delta \tilde{B}_{K^\ast}-Z^r_V)
+\frac{C_3^{r}}{8f^2} \} Q^2, \\
{\mathcal F}&=&-\{ \frac{g}{2 M_V^2}(1-\frac{M_V^2}{2 g^2 f^2})(\delta \tilde{B}_{K^\ast}-Z^r_V)
+\frac{C_3^{r}}{8f^2} \} 
        + \frac{M_V^2}{8 g f^4}\frac{1}{Q^2}
\{ \Sigma_{K \pi}( \frac{3}{4}\bar{J}_{K \pi}+\frac{1}{12}\bar{J}_{K \eta_8}) \nn \\
&+&c (Q^2-\Sigma_{K \pi})(\frac{5}{4} \bar{J}_{K \pi}+
\frac{1}{4} \bar{J}_{K \eta_8})
\}.
\eea
We obtain $\delta \tilde{B}_{K^\ast}$ and $\delta A_{K^\ast}$ in the above equations  
by taking the isospin limit of 
Eq.\ (\ref{DeltaAB_Kp}) and Eq.\ (\ref{deltaAB_Kp}) and they are given respectively as
follows,
\bea
\delta \tilde{B}_{K^\ast}&=& Z_V^{r}(\mu)+3
g_{\rho\pi\pi}^2 \Bigr{[} M^r_{K \pi}+ M^r_{K \eta_8} 
 \Bigl{]} , \nn \\
\delta A_{K^\ast}&=& \Delta A_{K^{\ast +}}+
C_1^r(\mu) M^2_{K^+} + C_2^r(\mu) (2\bar{M}^2_K+M^2_\pi)-Q^2 Z_V^r(\mu),
\label{eq:isospinAB}
\eea
where $\Delta A_{K^{\ast}}$ is given by,
\bea
\Delta A_{K^{\ast}}&=&- 3Q^2g_{\rho\pi\pi}^2 \Bigr{[} M^r_{K \pi}
+ M^r_{K \eta_8}\Bigl{]}+ 3g_{\rho\pi\pi}^2 
\Bigr{[} 
L_{K \pi}+L_{K \eta_8}
-\frac{f^2}{2}\{2 \mu_{K}+\mu_\pi+ \mu_{\eta_8}  \}\Bigr{]}.
\label{eq:isospinDA}
\eea
%%%%%%%%%%%%%%%%%%%%
%%%    Reference      %%%%%
%%%%%%%%%%%%%%%%%%%%

\end{document}